\documentclass[preprint]{aastex}

\def\micron{ $\mu$m}
\def\ISO{$ISO$}
\def\IRAS{$IRAS$}
\def\lesssim{\mathrel{\hbox{\rlap{\hbox{\lower4pt\hbox{$\sim$}}}\hbox{$<$}}}}
\def\gtrsim{\mathrel{\hbox{\rlap{\hbox{\lower4pt\hbox{$\sim$}}}\hbox{$>$}}}}
\def\arcsec{\hbox{$^{\prime\prime}$}}

\shorttitle{Optical identification of \ISO~FIR sources in the
Lockman Hole}
\shortauthors{Oyabu et al.}

\begin{document}
\title{Optical identification of \ISO~far-infrared sources in the
Lockman Hole using a deep VLA 1.4 GHz continuum survey}
\author{S. Oyabu\altaffilmark{1,2,*}, Min
  S. Yun\altaffilmark{2,3}, 
  T. Murayama\altaffilmark{4}, D.B. Sanders\altaffilmark{5},
  K. Kawara\altaffilmark{1}, 
  Y. Taniguchi\altaffilmark{4}, S. Veilleux\altaffilmark{6},
  H. Okuda\altaffilmark{7,8}, 
  H. Matsuhara\altaffilmark{7}, L.L. Cowie\altaffilmark{5},
  Y. Sato\altaffilmark{2,9},
   K. Wakamatsu\altaffilmark{10} \and
  Y. Sofue\altaffilmark{1}}
\altaffiltext{1}{Institute of Astronomy, the University of Tokyo, 2-21-1,
  Osawa, Mitaka, Tokyo, 181-0015, Japan}
\altaffiltext{2}{Visiting Astronomer, WIYN, National Optical
Astronomy Observatory, which is operated by the Association of
Observatories for Research in Astronomy, Inc. (AURA) under cooperative
agreement with the National Science Foundation.}
\altaffiltext{3}{Department of Astronomy, University of
  Massachusetts, 619 Lederle Graduate Research Center, Amherst, MA
  01003}
\altaffiltext{4}{Astronomical Institute, Tohoku University, Aoba,
  Sendai 980-77, Japan}
\altaffiltext{5}{Institute for Astronomy, University of Hawaii, 2680
  Woodlawn Drive, Honolulu, HI 96822, USA}
\altaffiltext{6}{Department of Astronomy, University of Maryland,
  College Park, MD 20742}
\altaffiltext{7}{Institute of Space and Astronautical Science, Japan
  Aerospace Exploration Agency, 
  3-1-1 Yoshinodai, Sagamihara, Kanagawa, 229-8510, Japan}
\altaffiltext{8}{Gunma Astronomical Observatory, Gunma 377-0702, Japan}
\altaffiltext{9}{National Astronomical Observatory of Japan, Mitaka,
  Tokyo 181-8588, Japan}
\altaffiltext{10}{Faculty of Engineering, Gifu University, Gifu
  501-1193, Japan}
\altaffiltext{*}{Present Address: Institute of Space and Astronautical
  Science, Japan Aerospace Exploration Agency,
    3-1-1 Yoshinodai, Sagamihara, Kanagawa 229-8510, Japan}
\email{oyabu@ir.isas.jaxa.jp}
\begin{abstract}
  By exploiting the far-infrared(FIR) and radio correlation, we have
  performed the
  Likelihood-Ratio analysis to identify optical counterparts to 
  the far-infrared
  sources that have been found in an area of $\sim 0.9$ deg$^2$ during
  the \ISO~deep far-infrared survey in the Lockman Hole. 
  New ground-based observations have been conducted to build up the catalogs 
  of radio and optical objects, which include a deep VLA observation at
  1.4 GHz, optical $R-$ \& $I$-band imaging on the Subaru 8m and UH 2.2m
  telescopes, and optical spectroscopy on the Keck II 10m and WIYN 3.5m
  telescopes.  
  This work is based on the FIR sample consisting of 116 and 20
  sources that are selected with the criteria of $F^C(90\mu\mathrm{m})
  \ge 43$ mJy or $F^C(170\mu\mathrm{m}) \ge 102$ mJy, respectively,
  where $F^C$ is the flux-bias corrected flux.  
  Using the likelihood ratio 
  analysis and the associated reliability, 44 FIR sources have been 
  identified with radio sources. Optical identification of the 44 FIR/radio 
  association is then conducted by using the accurate radio positions.
  Redshifts have been obtained for 29 out of 44 identified sources. 
  One hyper-luminous infrared galaxy (HyLIRG) with $L_{FIR} > 
  10^{13}\mathrm{L}_{\odot}$ and four ultraluminous infrared galaxies
  (ULIRGs) with $L_{FIR} = 10^{12-13}\mathrm{L}_{\odot}$ are
  identified in our sample while the remaining 24 FIR galaxies
  have $L_{FIR} < 10^{12}\mathrm{L}_{\odot}$.  The space density
  of the FIR sources at 
  $z = 0.3-0.6$ is $4.6\times 10^{-5}$Mpc$^{-3}$, which is 460 times  
  larger than the local value, implying a rapid evolution of the ULIRG 
  population. Most of \ISO~FIR sources 
  have $L(1.4 GHz)/L(90\mu\mathrm{m})$ similar to
  star-forming galaxies ARP 220 and M82, indicating that the star-formation is
  the dominant mechanism for their FIR and radio luminosity. 

  At least seven of our FIR sources show evidence for the presence of
  an active galactic nucleus (AGN)
  in optical emission lines, radio continuum excess, 
  or X-ray activity.  Three out of five (60\%) of the ULIRG/HyLIRGs 
  are AGN galaxies, suggesting that the AGN fraction among the 
  ULIRG/HyLIRG population may not change significantly between $z\sim 0.5$ 
  and the present epoch.
  Five of the seven AGN galaxies are within the ROSAT X-ray survey
  field, and two are within the XMM-Newton survey fields. X-ray
  emission has been detected in only one source, 1EX030, which is
  optically classified as a quasar. 
  The non-detection in the XMM-Newton 2-10 keV band suggests a very
  thick absorption column density of $3 \times 10^{24} \mathrm{cm}^2$
  or $A_V \sim 1200$ mag obscuring the central source of the two AGN galaxies.
  Several sources have an extreme FIR luminosity relative to 
  the optical $R$-band, $L(90\mu\mathrm{m})/L(R) > 500$, which 
  is rare even among the local ULIRG population.  While source
  confusion or blending might offer an explanation in some cases,
  they may represent a new population of galaxies with an extreme 
  activity of star formation in an undeveloped stellar system -- 
  i.e., formation of bulges or young ellipticals.
\end{abstract}
\keywords{infrared:galaxies --- galaxies:evolution ---  galaxies: starburst}


\section{Introduction}

The detection of the far-infrared Cosmic Infrared Background (CIRB)
radiation by FIRAS and DIRBE on COBE was the important step toward
understanding the physical property of the cosmological
far-infrared (FIR) radiation. 
The CIRB was interpreted as the integrated emission by dust in the
distant galaxies \citep{puget,fixsen,hauser} which set a relevant
constraint on the evolution of cosmic sources.
The CIRB is 10 times brighter than the expected intensity based on the
assumption that the infrared emissivity of galaxies does not change
with cosmic time \citep{takeuchi,franceschini} 
and has comparable contribution to the total intensity as expected
from the optical counts from the
Hubble deep field and the Subaru deep field\citep{totani2,totani}.
Therefore the CIRB's excess suggests that
galaxies in the past were much more {\it active} in the
FIR and that a substantial fraction of the total energy
emitted by high-redshift galaxies were absorbed by dust 
and re-emitted at long wavelengths.

The SCUBA bolometer camera\citep{holland} on the sub-millimeter telescope JCMT 
was able to resolve at least half of the CIRB 
at wavelength of 850\micron\ into
a population of very luminous infrared galaxies at $z\gtrsim 1$ 
\citep{smail97,hugh,barger,blain}.  ISOCAM
mid-infrared surveys \citep{rowancam,flores,mann} have also
reported a higher infrared luminosity density (thus a higher star 
formation rate) at $0.5\lesssim z \lesssim 1$ than estimated 
by previous optical studies. 

\IRAS~was able to detect infrared galaxies only to moderate redshifts
($z \sim 0.1$) with the exception of a few hyperluminous and/or
lensed objects such as FSC~10214+4724 \citep{rowan2}. 
The improvement of the sensitivity and the extension to the
longer wavelength (170\micron) in FIR with ISOPHOT instrument 
onboard the {\it Infrared Space Observatory} (\ISO)
provides us with a new tool to study FIR emission from galaxies at
the higher-redshifts than \IRAS, and 
the exploration of the ``optically dark side'' of
the star formation history through a deep FIR survey 
was the obvious next step.

The spectral energy distribution(SED) of actively star-forming galaxies
peaks at $\lambda \sim 100$\micron. 
The ``negative'' $k$-correction makes 
observations at wavelengths longer than this FIR peak 
advantageous for detecting high-redshift galaxies.
Furthermore, such measurements give the total luminosity without any 
model-dependent 
bolometric correction. Therefore several deep surveys were
undertaken with the ISOPHOT at 90\micron~and/or 170\micron.  
We performed a deep FIR survey of two fields in the Lockman Hole
region in both 90\micron~and 170\micron\ bands as a part of
the Japanese/UH cosmology project \citep{kawara}. 
A 170\micron~survey of two
fields in the southern Marano area and two fields in the northern
ELAIS fields with a combined area of 4 square degrees constitute the
FIRBACK program \citep{puget2}. A 90\micron~survey in ELAIS
\citep{efstathiou} covered 11.6 square degrees. 

In order to explore the nature of \ISO~FIR sources in the
Lockman Hole fields, we have identified counterparts to the
sources at optical and radio wavelengths. We obtained
their photometric characteristics and 
measured their redshifts in order to understand the 
genuine nature of \ISO~FIR sources in the Lockman Hole fields. 
Section 2 describes the observations and the data, 
Section 3 presents the method and results of the source
identification, and Section 4 describes the
discussions of \ISO~FIR sources in the Lockman Hole fields. 
The summary is presented in Section 5.  Appendix describes
the comparison of our catalogs with those in \citet{rodighiero1} and
\citet{rodighiero2}, which reduced same data with their own method.

Throughout the paper, a flat Universe with $H_0=70
\mathrm{km}\ \mathrm{s}^{-1}\ \mathrm{Mpc}^{-1}, \Omega_{M}=0.3$ and
$\Omega_{\lambda}=0.7$ is adopted.

\section{Data}

\subsection{ISO far-infrared catalogs}

\begin{figure}[htbp]
  \plottwo{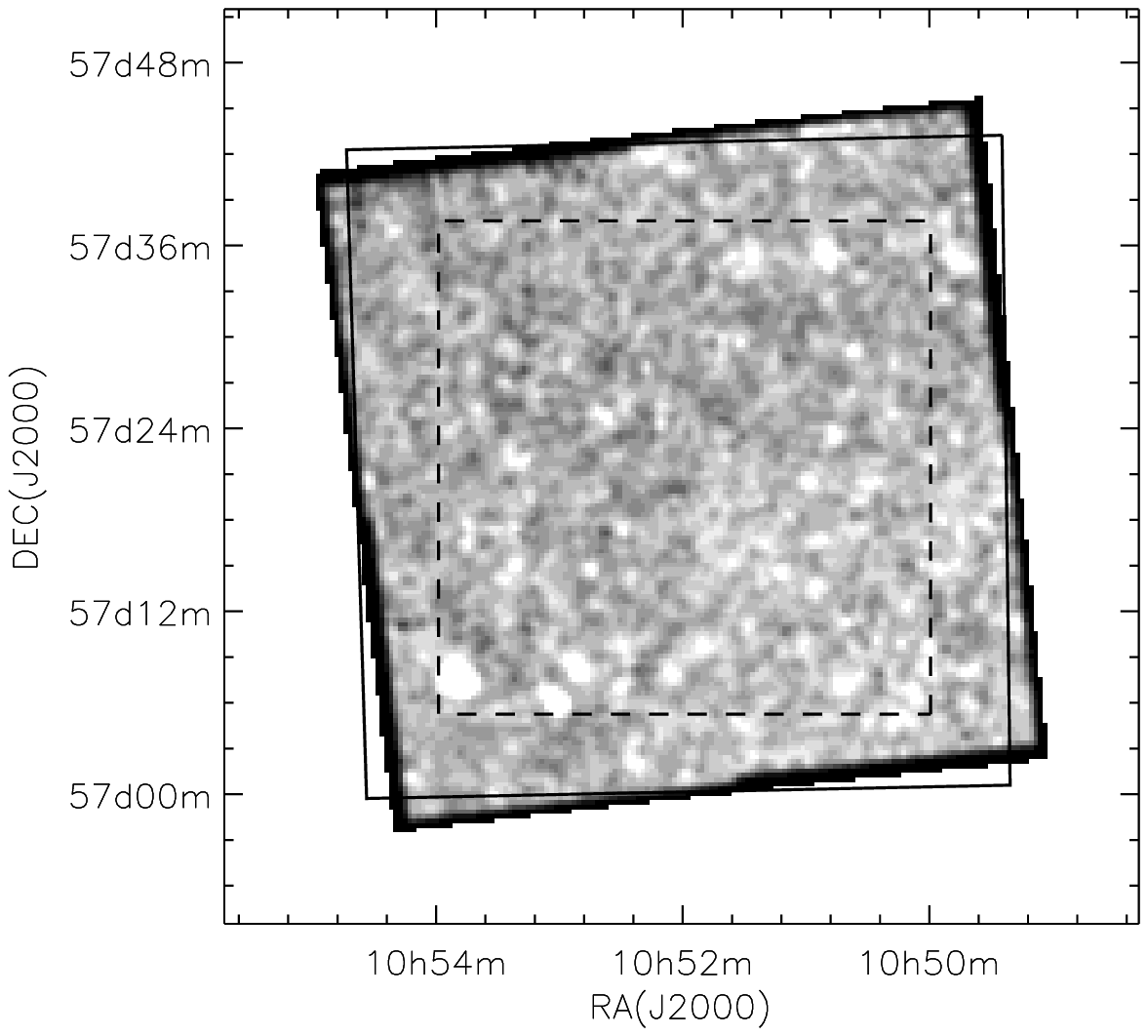}{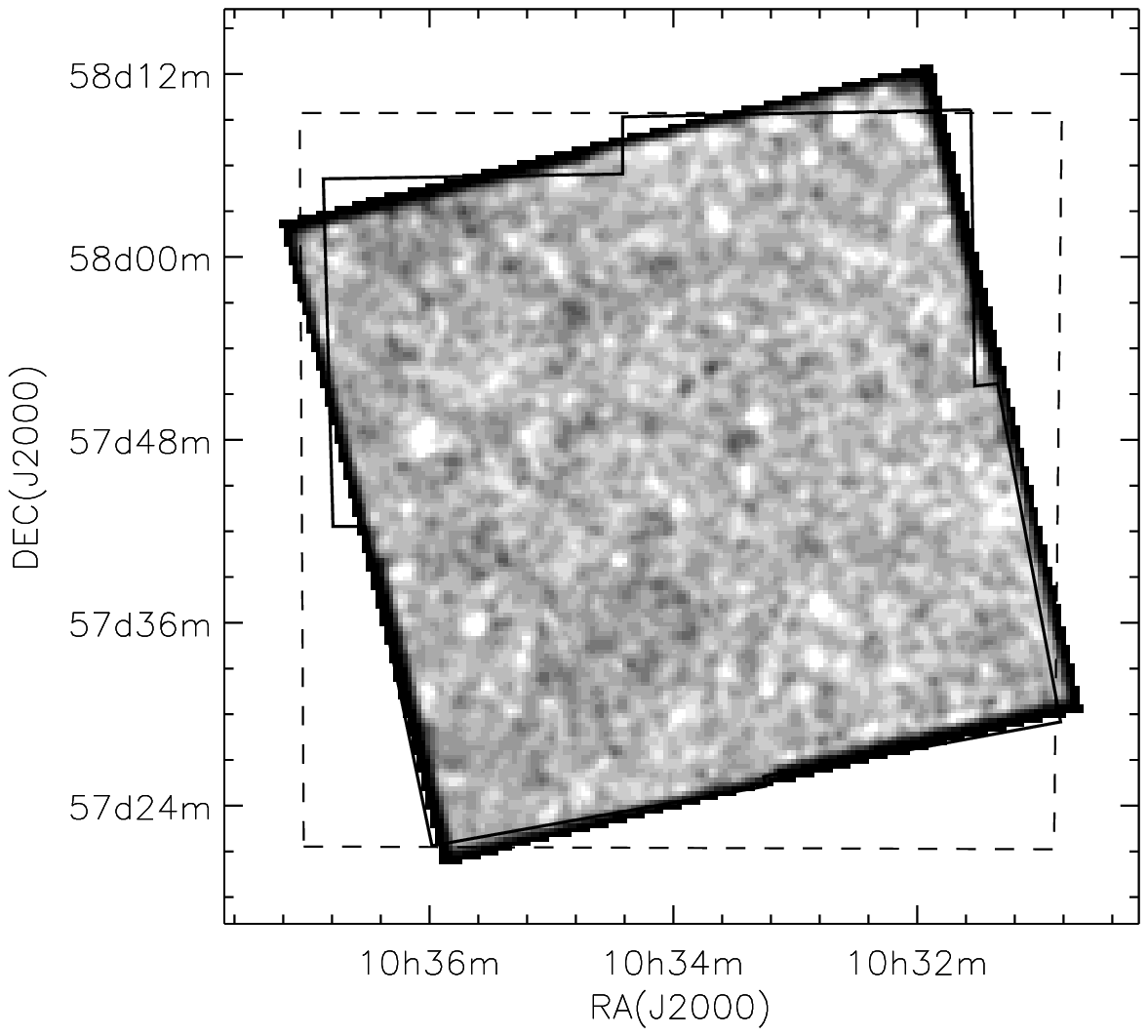}
  \caption{Survey areas superimposed on the \ISO~90\micron~images in the 
  Lockman Hole. The LHEX field is shown on the left panel and LHNW on the right. 
  The 170\micron~fields are shown by solid lines, while dashed lines
  present the areas of our 
  VLA 1.4 GHz sources which are used in this paper.}
  \label{fig:map}
\end{figure}

Our FIR survey was performed in the ISOPHOT bands C\_90 (90\micron~for
the reference wavelength) and
C\_160 \citep[170\micron; see ][]{kawara,matsuhara,kawara2} in the Lockman
Hole, where the HI column density is measured to be the lowest
in the sky \citep{lockman} and thus the confusion noise due to the 
infrared cirrus is expected to be the lowest. 
The survey includes
two fields named LHEX and LHNW, each of which covered approximately 
44\arcmin$\times$44\arcmin square area. One of the fields, LHEX, was
also the target of the ROSAT 
Lockman Hole ultra-deep HRI survey \citep{hasinger}. 
The 90\micron~and 170\micron~observations with \ISO \citep{kessler} and
the data reduction are described in detail by \citet{kawara,kawara2}. As shown
in Figure \ref{fig:map}, the ISOPHOT instrument \citep{lemke}
was used to map a total area of $\sim$0.9 square degrees, consisting of two
44\arcmin $\times$ 44\arcmin~fields.

The $IRAF$\footnote{IRAF is distributed by NOAO, which is
operated by AURA, Inc., under contract to the NSF.} DAOPHOT package,
which has been developed to perform stellar photometry in crowded
fields\citep{stetson}, 
was used to perform the source extraction from the FIR maps. The 
survey images are
very crowded, with two or more sources frequently appear blended.
The FWHM measurements show that most bright 
sources subtends no more than two detector pixels,
implying these sources are detected as point sources.  
The final 90\micron\ and 170\micron~source catalogs with
a signal to noise ratio of three or greater include 
223 and 72 sources, respectively \citep{kawara2}. 
UGC 06009\footnote{If more
  accurate fluxes are measured in UGC 06009, the fluxes of FIR
  sources should be re-scaled accordingly.}, 
which is the only IRAS source locating within our survey
fields, was used for the flux scaling assuming
F(90\micron)=1218 mJy and F(170\micron)=1133 mJy.
The flux density of the cataloged sources ranges between 40 mJy 
and 400 mJy at 90\micron~and between 90 mJy and 410 mJy
at 170\micron.

The surface number densities of \ISO~FIR sources are sufficiently high 
(10-20 beams per source)
that the derived quantities such as flux, position, completeness, and
detection limits are significantly affected by source
confusion. To evaluate these effects, \citet{kawara2} performed
a set of simulations by adding artificial sources to the 
observed FIR maps. These simulations have shown
that the measured fluxes of faint sources are indeed significantly
overestimated and thus the correction for the flux bias is
important.  
The following expressions are used for the correction: 
\begin{eqnarray}
  \label{equ:flux}
  F(90\mu\mathrm{m}) = 66+0.78 \times F^C(90\mu\mathrm{m}) + 0.0001
  \times F^C(90\mu\mathrm{m}) ^2 \\
  \label{equ:flux2}
  F(170\mu\mathrm{m}) = 160+0.32 \times F^C(170\mu\mathrm{m}) + 0.0007
  \times F^C(170\mu\mathrm{m}) ^2 
\end{eqnarray}
where $F$ and $F^C$ denote the observed flux and the flux after 
correction for the bias effect, respectively. These expressions represent
the results from the simulations by \citet{kawara2} with 2-3\% accuracy for
$F(90\mu\mathrm{m}) \ge 100 \mathrm{mJy}$ or $F(170\mu\mathrm{m}) \ge
250 \mathrm{mJy}$, and 10\% for $F(170\mu\mathrm{m}) < 250 \mathrm{mJy}$. Thus,
the errors resulting from the expressions are small enough to be ignored. 
In this paper, we refer these corrected flux with Equ.~(\ref{equ:flux})
and (\ref{equ:flux2}).

Positional uncertainties are also estimated from these simulations. 
Position error depends on the brightness of FIR sources, and 
the dispersion in the measured position relative to the input
position of the artificial sources, $\sigma_{pos}$, is larger 
toward fainter sources.  The derived positional uncertainties
from the simulations are 
$\sigma_{pos}(90\mu\mathrm{m}) \sim 20\arcsec$ for 
$F(90\mu\mathrm{m})\sim 100\mathrm{mJy}$ and 
$\sigma_{pos}(170\mu\mathrm{m}) \sim 35\arcsec$
for $F(170\mu\mathrm{m}) \sim 200\mathrm{mJy}$.
Their simulation also show that the completeness rapidly decrease as the
flux decrease below 200 mJy; 
for example, the detection rate is 73\% for sources with $F^C(90\mu
\mathrm{m}) = 43\mathrm{mJy}$, and 64\% for $F^C(170\mu \mathrm{m}) =
102 \mathrm{mJy}$.


\clearpage
\clearpage

\subsection{Radio observations}


The \ISO~positional uncertainties of 20\arcsec~- 35\arcsec~are too large to
search for optical counterparts directly because there are always
several faint optical sources within each error circle. Thus we
have taken a two-steps approach to identify FIR sources. In the
first step, deep Very Large Array (VLA) 
1.4 GHz imaging was performed and radio counterparts
were found utilizing the well-known FIR-radio correlation 
\citep{condon92,yun01}.
This approach may introduce a slight bias in favor of star-forming 
galaxies, but the AGN dominated sources are rare in general among the FIR
selected sample \citep[less than a few per cent, see ][]{yun01}. Once
radio counterparts are known, then the positional uncertainty 
is reduced greatly to $\le1\arcsec$, and the second step of searching 
for optical counterparts becomes straight forward. 


Deep 1.4 GHz continuum images of the LHNW and LHEX fields were obtained
using the NRAO\footnote{The National Radio Astronomy 
Observatory is a facility of the National Science
Foundation operated under cooperative agreement by Associated 
Universities, Inc.} VLA in the B-configuration in February 2000 and
March-April 2001, respectively.  
Four separate pointings cover a $45' \times 45'$ square region
centered on the two \ISO~survey fields.
The angular resolution of the final images was about $5''$ (FWHM).

The rms noise achieved in the central $32' \times 32'$ mapping
region of the LHEX field is $1\sigma \sim 15 \mu$Jy.  After combining
additional
data from the archive, the rms noise in the northeast quadrant
centered near the ROSAT Deep Survey field \citep{hasinger}
has been improved to $1\sigma \sim 10 \mu$Jy.  Our new VLA 1.4 GHz
continuum image has about 2-3 times higher angular resolution and
improved sensitivity over the previous
observations by \citet[][ $\theta \sim 12''$ and
$1\sigma \sim 30 \mu$Jy]{ruiter}.

In the LHNW field, four separate pointings were also used. However,
the imaging dynamic range in the LHNW field was severely limited
by the presence of a bright radio source 3C 244.1 (4.2 Jy at 1.4 GHz)
just outside the primary beam. Incomplete subtraction of the
time varying
PSF response resulted in a strong gradient of additional ``noise''
across the radio continuum image with a median value of 
$1\sigma \sim 30 \mu$Jy in the image center.
Therefore the radio source identification is dependent entirely
on the local noise level, and the radio source catalog is
complete only at the highest flux density level ($\sim 200~\mu$Jy).
A more detailed descriptions of these VLA observations will be
presented in a later paper (Yun et al., in preparation).

In this paper, we use 387 radio sources brighter than 60 $\mu$Jy($\ge 4
\sigma$) in the central $32' \times 32'$ region of the LHEX field and 76
sources in the entire LHNW field. 

\subsection{Optical $R$ and $I-$band imaging}

The $I-$band images of the two Lockman Hole fields
were obtained using the 8K CCD Camera \citep{Luppino}
attached to the f/10 Cassegrain focus of the University of
Hawaii 88$''$ telescope on May 19-24 in 1999.  This camera has 
a 18\arcmin$\times$18\arcmin~field of view. A $2\times2$-pixel
binning was used yielding a sampling resolution of 0\farcs26 pixel$^{-1}$. 
Nine pointings were used in each field to cover each field completely.
The total exposure times were 14 and 21 minutes for
the LHEX and LHNW fields, respectively.  The seeing was about 
$\sim 0\farcs6$, and the $5\sigma$ detection limit reaches 
$I\sim 22 - 23$ depending on the location on the camera.
Flux calibration was performed using the observations of
SA103 and SA104 
\citep{landolt} with a systematic uncertainty of $\sim0.05$ mag.


The deep $R-$band imaging of the both fields was carried out using the
prime-focus camera, Suprime-Cam \citep{suprime} on the Subaru Telescope on 
March 19 in 2001. This camera has ten 4K$\times$2K CCDs which
provides a 34\arcmin$\times$27\arcmin~field of
view with 0\farcs2/pixel sampling.  During our observing run, one CCD
on the corner of the focal plane was not available. 
To cover entire \ISO\ survey fields, 
four different pointings were used in each field. 
The exposure times were 30 minutes for the two pointings 
on the west and 25 minutes for the two pointing on the east.
The wide-field optical corrector unit of the Suprime-Cam
introduces the significant image distortion on the focal
plane (e.g., 18\arcsec~distortion at 20\arcmin away from the center).
This distortion is corrected using {\it IRAF} task GEOTRAN.
The $5\sigma$ detection limit reaches
26.5 mag under typical seeing of $\sim$ 0\farcs8. Flux calibration was 
performed using the observations of SA101 \citep{landolt} and
SA57 \citep{majewski} fields, with an estimated systematic 
uncertainty of $\sim 0.08$ mag.

The astrometry calibration of all optical images was obtained
by comparing with the positions of stars in the United States Naval
Observatory (USNO) A2.0 catalog \citep{monet}.  
The astrometric uncertainty is estimated to be less
than 1\farcs0 for both $R$ and $I-$band imaging.

\subsection{Optical Spectroscopy}

We performed spectroscopy of optical
objects identified as radio counterparts to the
FIR sources.  Keck II telescope at the Mauna Kea Observatory
and WIYN\footnote{The WIYN Observatory is a joint facility of the
University of Wisconsin-Madison, Indiana University, Yale 
University, and the National Optical Astronomy Observatories.} 
3.5m telescope at the Kitt Peak National Observatory were used with wide
wavelength coverage, allowing many emission features for line
identification. 


\subsubsection{Keck Observations}

Spectroscopic observations with Keck II telescope were performed
preferentially in order of optical brightness on March 30-31 in 2000
and on January 23-24 in 2001, using the Echelle Spectrograph and
Imager (ESI:\citealt{esi}) in the low-dispersion mode. 
The long slit was set to the optical center of the radio sources.
The position angle of the slit was adjusted so that radio sources can
be observed simultaneously from 3,900\AA~to 11,000\AA. 
The slit width and length were 1\arcsec (6.5 pixels) and 8\arcmin, 
respectively. The spectral resolution ranged from 800 at the blue part
to 300 per pixel at the red part.  Exposure times used were
between 300 sec and 5400 sec, depending on the optical brightness 
of the target sources. Standard data reductions were carried out 
using {\it IRAF}. 
Presence of emission or absorption lines was searched by eye. 
In this paper, we present 15 Keck II spectra of \ISO~FIR galaxies. 

\subsubsection{WIYN/HYDRA Observations}

Spectroscopic observations were performed with the HYDRA multi-object
spectrometer on
the WIYN
3.5m
telescope on February 19 in 2002 and February 6-7 in 2003, using the red
cables (2\arcsec~diameter), the 316 line mm$^{-1}$ grating with
G5 filter, and the bench camera. The spectra
cover the wavelength region 5,020$-$10,000 \AA~with a
spectral resolution of 2.64 \AA~/pixel. 

To cope with faint spectra in fiber-based spectroscopy, beam switching
was used to efficiently remove sky emission features such as fringes
and time-dependent airglow emission lines which are heavily blended in
some cases.  The beam switching requires two different 
pointings of the telescope.  
At pointing A, one half of 96 fibers
available for HYDRA were centered on the targets while the others were
set to the sky. At pointing B, the roles of fibers were switched. Beam
switching was repeated 7 times every 30-minute, for a total
on-source integration time of 210 minutes. 

A specialized IRAF package DOHYDRA \citep{valdes} was used to perform 
scattered light removal, spectra extraction, flat fielding,  
fiber throughput correction, and wavelengths calibration.  Sky subtraction 
and subsequent spectrum co-adding require a careful treatment
and are not fully supported by DOHYDRA. We thus developed special IDL
routines to handle these tasks. 
Sky brightness, which is dominated by airglow emission 
lines, varies with airmass and time. The sky spectra were removed from the 
``on-target'' fibers by using the sky background determined from the preceding 
and following ``sky'' fibers. Some scaling of the sky background were required
to subtract time-dependent OH airglow emission. 
After subtracting the sky, the CCD fringes on ``on-target'' fibers
reduced greatly. The resultant sky-subtracted spectra were then
coadded to improve the signal-to-noise ratio.
Extra-galactic emission lines were searched for by eye and by comparing the
source spectra with the airglow spectra and distinguishing the
residual airglow from real emission lines. 

Spectra were obtained for a total of 50 objects in the two
fields.  The brightness of the target objects range between
$R=17$ and $R=21$.  Nine of the 50 objects are \ISO\ FIR sources
identified as radio counterparts.  The remainder are other
faint radio sources identified by the deep radio survey.


\section{Radio Counterpart Identification}

In this section, we illustrate the processes of identifying FIR
sources with 1.4 GHz radio sources. Our 1.4 GHz survey has been carried
out within the two \ISO~fields in the Lockman Hole. The areas where the 
present identification work was performed, are 1089 arcmin$^2$ in LHEX and 1552
arcmin$^2$ in LHNW. 
A typical position uncertainty of the radio source is $\sim 1$\arcsec,
negligibly smaller than the typical \ISO~FIR position uncertainty.

\begin{deluxetable}{lrrrrrrrrr}
  \tabletypesize{\small}
  \tablewidth{400pt}
  \tablecaption{Radio sources within a 3$\sigma$ FIR error
    circle: a 54\arcsec~radius at
    90\micron~and 90\arcsec~at 170\micron \label{tab:rid}}
  \tablehead{
 & Area & & & \multicolumn{3}{c}{Association} & & \multicolumn{2}{c}{Identification\tablenotemark{f}} \\
\cline{5-7}   \cline{9-10}
 & \colhead{(arcmin$^2$)}& \colhead{$N_{all}$\tablenotemark{a}} & \colhead{$N_R$\tablenotemark{b}} &
    \colhead{{$N_S$}\tablenotemark{c}} &
    \colhead{$N_M$}\tablenotemark{d} & $\frac{N_S +
      N_M}{N_{all}}$\tablenotemark{e} & & $N_{id}$ & $N_{id}/N_{all}$}
  \startdata
  LHEX 90\micron  & 1089 & 49 & 387 & 15 & 25 &  0.82 & & 27 & 0.55 \\
  LHEX 170\micron & 1089 & 9  & 387 & 0 & 9   &  1.00 & & 8  & 0.89 \\
  LHNW 90\micron  & 1552 & 67 & 76  & 17 & 2 &  0.28 & & 15 & 0.22 \\
  LHNW 170\micron & 1552 & 11 & 76  & 5 & 2  &  0.63 & &  5 & 0.45 \\
  \enddata
  \tablenotetext{a}{The number of FIR sources for this study}
  \tablenotetext{b}{The number of radio sources for
    this study.}
  \tablenotetext{c}{The number of FIR sources with a single radio
    source within its FIR error circle.}
  \tablenotetext{d}{The number of FIR sources with multiple radio
    sources within its FIR error circle.}
  \tablenotetext{e}{Fraction of FIR sources with one or more radio
    counterpart candidates.}
  \tablenotetext{f}{Results of the source identification. See section \S~\ref{sec:32}.}
\end{deluxetable}

The original catalogs by \citet{kawara2} contain sources with
signal-to-noise ratios of 3$\sigma$ or better. 
In the present work, we exclude faint sources
which have $F^C(90\mu\mathrm{m}) < 43
\mathrm{mJy}$ or $F^C(170\mu\mathrm{m}) < 102
\mathrm{mJy}$, which correspond to the catalog fluxes, 
$F(90\mu\mathrm{m}) < 100 \mathrm{mJy}$ or $(170\mu\mathrm{m}) < 200
\mathrm{mJy}$. Accordingly, we adopt the position uncertainty of
$1\sigma_{POS}(90\mu\mathrm{m})=18\arcsec$ and
$1\sigma_{POS}(170\mu\mathrm{m})=30\arcsec$ 
\citep[see Figure~4 of ][]{kawara2}.
Excluding UGC~06009, which is the only \IRAS~source within the survey
areas and is used for the flux-calibration, the total numbers of 
90\micron~and 170\micron~sources included in this study are 116 and 20,
respectively.

Large FIR error circles and a high number density of faint radio
sources ensures that there is a high probability of having confusing 
multiple radio sources within each FIR error circle. 
For example, 
25 out of 49 ($\sim 50$\%) of the 90\micron~sources in the 
LHEX field have two or more radio sources with in a $3\sigma$ error
circle (54\arcsec~radius; see Table \ref{tab:rid}). Thus the task at 
hand is to reject confusing radio sources which are within the
error circle by chance, by quantifying the reliability of every
identification through a statistical approach as discussed below.

\subsection{Likelihood ratio analysis}


The likelihood ratio analysis using cross-association
\citep{ruiter2,LR,Mann,xid} is a commonly used technique
for identifying sources in crowded 
fields. Here, we adopt the prescription given by \citet{xid}.
It is assumed that each FIR source is physically associated with either
one radio source or none.  It is also assumed that there are 
two types of FIR sources -- one type has one real radio counterpart 
while the second type has no real radio counterpart.  Because
of the source confusion due to the high number density of radio
sources, in many cases there are one or more radio sources found
within a 3$\sigma$ error circle centered at the FIR position. 

To assess the reliability of individual identifications, we begin 
with the likelihood ratio which is described by \citet{ruiter2} and 
\citet{wolstencroft}.  Position uncertainty and number density of 
radio sources are known, and it is assumed that each source has
one or more candidate radio counterparts within a $3\sigma$ error circle. A
dimensionless angular distance $r$, between the FIR and radio positions,
is defined as, 
\begin{equation}
r^2=\left( \frac{\Delta \theta^2}{\sigma_{pos}^2+\sigma_R^2}\right)
\end{equation}
where $\Delta \theta$ is the positional differences between the
FIR and radio sources, and $\sigma_R$ is the standard deviation of
the radio positions. As already discussed, $\sigma_R$ is typically
$\sim$1\arcsec\ and negligible when compared with FIR error $\sigma_{pos}
\sim 18$\arcsec~or 30\arcsec. Thus we adopt $r = \Delta\theta/\sigma_{pos}$. 
The surface density of radio sources, $n(f)$, is the number density
of radio sources with 1.4 GHz flux equal to or greater than $f$, 
and it is adopted from \citet{richards}. 

Assuming that a FIR source and its {\it true} radio counterpart are  
located at the same position, the measured separation $r$ is due to the 
FIR position uncertainty. Then, the probability, $dp_{id}$, of having a real 
radio counterpart at a distance between $r$ and $r+dr$ from the FIR
sources, is $dp_{id} = r \exp(-r^2/2) dr$.  The probability, $dp_c$,  
of finding a first confusing object between $r$ and $r+dr$ from the
FIR source is a product of two probabilities; the probability of not
having a counterpart within $r$, which is given by $e^{-\pi r^2
  \sigma^2 n(f)}$, 
and the probability of a confusing object between $r$ and $r+dr$, which is 
$1-e^{-2\pi r dr \sigma^2 n(f)} \sim 2\pi r \sigma^2 n(f) dr$.
Thus the probability, $dp_c$, is expressed as 
$dp_{c}=2\pi\sigma^2 n(f) r dr \times \exp \left(-\pi r^2 \sigma^2 n(f)\right) $.

The likelihood ratio is defined as the ratio of $dp_{id}/dp_{c}$, and 
\begin{equation}
LR(r,f)=\frac{1}{2\pi \sigma^2 n(f)} \exp{ \left(-\frac{r^2}{2}+\pi r^2
    \sigma^2 n(f) \right) }.
\label{eqa:lr}
\end{equation}

\begin{figure}
  \plotone{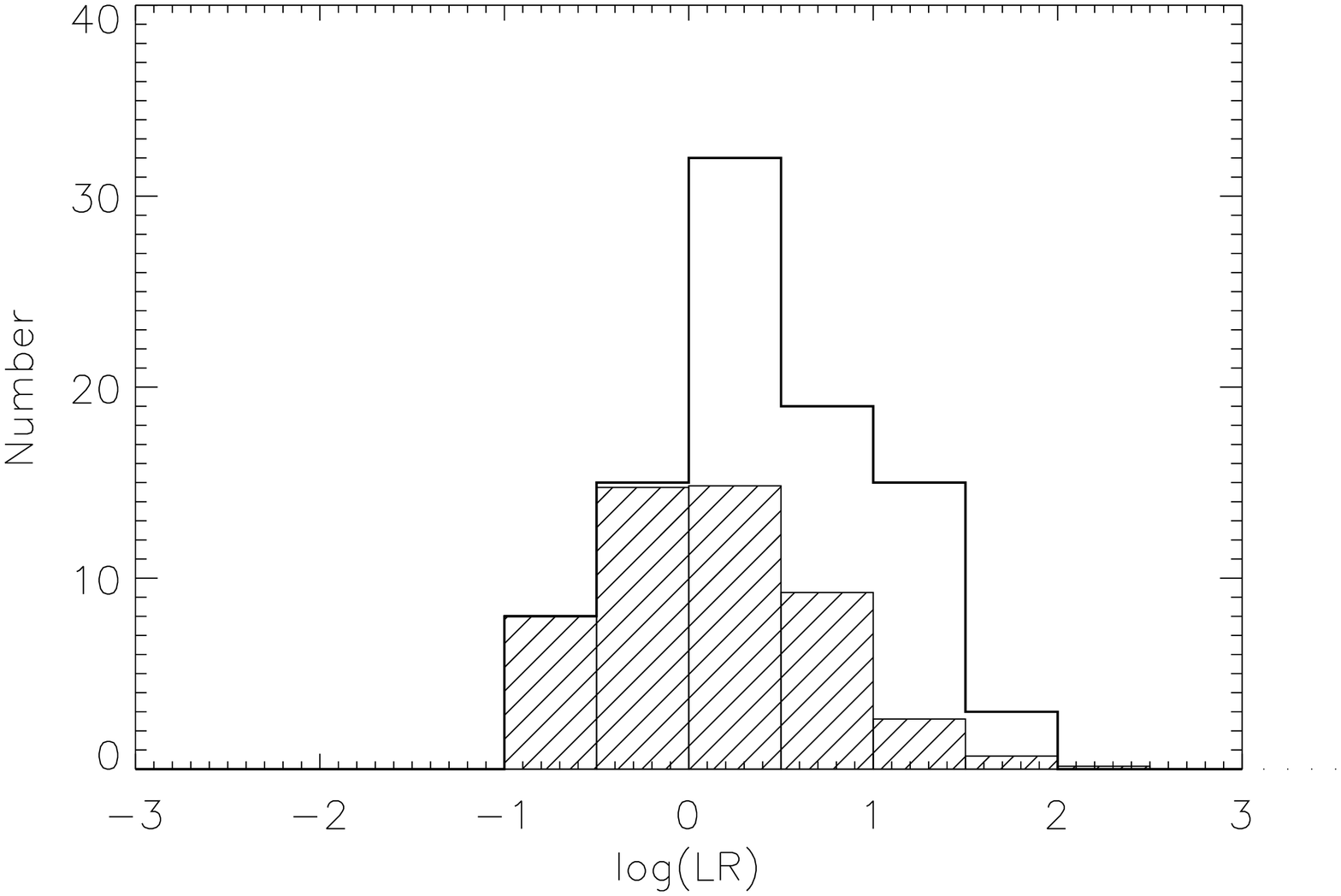}
  \caption{Number distributions as a function of $LR$ for radio sources lying 
   within the $3\sigma$ error circle centered at the \ISO~90\micron\ source
   position. 
   The empty histogram shows the observed distribution $N_{source}(LR)$ while 
   the shaded histogram shows the background distribution obtained from 
   assigning randomly generated positions to the radio sources 
   $N_{ran}(LR)$.\label{fig:lr90}}
\end{figure}

\begin{figure}
  \plotone{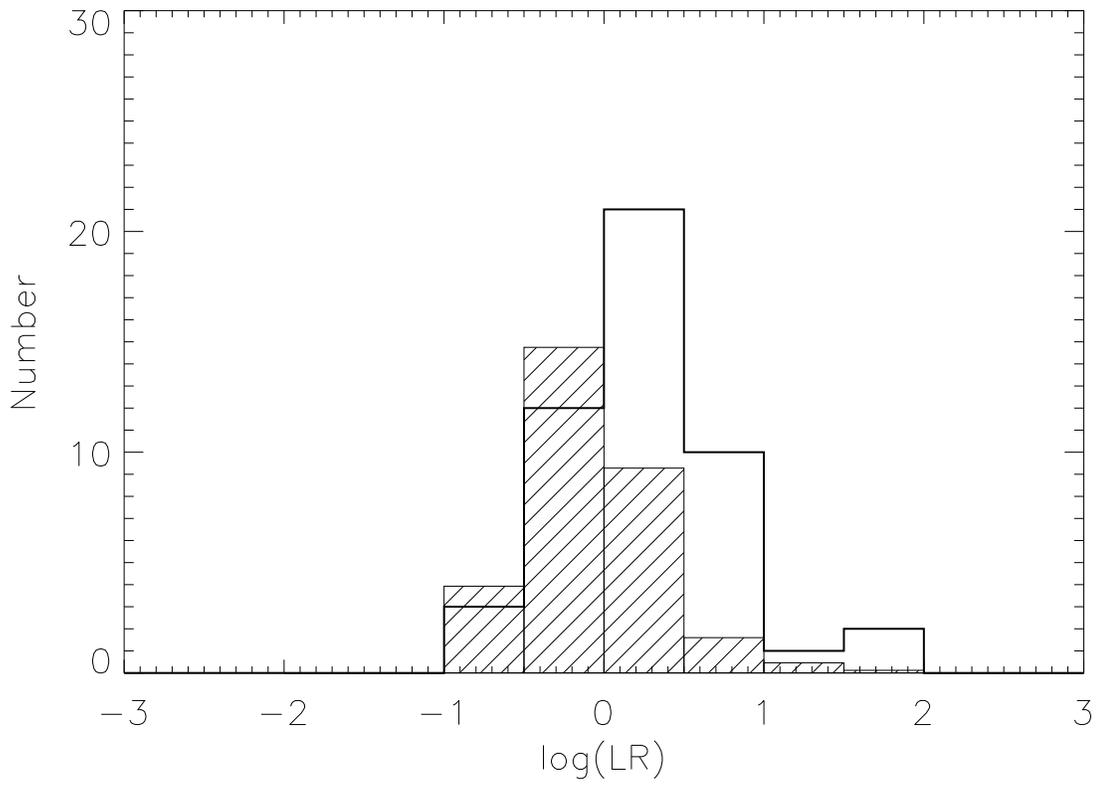}
  \caption{Same as Fig.~\ref{fig:lr90} but for the 170\micron~sources.
           \label{fig:lr170}}
\end{figure}

For each FIR source, $LR$ is calculated for all radio sources 
within the 3$\sigma$ error circle. Then the frequency 
distribution $N(LR)$ is computed.  Figures~\ref{fig:lr90} and \ref{fig:lr170} 
show the distribution of $LR$ as thick-lined histograms for 90\micron~and 
170\micron. The $LR$ distribution was derived only in 
the LHEX field because the LHNW field radio data suffers from a
non-uniform noise and complex source statistics.

Once the distribution of the likelihood ratio $LR$ is determined, 
a reliability $R$ for identification with $LR$ can be quantified as
\begin{equation}
  R(LR)=\frac{N_{true}(LR)}{N_{true}(LR)+N_{false}(LR)},
\label{eqa:r}
\end{equation}
where $N_{true}(LR)$ and $N_{false}(LR)$ are the numbers of $true$ and $false$ 
radio counterparts, respectively. In practice, however, 
it is generally unknown which radio source is the real counterpart or 
what $N_{true}$ and $N_{false}$ are.

The denominator in Equ.~(\ref{eqa:r}) can be re-written as 
\begin{equation}
  N_{true}(LR)+N_{false}(LR)=N_{source}(LR),
\end{equation}
where $N_{source}(LR)$ is the total number of radio sources with 
$LR$ in the field.  In general $N_{true}$ is not known, but we can
estimate $N_{true}$ from the difference between $N_{source}$ and 
$N_{ran}$, where 
$N_{source}$ and $N_{ran}$ are the numbers of radio sources in  
regions containing FIR sources and containing no FIR sources, respectively. 
There is not enough area free from \ISO~FIR sources because of the 
high source density and large positional uncertainty of the \ISO~FIR sources. 
We estimated $N_{ran}(LR)$ by randomly assigning positions to 
radio sources in the catalog. 
These simulations are repeated 100 times and averaged.  The resultant 
$N_{ran}(LR)$ is shown as shaded histograms in Figures \ref{fig:lr90} and
\ref{fig:lr170}. Comparing the observed distributions $N_{source}(LR)$  
with the randomly generated  $N_{ran}(LR)$, it is clear that 
$N_{source}(LR)$ has excess over $N_{ran}(LR)$ for $\log(LR) \gtrsim 0$.  
 
\begin{figure}
  \plotone{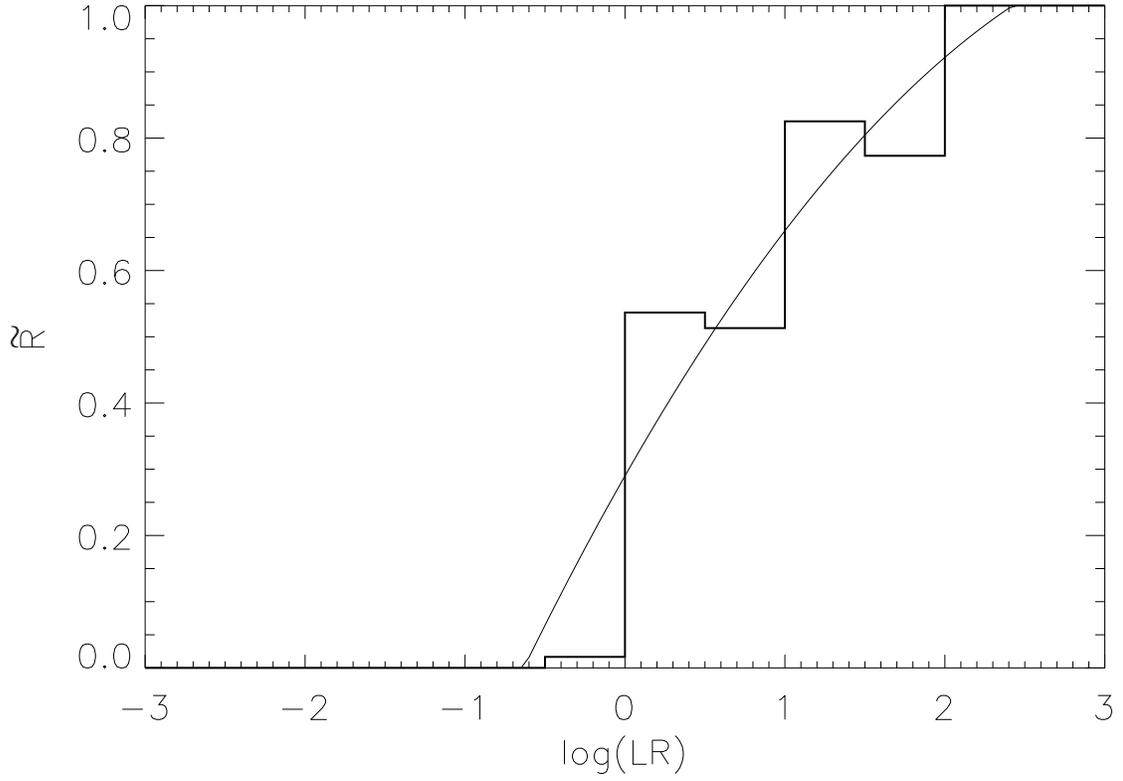}
  \caption{Modified reliability $\tilde{R}$ as a function of $LR$ for
    the 90\micron\ sources. The thin line is a fit to $\tilde{R}$: 
    $\tilde{R} = 0.29 + 0.42 x - 0.05 x^2$ where $x=\log(LR)$. 
    This function is used as an input for deriving the probability 
    for source identification $P_{id,i}$ and 
    unidentification $P_{no-id}$. 
    \label{fig:r90}}
\end{figure}
\begin{figure}
  \plotone{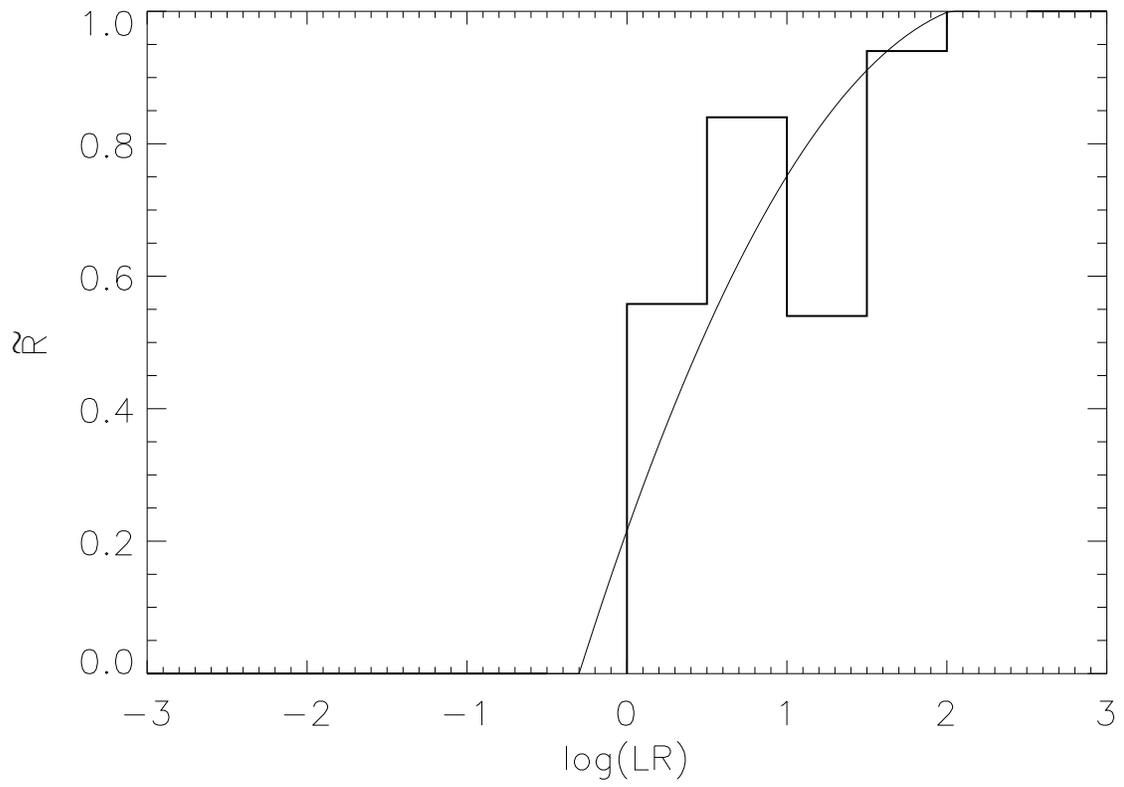}
  \caption{Same as Fig.~\ref{fig:r90} but for the 170\micron\
    sources. The thin curve is $\tilde{R} = 0.21 + 0.68 x
    - 0.14 x^2$ where $x=\log(LR)$.\label{fig:r170}}
\end{figure}

Our radio catalog consists of both kinds of radio sources: ``real'' and ``false'' 
counterparts of FIR sources. Thus, $N_{ran}(LR)$ should be greater than 
$N_{false}(LR)$. Using $N_{ran}(LR)$, $N_{true}(LR)$ is represented as;
\begin{equation}
  N_{true}(LR)\ge N_{source}(LR)-N_{ran}(LR).
\end{equation}
Here we introduce the modified reliability $\tilde{R}(LR)$, as defined by
\begin{equation}
  \tilde{R}(LR)=\frac{N_{source}(LR)-N_{ran}(LR)}{N_{source}(LR)}
  \le
  \frac{N_{true}(LR)}{N_{source}(LR)}=R(LR).
\label{eqa:modified}
\end{equation}
$\tilde{R}$ is plotted on Figures~\ref{fig:r90} and \ref{fig:r170} together 
with quadratic approximations of $\tilde{R}$ as a function of log($LR$).

As seen in Eqa.~(\ref{eqa:modified}), $\tilde{R}(LR)$ is less than the  
true reliability $R(LR)$. The typical ratio $R/\tilde{R}$ for 
$log(LR)>0$ can be estimated as follows. According to Table~\ref{tab:rid}, 
the number of radio sources is 387, and 27 of them are identified with 
90\micron\ sources. Thus, the fraction of true radio counterpart is 
27/387. Because of randomly assigned radio positions 
in the simulation, this fraction should be independent of $LR$. 
In Figure \ref{fig:lr90}, $N_{source}(log(LR)>0)$ = 69 and 
$N_{ran}(log(LR)>0)$ = 28. Then, $N_{false}(log(LR)>0)$ is 
$28\times(1.-27/387)$.
Substituting these into Eqa.~(\ref{eqa:r}) and (\ref{eqa:modified}), 
$R = [69-28\times(1.-27/387)]/69 = 0.62$ and 
$\tilde{R} = [69-28]/69 = 0.59$.  Thus, 
$R(LR)/\tilde{R}(LR)$ = 1.05 for $log(LR)>0$, and the difference 
between the true and modified reliability $R$ and $\tilde{R}$ is small.

Where more than one radio sources are found within a 3$\sigma$ error 
radius, the sum of $R$ of individual sources frequently exceeds unity. 
Thus, it is necessary to normalize $R$ so that the sum of $R$ plus the 
probability of having no radio counterpart is unity. Suppose that 
there are $M$ radio candidates, one of which or none of which 
is a true radio counterpart.  
$P_{id,i}$ and $P_{no-id}$ denote the probability that the $i-th$ 
candidate is the uniquely true radio counterpart and 
none of radio candidates is the real radio counterpart.
Then the probability of the
$i-th$ candidate to be ``true'' and ``false'' 
are $\tilde{R}_i$ and $(1-\tilde{R}_i)$, respectively. 
Thus, $P_{no-id}$ and $P_{id,i}$ are:
\begin{equation}
  P_{no-id}=\frac{\Pi_{j=1}^M(1-\tilde{R}_j)}{S},
\end{equation}
and
\begin{equation}
  P_{id,i}=\frac{\left[\tilde{R}_i\Pi_{j\ne i}^{M}(1-\tilde{R}_j)\right]}{S},
\end{equation}
where $S$ is a normalization factor, specific to each \ISO~FIR source. 
Setting the sum of all the probabilities to unity gives
\begin{equation}
  P_{no-id}+\sum_{i=1}^M P_{id,i}=1.
\end{equation}
The normalization factor $S$ is derived as,
\begin{equation}
  S=\sum_{i=1}^M
  \tilde{R}_i\Pi_{j\ne i}^{M}(1-\tilde{R}_j)+\Pi_{j=1}^M(1-\tilde{R}_j).
\end{equation}

\subsection{Identification\label{sec:32}}

\begin{figure}
  \plotone{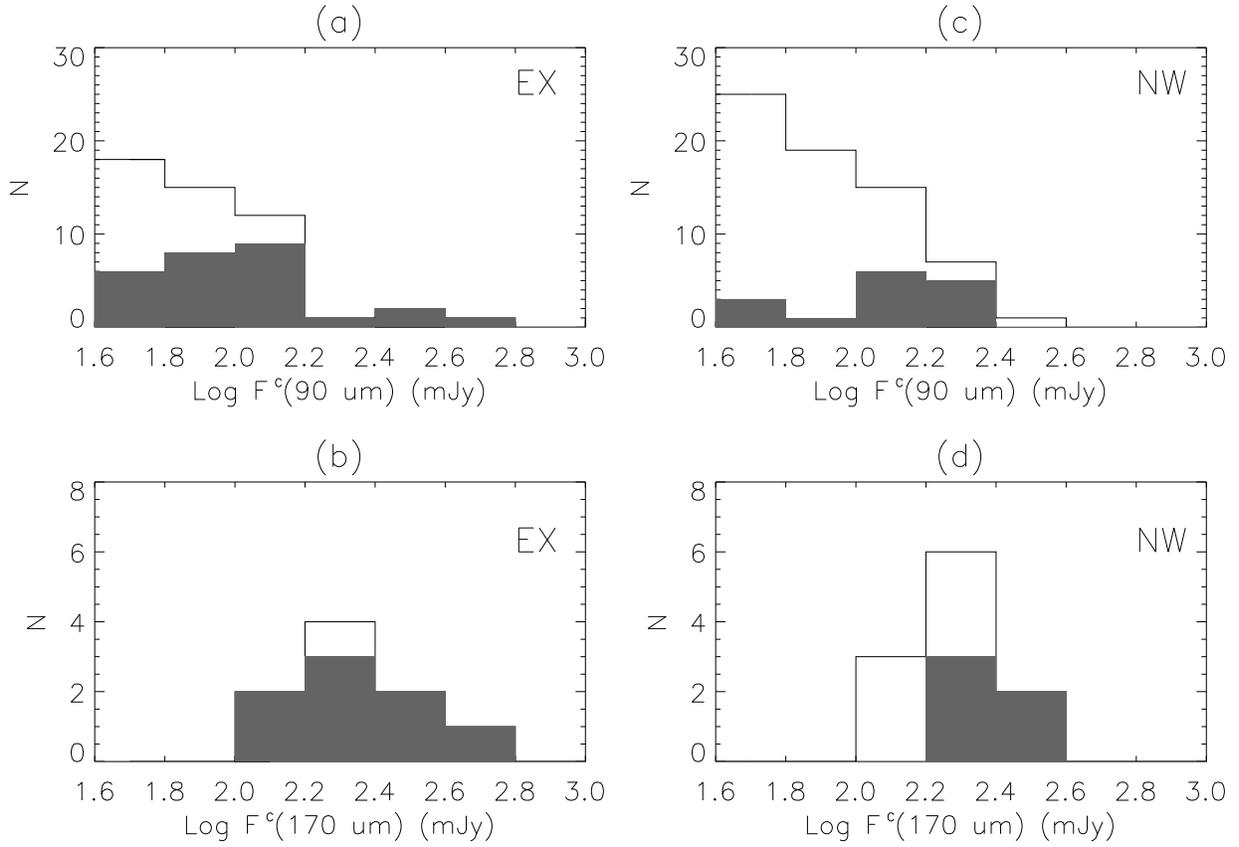}
  \caption{Histograms of FIR sources. The empty histograms show the distributions of 
   all FIR sources used for the identification analysis, while the shaded histograms 
   represent the distribution of identified FIR sources. Panel (a) is for 
   90\micron~sources in the LHEX field, (b) for 170\micron~sources in LHEX, 
   (c) for 90\micron~sources in LHNW, and (d) for 170\micron~sources in LHNW. 
   \label{fig:histo}}
\end{figure}

The main criterion adopted for a true radio counterpart identification
is the condition  $P_{id} > P_{no-id}$. The results are summarized in 
Table \ref{tab:rid}. In the LHEX field, 27 of 49 90\micron~sources are
identified with radio sources. The identification rate is higher 
in bright sources than in 
fainter sources as shown in Fig \ref{fig:histo}. 
The rate is 81\% (13/16) for 90\micron~sources brighter than 
$100\mathrm{mJy}$ while it is reduced to 42\% (14/33) for 
sources fainter than $100\mathrm{mJy}$.
This is due to poorer positional accuracy \citep{kawara2} for fainter sources. 
Seven of nine 170\micron~sources in the LHEX field meet our criterion. 
Five of them have the same identification as their 90\micron~counterpart. 
One exception is 1EX085/2EX068, for which the 90\micron~identification
is given priority for the reason described at the end of this section. 
While one additional source, 2EX016, does not meet our identification
criterion, the identified radio source for the associated 
90\micron~source, 1EX034, is regarded as the true counterpart. 
In summary, a total of eight 170\micron~sources are identified.

Because of the higher effective noise in the radio continuum image of the
LHNW field, there are only a few cases where two or more radio candidates 
are found.  To proceed with the same source identification criterion, 
the reliability functions derived from the LHEX field 
were applied to the LHNW field as well. As a result, 15 of 67 sources and 
5 of 11 sources in 90\micron~and 170\micron~are identified with radio 
sources. The identification rate is 48\% (11/23) for 90\micron~sources 
brighter than $100\mathrm{mJy}$ and 9\% (4/44) for fainter than 
$100\mathrm{mJy}$.

Optical identification was made by searching optical objects within a
2\arcsec~radius of the radio counterpart on the Subaru $R$-band image. 
If no objects are found, bright objects ($R<20$) are searched within 
a 5\arcsec~ radius from the radio counterpart. All of the 
radio counterparts are identified with an optical object. The dispersion of 
separation between radio and optical sources is only 0\farcs6. All of 
the optical counterparts are galaxies except for 1EX030, which is a 
point-like source showing broad emission lines characteristic
of a quasar. The $R$-band images 
are shown together with 1.4 GHz contours in Figure \ref{fig:opt}.

Individual radio counterparts are listed together with the optical data 
in Table~\ref{tab:tab1} and \ref{tab:tab2} for the 90\micron\ 
and 170\micron~sources.  In columns 1 \& 2, names of the 90 and 
170\micron~sources are given. In column 3, a name of radio sources
are given in order of appearance. In columns 4 \& 5, the radio 
coordinates are given in the J2000 system. In columns 6 \& 7, the FIR fluxes 
are given including the correction for the bias
effect(Equ. (\ref{equ:flux}) and (\ref{equ:flux2})).
In column 8, the 1.4 GHz flux is given in $\mu$Jy. In columns 9 \& 10, $R$ and 
$I-$band magnitudes are given in Vega system. Column 11 lists the
angular distance of 
the radio counterpart from the position of the FIR source in arcsec. 
In columns 12-14, $LR$, $P_{id}$, and $P_{no-id}$ are given. 
If more than one radio sources meet the identification criterion 
$P_{id} > P_{no-id}$, then all of radio sources are listed in the 
tables. 
Table~\ref{tab:id} summarizes the optical and radio 
properties of the identified \ISO~FIR sources.  

Source identification is complicated in several cases, and additional
details are discussed below. 

\begin{description}
\item[1EX076.] 1EX076 has three radio candidates: two bright galaxies
(REX11,REX13) and one bright radio source (REX12).  REX11, REX12 and
REX13 are shown at
the $R-$band image in Figure \ref{fig:opt2}(a)
REX11 and REX13 are a
interacting galaxies pair at the same redshift of $z=0.073$. 
Moreover, REX11 is the closest to 1EX076 and its reliability 
is the highest among the three objects. REX11 is thus the most 
likely optical counterpart of this FIR source although there might 
also be a significant contribution from REX13.

\item[1EX034/2EX016.] Although 2EX016 does not have any radio 
candidate that meets our identification criterion, 1EX034 is 
identified with a radio source. This radio source is 
thus regarded as the counterpart of 1EX034/2EX036.

\item[1EX269/2EX047.] As shown in Figure \ref{fig:opt2}(b), 1EX269 and
    2EX047 have three common candidates
(REX21, 22, \& 23). REX21 is regarded as the counterpart because REX21 has 
the highest reliability among the three.

\item[1EX085/2EX068.] 2EX068 has two radio candidates (REX07, REX32)
meeting the criterion, while 1EX085 has only one candidate (REX07) 
meeting the identification criterion(Figure \ref{fig:opt2}(c)). Thus, 2EX068 is 
identified with REX07. 

 
\item[1NW025/2NW006.] 1NW025($F^c(90\mu\mathrm{m})=91$mJy; RA=10:35:16.2, Dec.=+57:33:19 at J2000)
    does not have any candidates meeting the  
  criterion while 2NW006($F^c(170\mu\mathrm{m})=158$mJy; RA=10:35:17.0,
  Dec.=+57:33:22 at J2000)
  is identified with a relatively bright radio 
  source (F(1.4 GHz)=2587 $\mu$Jy; RA=10:35:23.30, Dec.=+57:32:49.6 at J2000). 
  The separation between 1NW025 and 2NW006 is 
  only 8\arcsec. Since the identified source is too far (3.6
  $\sigma_{pos}$) from the 1NW025 position. We regard 1NW025/2NW006 as
  unidentified.
\end{description}

\clearpage

\clearpage
\clearpage
\clearpage
\clearpage

\begin{deluxetable}{lllrrrrrrrrrrr}
\rotate
\tabletypesize{\scriptsize}
\tablecolumns{14}
\tablewidth{610pt}
\tablecaption{A list of the identified far-infrared source at 90$\mu\mathrm{m}$.\label{tab:tab1}}
\tablehead{
\multicolumn{3}{c}{Name} & \colhead{RA\tablenotemark{b}} & \colhead{Dec.\tablenotemark{b}} & \colhead{$F^C(90\mu\mathrm{m})\tablenotemark{c}$} & \colhead{$F^C(170\mu\mathrm{m})$\tablenotemark{c}} & \colhead{$F_{1.4 GHz}$} & \colhead{$R$} & \colhead{$I$} & \colhead{$dist.$} & \colhead{$LR$} & \colhead{$P_{id}$}  & \colhead{$P_{no-id}$}  \\
\colhead{(C$\_$90)} & \colhead{(C$\_$160)\tablenotemark{a}} &
\colhead{(1.4 GHz)}& \colhead{(J2000)}
& \colhead{(J2000)} & \colhead{(mJy)} & \colhead{(mJy)} & \colhead{($\mu$Jy)} &  & & \colhead{(arcsec)}
}
\startdata
\multicolumn{14}{c}{LHEX}\\
\tableline
 1EX153 &  ...    &   REX01 & 10 50 02.13&   +57 16 21.3 &       56 $\pm$ 16 &    ... &      143 &  21.8 &  21.5 &   13 &    6.88 &   0.48 &    0.31 \\
 1EX116 &  ...    &   REX02 & 10 50 12.54&   +57 11 36.8 &      107 $\pm$ 12 &    ... &      377 &  24.1 &  $>$22.0 &   34 &    4.41 &   0.41 &    0.35 \\
 1EX117 &  ...    &   REX03 & 10 50 19.73&   +57 28 13.2 &       66 $\pm$ 12 &    ... &      227 &  19.1 &  18.5 &   16 &    9.99 &   0.43 &    0.22 \\
 1EX088 &  ...    &   REX04 & 10 50 37.55&   +57 28 44.9 &      104 $\pm$ 23 &    ... &      392 &  19.8 &  19.0 &    9 &   21.48 &   0.67 &    0.21 \\
 1EX048 &  2EX004 &   REX05 & 10 50 41.99&   +57 07 06.4 &      254 $\pm$ 19 &      407$\pm$ 30  &      346 &  16.3 &  15.8 &   12 &   17.15 &   0.67 &    0.25 \\
 1EX041 &  2EX013 &   REX06 & 10 50 52.39&   +57 35 07.5 &      310 $\pm$ 80 &      343$\pm$ 88  &      182 &  $<$15.3\tablenotemark{d} &  14.7 &    9 &   10.01 &   0.42 &    0.22 \\
 1EX085 &  2EX068 &   REX07 & 10 50 56.57&   +57 16 31.2 &       73 $\pm$ 24 &      125$\pm$ 14  &      242 &  20.5 &  19.8 &   14 &   11.53 &   0.54 &    0.26 \\
 1EX084 &  ...    &   REX08 & 10 51 00.90&   +57 20 36.0 &      114 $\pm$ 21 &    ... &      261 &  24.3 &  22.0 &   17 &   10.69 &   0.57 &    0.28 \\
 1EX081 &  2EX115 &   REX09 & 10 51 13.44&   +57 14 26.2 &       88 $\pm$ 16 &      212$\pm$ 47  &      611 &  20.0 &  19.1 &    5 &   34.36 &   0.72 &    0.16 \\
 1EX038 &  ...    &   REX10 & 10 51 19.66&   +57 28 04.0 &       43 $\pm$ 13 &    ... &       94 &  24.5 &   $>$22.0 &   18 &    3.65 &   0.51 &    0.49 \\
 1EX076 &  ...    &   REX11 & 10 51 28.07&   +57 35 02.4 &      152 $\pm$ 17 &    ... &      303 &  16.6 &  16.3 &   18 &   11.53 &   0.34 &    0.16 \\
       &        &  REX12 & 10 51 20.86& +57 35 32.6 &         &         &    7177 & 24.9 & $>$22.0 &  47 &   7.99 &  0.27 &   0.16  \\
       &        &  REX13 & 10 51 25.68& +57 35 44.2 &         &         &     497 & 15.5 & 15.2 &  32 &   6.59 &  0.24 &   0.16 \\
 1EX075 &  ...    & REX14 & 10 51 34.41&   +57 33 59.9 &       53 $\pm$ 18 &    ... &      110 &  16.6 &  16.0 &   21 &    3.72 &   0.41 &    0.39 \\
 1EX100 &  ...    & REX15 & 10 51 37.01&   +57 29 40.8 &       54 $\pm$ 14 &    ... &     1901 &  22.3 &  21.7 &   34 &   15.00 &   0.57 &    0.23 \\
 1EX047 &  2EX036 & REX16 & 10 51 52.38&   +57 09 50.1 &      113 $\pm$ 21 &      252$\pm$ 46 &     2642 &  22.2 &  21.2 &   23 &   50.76 &   0.76 &    0.13 \\
 1EX034 &  2EX016 & REX17 & 10 52 07.19&   +57 07 44.8 &      124 $\pm$ 18 &      184$\pm$ 27 &      159 &  17.1 &  16.5 &    7 &    9.13 &   0.60 &    0.33 \\
 1EX130 &  ...    & REX18 & 10 52 31.82&   +57 09 27.1 &       54 $\pm$ 12 &    ... &      123 &  22.9 &  21.7 &   17 &    5.07 &   0.31 &    0.24 \\
 1EX030 &  ...    & REX19 & 10 52 39.55&   +57 24 31.0 &      132 $\pm$ 29 &    ... &      136 &  17.7 &  17.6 &    5 &    8.01 &   0.46 &    0.27 \\
 1EX028 &  ...    & REX20 & 10 52 52.76&   +57 07 53.7 &      150 $\pm$ 33 &    ... &      260 &  17.5 &  16.8 &   32 &    3.80 &   0.52 &    0.48 \\
 1EX269 &  2EX047 & REX21 & 10 52 56.82&   +57 08 25.7 &      206 $\pm$ 53 &      163$\pm$ 42 &      439 &  16.8 &  16.2 &   22 &   13.29 &   0.39 &    0.17 \\
       &        &  REX22 & 10 52 52.76& +57 07 53.7 &         &         &     260 & 17.5 & 16.8 &  30 &   4.55 &  0.20 &   0.17 \\
       &        &  REX23 & 10 52 58.00& +57 08 34.8 &         &         &     337 & 22.7 & 21.0 &  34 &   3.94 &  0.18 &   0.17 \\
 1EX062 &  ...    &  REX24 & 10 53 01.36&   +57 05 42.9 &      399 $\pm$ 29 &    ... &      705 &  16.9 &  16.3 &    4 &   39.33 &   0.83 &    0.17 \\
 1EX179 &  ...    &  REX25 & 10 53 18.96&   +57 21 40.4 &      129 $\pm$ 29 &    ... &      247 &  17.1 &  16.6 &    7 &   14.48 &   0.60 &    0.25 \\
 1EX126 &  ...    &  REX26 & 10 53 22.83&   +57 15 00.0 &       74 $\pm$ 16 &    ... &      261 &  21.4 &  20.2 &    5 &   15.77 &   0.57 &    0.22 \\
 1EX026 &  ...    &  REX27 & 10 53 25.30&   +57 29 11.4 &       90 $\pm$ 20 &    ... &      507 &  17.6 &  17.4 &    8 &   27.18 &   0.76 &    0.21 \\
 1EX025 &  ...    &  REX28 & 10 53 28.01&   +57 11 15.5 &       67 $\pm$ 15 &    ... &      300 &  18.2 &  17.6 &   10 &   15.91 &   0.72 &    0.28 \\
 1EX125 &  ...    &  REX29 & 10 53 26.68&   +57 14 05.9 &       83 $\pm$ 12 &    ... &      226 &  19.3 &  18.7 &   15 &   10.25 &   0.66 &    0.34 \\
 1EX112 &  ...    &  REX30 & 10 53 42.08\tablenotemark{e} &   +57 30 26.1\tablenotemark{e} &       74$\pm$ 25 &    ... &      335 &  24.6 &  $>$22.0 &   18 &   13.05 &   0.70 &    0.30 \\
 1EX055 &  ...    &  REX31 & 10 53 57.76&   +57 23 47.1 &       61$\pm$ 20 &    ... &      119 &  22.5 &  21.3 &   18 &    4.59 &   0.37 &    0.31 \\
\tableline
\multicolumn{14}{c}{LHNW}\\
\tableline
1NW130 &  2NW013 &  RNW01 & 10 31 23.03&   +57 42 26.4 &      172 $\pm$ 19 &      359 $\pm$ 53&      677 &  20.0 &  19.1 &   10 &   32.89 &   0.81 &    0.19 \\
 1NW034 &  ...    &  RNW02 & 10 31 47.58&   +57 49 27.5 &      119$\pm$ 22  &      ... &      200 &  20.5 &  19.7 &   24 &    5.58 &   0.58 &    0.42 \\
 1NW077 &  ...    &  RNW03 & 10 32 05.80&   +58 02 39.2 &      122$\pm$ 22  &      ... &      447 &  20.2 &  19.0 &    9 &   24.22 &   0.77 &    0.23 \\
 1NW192 &  2NW003 &  RNW04 & 10 32 49.52&   +57 37 07.9 &      226$\pm$ 33  &      262 $\pm$ 39&      420 &  17.1 &  16.5 &    9 &   22.94 &   0.77 &    0.23 \\
 1NW031 &  ...    &  RNW05 & 10 33 15.43&   +57 31 01.8 &      128$\pm$ 28  &      ... &      408 &  21.6 &  20.6 &   34 &    4.50 &   0.54 &    0.46 \\
 1NW021 &  2NW005 &  RNW06 & 10 33 20.20&   +57 49 13.1 &      127$\pm$ 23  &      192 $\pm$ 50&      343 &  18.7 &  18.0 &   11 &   17.71 &   0.74 &    0.26 \\
 1NW062 &  ...    &  RNW07 & 10 33 30.28&   +57 42 25.0 &       57$\pm$ 15  &      ... &      143 &  19.9 &  19.1 &   15 &    6.48 &   0.60 &    0.40 \\
 1NW100 &  ...    &  RNW08 & 10 33 52.36&   +57 27 01.4 &       62$\pm$  9  &      ... &      292 &  17.7 &  17.3 &   19 &   10.62 &   0.67 &    0.33 \\
 1NW030 &  ...    &  RNW09 & 10 33 58.71&   +57 43 17.3 &      105$\pm$ 19  &      ... &      915 &  17.3 &  16.6 &   11 &   41.41 &   0.83 &    0.17 \\
 1NW272 &  2NW009 &  RNW10 & 10 33 59.09&   +57 29 52.4 &       91$\pm$ 20  &      199 $\pm$ 29&     1493 &  20.8 &  19.1 &   15 &   53.34 &   0.86 &    0.14 \\
 1NW026 &  ...    &  RNW11 & 10 35 13.69&   +57 34 44.6 &       51$\pm$ 13  &      ... &      434 &  21.9 &  21.0 &    9 &   23.56 &   0.77 &    0.23 \\
 1NW092 &  ...    &  RNW12 & 10 36 04.05&   +57 48 12.2 &      205$\pm$ 38  &      ... &      267 &  19.5 &  18.9 &   20 &    9.41 &   0.65 &    0.35 \\
 1NW023 &  ...    &  RNW13 & 10 36 06.52&   +57 47 02.1 &      169$\pm$ 31  &      ... &      670 &  16.7 &  16.2 &   23 &   17.69 &   0.72 &    0.26 \\
 1NW022 &  ...    &  RNW14 & 10 36 11.61&   +57 43 21.1 &      101$\pm$ 22  &      ... &      262 &  16.3 &  15.6 &    3 &   16.19 &   0.72 &    0.28 \\
 1NW133 &  ...    &  RNW15 & 10 36 15.41&   +57 39 14.7 &      169$\pm$ 37  &      ... &      338 &  16.8 &  16.1 &   13 &   16.31 &   0.72 &    0.28 \\
\enddata
\tablenotetext{a}{The positional association of FIR sources in both bands come from \citet{kawara2}.}
\tablenotetext{b}{The J2000 coordinates come from the VLA 1.4 GHz radio data.}
\tablenotetext{c}{The bias corrected flux density using
  Equ.~(\ref{equ:flux}.)}
\tablenotetext{d}{$R-$band photometry is not available due to saturation.}
\tablenotetext{e}{The positional difference between optical
  and radio is 3\farcs0, which is larger than the typical error
  0\farcs6.}
\end{deluxetable}

\begin{deluxetable}{lllrrrrrrrrrrr}
\rotate
\tabletypesize{\scriptsize}
\tablecolumns{14}
\tablewidth{610pt}
\tablecaption{A list of the identified far-infrared source at
  170\micron~.\label{tab:tab2}}
\tablehead{
\multicolumn{3}{c}{Name} & \colhead{RA\tablenotemark{b}} &
\colhead{Dec.\tablenotemark{b}} &
\colhead{$F^C(170\mu\mathrm{m})$\tablenotemark{c}} &
\colhead{$F^C(90\mu\mathrm{m})$\tablenotemark{c}} & \colhead{$F_{1.4 GHz}$}
& \colhead{$R$} & \colhead{$I$} & \colhead{dist.} & \colhead{$LR$} &
\colhead{$P_{id}$}  & \colhead{$P_{no-id}$}  \\
\colhead{(C$\_$160)} & \colhead{(C$\_$90)\tablenotemark{a}} &
\colhead{(1.4 GHz)} &
\colhead{(J2000)} & \colhead{(J2000)} & \colhead{(mJy)} &
\colhead{(mJy)} & \colhead{($\mu$Jy)} &  & & \colhead{(arcsec)}}
\startdata
\multicolumn{14}{c}{LHEX}\\
\tableline
  2EX004 &   1EX048 &    REX05 & 10 50 41.99&   +57 07 06.4 &       407$\pm$ 30 &       254 $\pm$ 19&       346 &   16.3 &   15.8 &    16 &     6.79 &    0.51 &     0.25 \\
  2EX013 &   1EX041 &    REX06 & 10 50 52.39&   +57 35 07.5 &       343$\pm$ 88 &       310 $\pm$ 80&       182 &   15.3 &   14.7 &    22 &     3.34 &    0.39 &     0.35 \\
  2EX068 &   1EX085 &    REX32 & 10 50 56.54&   +57 15 32.7 &       125$\pm$ 14 &        73 $\pm$ 24&       462 &   22.8 &   22.3 &    27 &     7.03 &    0.45 &     0.21 \\
       &        &        REX07 & 10 50 56.57& +57 16 31.2 &         &     ... &     242 & 20.5 & 19.8 &  34 &   3.19 &  0.22 &   0.21 \\
  2EX115 &   1EX081 &    REX09 & 10 51 13.44&   +57 14 26.2 &       212$\pm$ 47 &        88 $\pm$ 16&       611 &   20.0 &   19.1 &    40 &     5.74 &    0.50 &     0.28 \\
  2EX036 &   1EX047 &    REX16 & 10 51 52.38&   +57 09 50.1 &       252$\pm$ 46 &       113 $\pm$ 21&      2642 &   22.2 &   21.2 &    37 &    19.85 &    0.77 &     0.14 \\
2EX016 & 1EX034 &  REX17 & 10 52 07.19& +57 07 44.8 &               184$\pm$ 27 &    124$\pm$ 18  &     159 & 17.1  & 16.5 &  32 &  2.36 &  0.30\tablenotemark{f} &   0.39\tablenotemark{f} \\    
  2EX015 &  ...   & REX33 & 10 52 37.38\tablenotemark{g}&   +57 31 03.5\tablenotemark{g} &       151 $\pm$ 38 &        ... &      14667 &  20.7 &   19.1 &    33 &    82.74 &    0.96 &     0.01 \\
  2EX047 &   1EX269 & REX21 &  10 52 56.82&   +57 08 25.7 &       163 $\pm$ 42&       206 $\pm$ 53&       439 &   16.8 &   16.2 &    29 &     6.25 &    0.33 &     0.18 \\
       &        &  REX22 & 10 52 52.76& +57 07 53.7 &         &     ... &     260 & 17.5 & 16.8 &  28 &   4.11 &  0.23 &   0.18 \\
       &        &  REX23 & 10 52 58.00& +57 08 34.8 &         &     ... &     337 & 22.7 & 21.0 &  41 &   3.37 &  0.19 &   0.18 \\
\tableline
\multicolumn{14}{c}{LHNW}\\
\tableline
 2NW013 &  1NW130 &  RNW01 &  10 31 23.03&   +57 42 26.4 &      359 $\pm$ 53 &      172 $\pm$ 19&      677 &  20.0 &  19.1 &   19 &   11.50 &   0.76 &    0.24 \\
 2NW003 &  1NW192 &  RNW04 &  10 32 49.52&   +57 37 07.9 &      262 $\pm$ 39 &      226 $\pm$ 33 &      420 &  17.1 &  16.5 &   27 &    6.43 &   0.66 &    0.34 \\
 2NW005 &  1NW021 &  RNW06 &  10 33 20.20&   +57 49 13.1 &      192 $\pm$ 50 &      127 $\pm$ 23 &      343 &  18.7 &  18.0 &   13 &    6.99 &   0.62 &    0.30 \\
 2NW004 &  ...    &  RNW16 &  10 33 41.32&   +58 02 21.1 &      165 $\pm$ 41 &      ... &      380 &  18.0 &  17.2 &   26 &    6.00 &   0.65 &    0.35 \\
 2NW009 &  1NW272 &  RNW10 &  10 33 59.09&   +57 29 52.4 &      199 $\pm$ 29 &       91 $\pm$ 20 &     1493 &  20.8 &  19.1 &   31 &   16.01 &   0.82 &    0.18 \\
\enddata
\tablenotetext{Note}{Same as Table \ref{tab:tab1}.}
\tablenotetext{f}{The radio source of 2EX016 does not meet our criterion,
  $P_{id}>P_{no-id}$. However the associated 90\micron~source, 1EX034,
  is identified with a radio source, which is
  regarded as the counterpart of 1EX034/2EX016.}
\tablenotetext{g}{The position comes from the optical image because the
associated radio source position is dominated by the bright,
extended radio lobes.}
\end{deluxetable}

\begin{deluxetable}{llrrrrrrrr}
\tablecaption{A list of identified FIR sources\label{tab:id}}
\tabletypesize{\scriptsize}
\tablewidth{400pt}
\tablecolumns{8}
\tablehead{\multicolumn{2}{c}{Name} & \colhead{$F^{C}(90 \mu \rm{m})$}
  & $F^{C}(170 \mu \rm{m})$ & \colhead{$\frac{L(1.4 GHz)}{L(90\mu
      \rm{m})}$} & \colhead{$\frac{L(90 \mu \rm{m})}{L(R)}$} &
  \colhead{$z$\tablenotemark{a}} & \colhead{log($L_{FIR}$)} &
    \colhead{Nature\tablenotemark{b}} & \\
\colhead{(C$\_$90)} & \colhead{(C$\_$160)} & (mJy) & (mJy) &
($10^{-7}$) &  &  & ($L_\odot$) & & 
}
\startdata
 1EX153&    ...&     56&    ...&      11.5&   69.2&    ...&    ...&    ...&\\
 1EX116&    ...&    107&    ...&      16.0& 1063.3&    ...&    ...&    ...&\\
 1EX117&    ...&     66&    ...&      15.6&    6.4&    ...&    ...&    ...&\\
 1EX088&    ...&    104&    ...&      17.0&   20.3& 0.363K&   11.7&    ...&\\
 1EX048& 2EX004&    254&    407&       6.1&    2.0& 0.091K&   10.9&    ...&\\
 1EX041& 2EX013&    310&    343&       2.6& $<$0.9& 0.028K&    9.8&    ...&\\
 1EX085& 2EX068&     73&    125&      15.0&   26.3& 0.396K&   11.7&      i&\\
 1EX084&    ...&    114&    ...&      10.3& 1360.4&    ...&    ...&    ...&\\
 1EX081& 2EX115&     88&    212&      31.1&   20.2& 0.362K&   11.8&    ...&\\
 1EX038&    ...&     43&    ...&       9.8&  626.1&    ...&    ...&    ...&\\
 1EX076&    ...&    152&    ...&       9.0&    1.5& 0.073H&   10.3&  i,HII&\\
 1EX075&    ...&     53&    ...&       9.2&    0.5& 0.074H&   10.1&    ...&\\
 1EX100&    ...&     54&    ...&     157.9&  106.9&    ...&    ...&   AGN&\\
 1EX047& 2EX036&    113&    252&     105.0&  200.2&    ...&    ...&    AGN&\\
 1EX034& 2EX016&    124&    184&       5.8&    2.0& 0.123K&   10.8&    ...&\\
 1EX130&    ...&     54&    ...&      10.3&  187.1&    ...&    ...&    ...&\\
 1EX030&    ...&    132&    ...&       4.6&    3.8& 1.110H&   13.1& Quasar&\\
 1EX028&    ...&    150&    ...&       7.8&    3.4& 0.163K&   11.0&    ...&\\
 1EX269& 2EX047&    206&    163&       9.6&    2.6& 0.080K&   10.5&    ...&\\
 1EX062&    ...&    399&    ...&       7.9&    5.3& 0.080K&   10.8&    ...&\\
 1EX179&    ...&    129&    ...&       8.6&    2.2& 0.133H&   10.8&    ...&\\
 1EX126&    ...&     74&    ...&      15.9&   61.3&    ...&    ...&    ...&\\
 1EX026&    ...&     90&    ...&      25.3&    2.4& 0.029H&    9.3&   HII&\\
 1EX025&    ...&     67&    ...&      20.3&    3.0& 0.162H&   10.8&   i,AGN&\\
 1EX125&    ...&     83&    ...&      12.2&   10.6& 0.231H&   11.2&    ...&\\
 1EX112\tablenotemark{c}&    ...&     74&    ...&      20.3& 1156.1&    ...&    ...&    ...&\\
 1EX055&    ...&     61&    ...&       8.8&  145.1&    ...&    ...&    ...&\\
    ...& 2EX015&    ...&    151&$>$ 1527.9&$<$19.2& 0.710K&   12.3&    AGN&\\
 1NW130& 2NW013&    172&    359&      17.7&   40.0& 0.502K&   12.4&    ...&\\
 1NW034&    ...&    119&    ...&       7.5&   45.7&    ...&    ...&    ...&\\
 1NW077&    ...&    122&    ...&      16.5&   32.8&    ...&    ...&     i?&\\
 1NW192& 2NW003&    226&    262&       8.4&    3.6& 0.115K&   11.0&    ...&\\
 1NW031&    ...&    128&    ...&      14.3&  125.6&    ...&    ...&    ...&\\
 1NW021& 2NW005&    127&    192&      12.2&    8.7& 0.240K&   11.5&      i&\\
 1NW062&    ...&     57&    ...&      11.3&   12.3& 0.498B&   11.9&    ...&\\
 1NW100&    ...&     62&    ...&      21.3&    1.7& 0.182B&   10.9&    ...&\\
 1NW030&    ...&    105&    ...&      39.3&    2.0& 0.263K&   11.4&    ...&\\
 1NW272& 2NW009&     91&    199&      73.5&   43.7& 0.469K&   12.1&    AGN&\\
 1NW026&    ...&     51&    ...&      38.1&   68.8&    ...&    ...&     i?&\\
 1NW092&    ...&    205&    ...&       5.9&   29.5& 0.511B&   12.4&    ...&\\
 1NW023&    ...&    169&    ...&      17.8&    1.9& 0.044B&    9.9&    ...&\\
 1NW022&    ...&    101&    ...&      11.7&    0.8& 0.093K&   10.4&    ...&\\
 1NW133&    ...&    169&    ...&       9.0&    2.1& 0.114H&   10.7&    AGN&\\
    ...& 2NW004&    ...&    165&   $>$39.5& $<$1.6& 0.075H&   10.2&    ...&\\

\enddata
\tablenotetext{a}{Characters represent the used telescope; (K)
  Keck II/ESI data, (H) WIYN/HYDRA data and (B) from
  A. Barger (private communication)}
\tablenotetext{b}{(i) represents interacting galaxy system 
  from Figure \ref{fig:opt}. (HII) shows
  the sources with HII-region-like spectra. (AGN) shows the
  sources with AGN/LINER-like spectra or radio excess. See 
  section \S~\ref{sec:nat} for details.}
\tablenotetext{c}{An alternative ID is a $R=19.6$ galaxy at
  $z=0.586$ which is known to be an X-ray sources at RA=10:53:39.70
  and Dec=+57:31:05.0(J2000)\citep{lehmann}.}
\end{deluxetable}

\clearpage

\clearpage
\begin{figure}
     \plotone{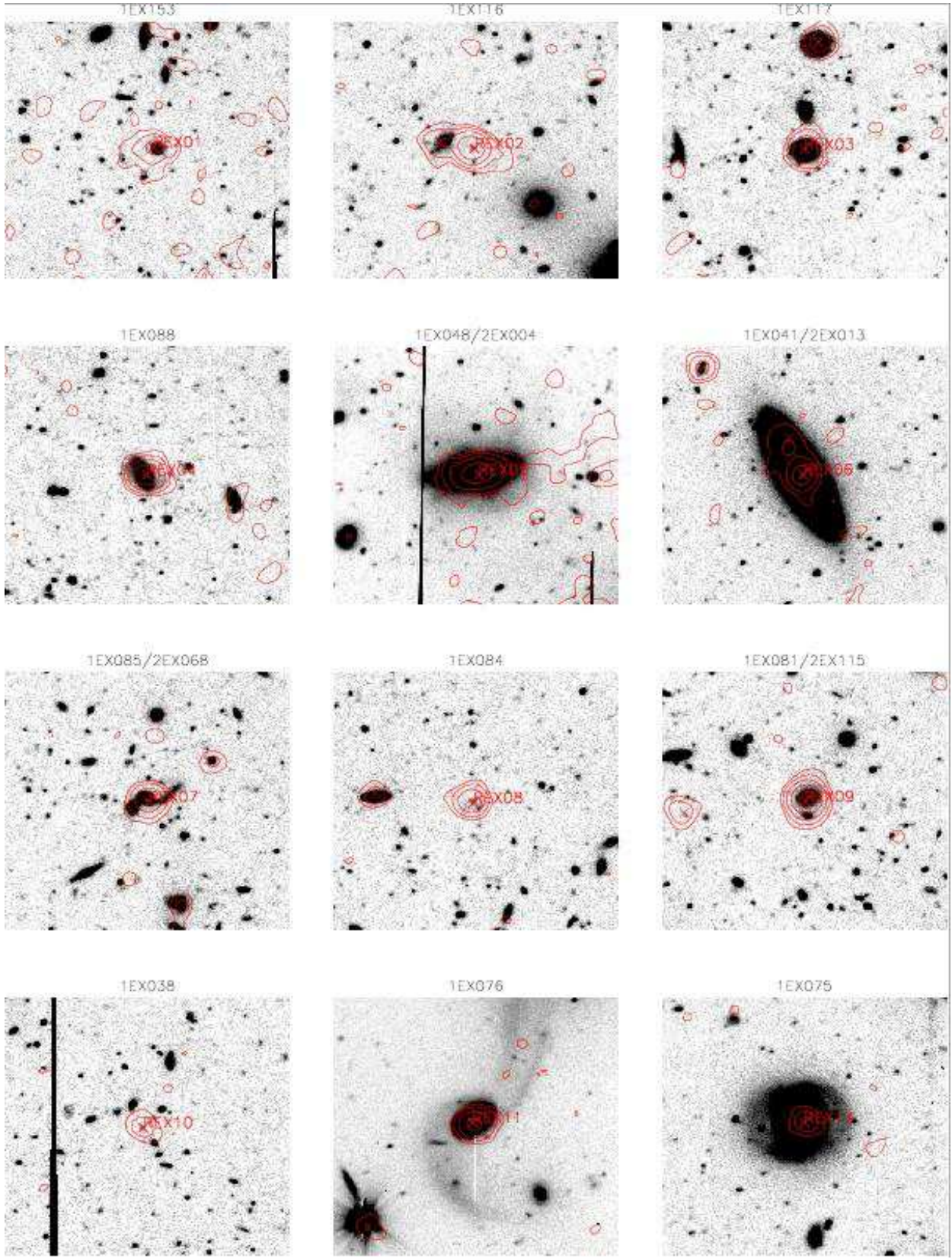}
    \caption{R-band images centered at the positions of the 
      radio counterparts. The 
      image size is 60\arcsec $\times$ 60\arcsec. North is up, and
      east to the left.
     Red 1.4 GHz contours are 
      plotted at a step of 2, 4, 8, 16, 32 $\times \sigma$. Crosses
      present the positions of the 1.4 GHz radio sources.}
    \label{fig:opt}
\end{figure}

\clearpage
\begin{figure}
  \figurenum{7}
  \plotone{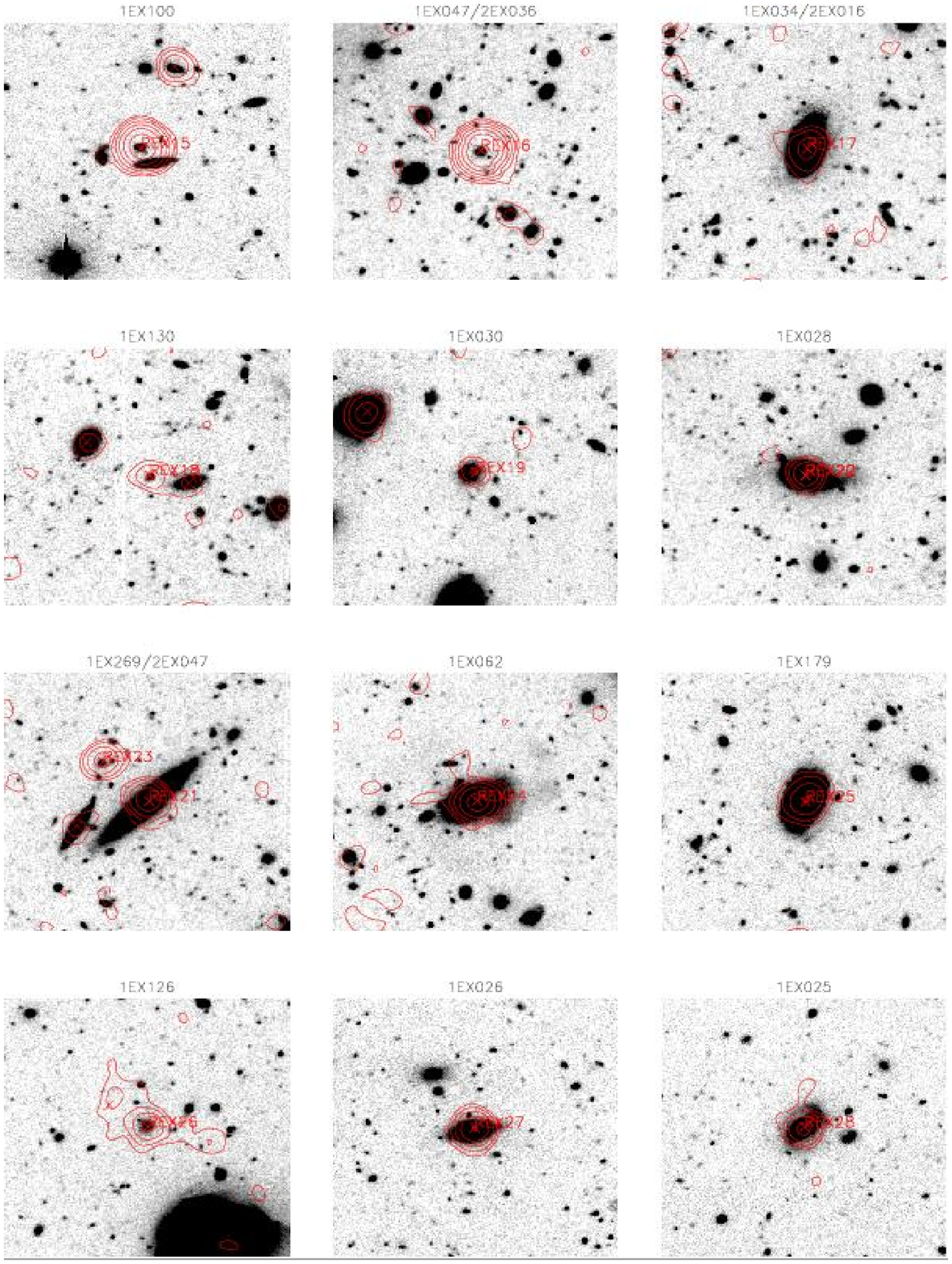}
    \caption{Continued}
\end{figure}

\clearpage
\begin{figure}
  \figurenum{7}
  \plotone{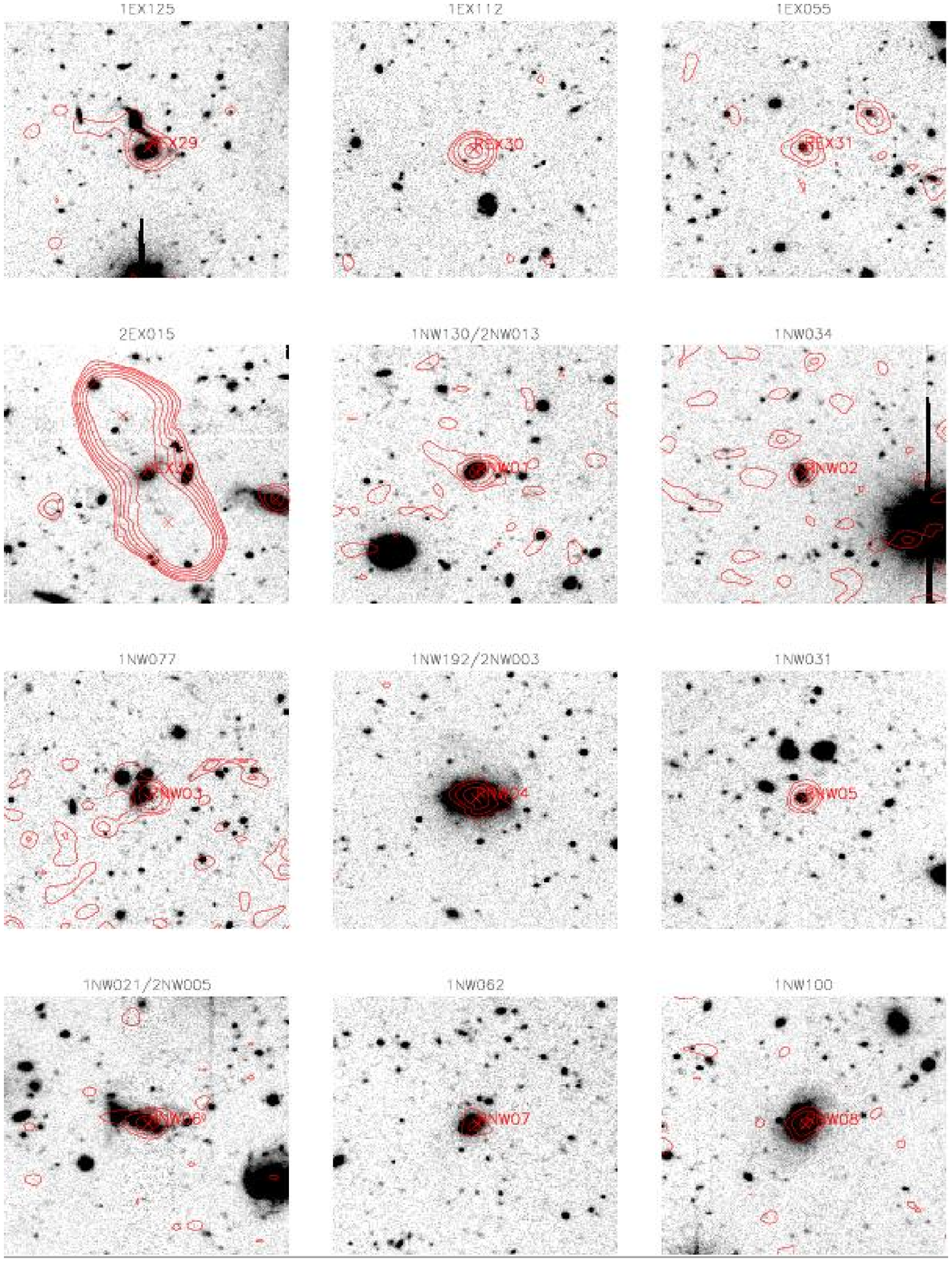}
    \caption{Continued}
\end{figure}

\clearpage
\begin{figure}
  \figurenum{7}
  \plotone{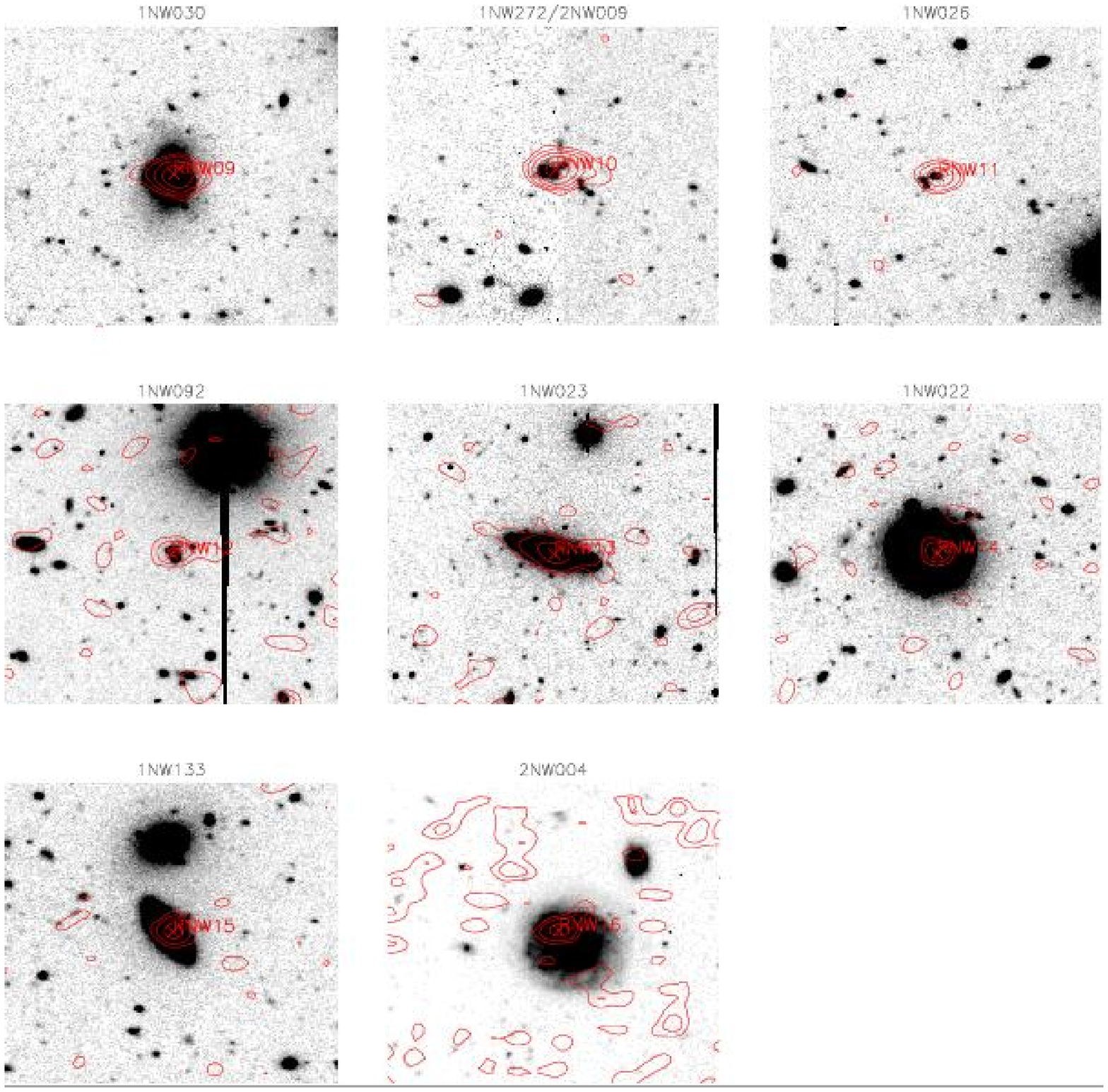}
    \caption{Continued}
\end{figure}
\clearpage

\begin{figure}
    \plotone{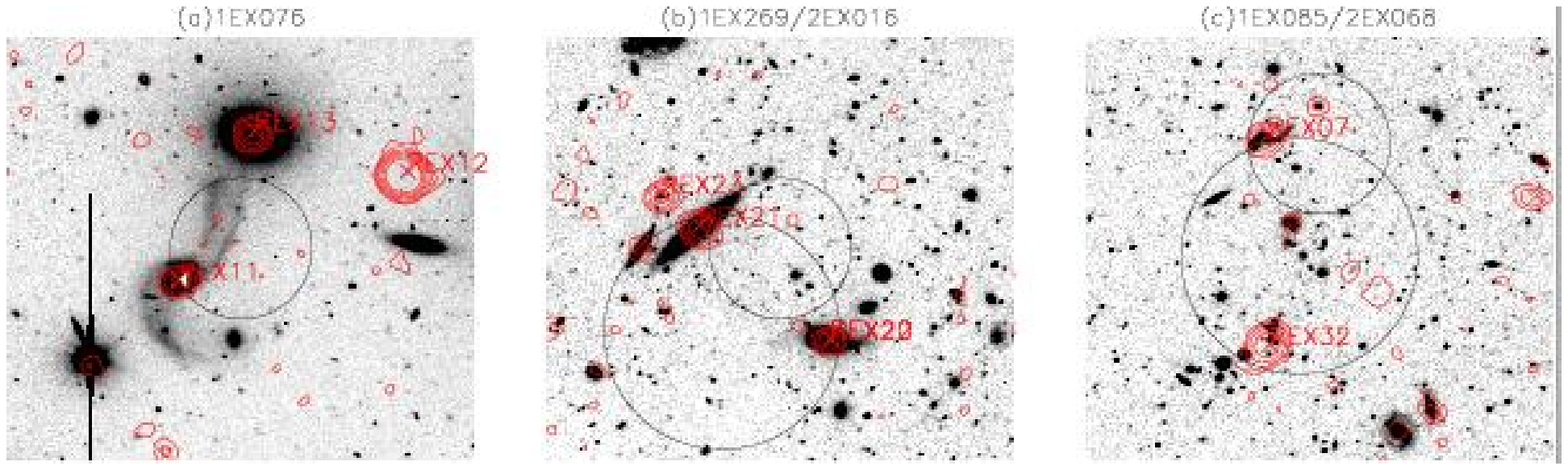}
    \caption{Wide R-band images(2\arcmin~
      $\times$ 2\arcmin) for detail explanation of the identification
      of (a)1EX076,
      (b)1EX269/2EX016 and (c)1EX085/2EX068. Red Contours and crosses
      are same as Fig. 7. 
      Small and large circles show
      1$\sigma$ positional error of the 90\micron~sources and the
      170\micron~sources, respectively.}
    \label{fig:opt2}
\end{figure}

\clearpage

\section{Discussion}

\subsection{Redshift and FIR Luminosity}
\label{sec:fir}

Spectroscopic redshifts of 29 out of 44 FIR sources in Table~\ref{tab:id}
are available.  
Twenty five optical spectra of FIR sources were obtained with the KECK II
and WIYN telescopes while redshifts of four objects 
(1NW062, 1NW100, 1NW092 and 1NW044) were kindly provided to us 
by A. Barger (private communication).
The 25 optical spectra are shown in Figure \ref{fig:spe}.

\clearpage
\begin{figure}
  \begin{center}
    \begin{tabular}{ccc}
      \resizebox{65mm}{!}{\includegraphics{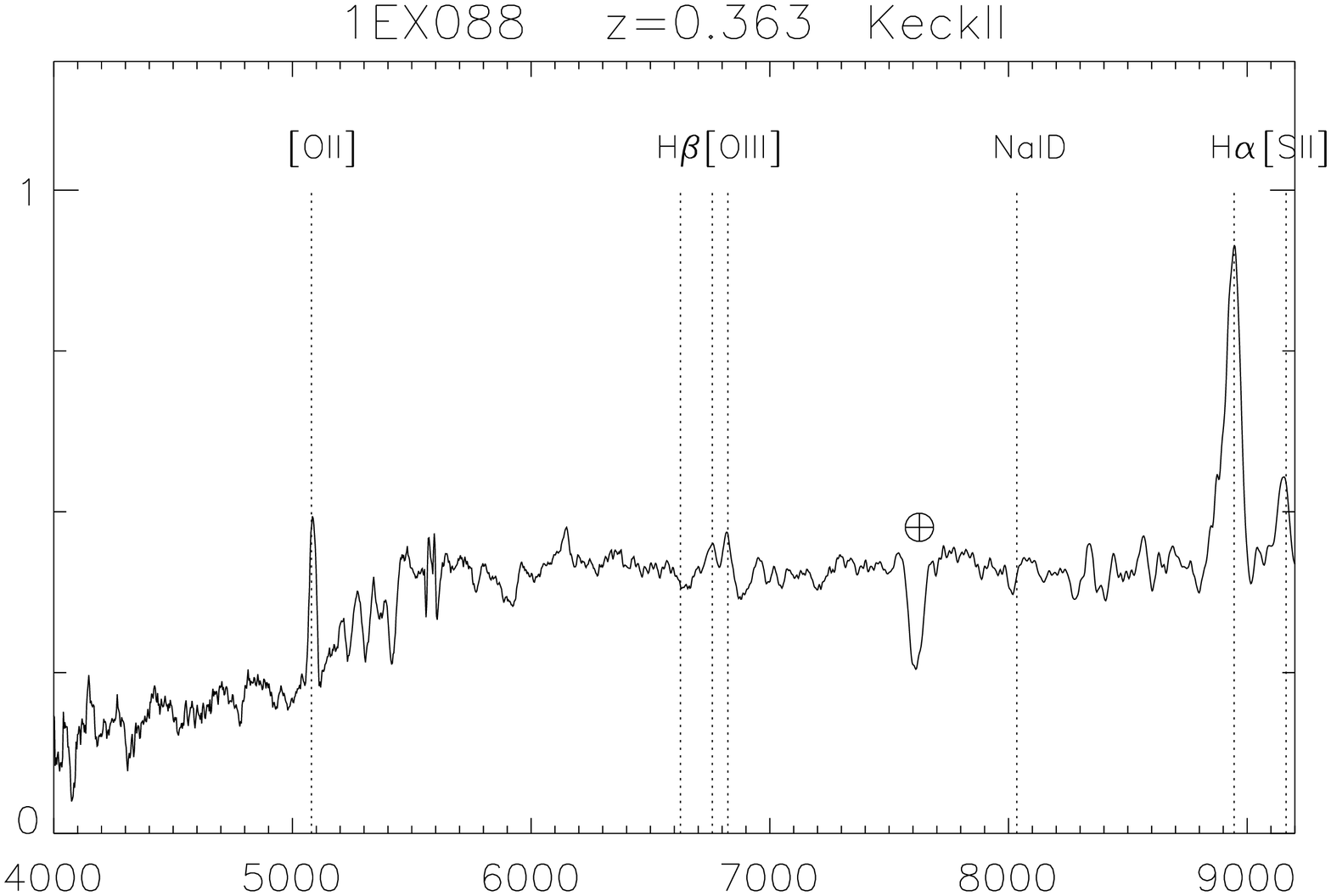}} &
      \resizebox{65mm}{!}{\includegraphics{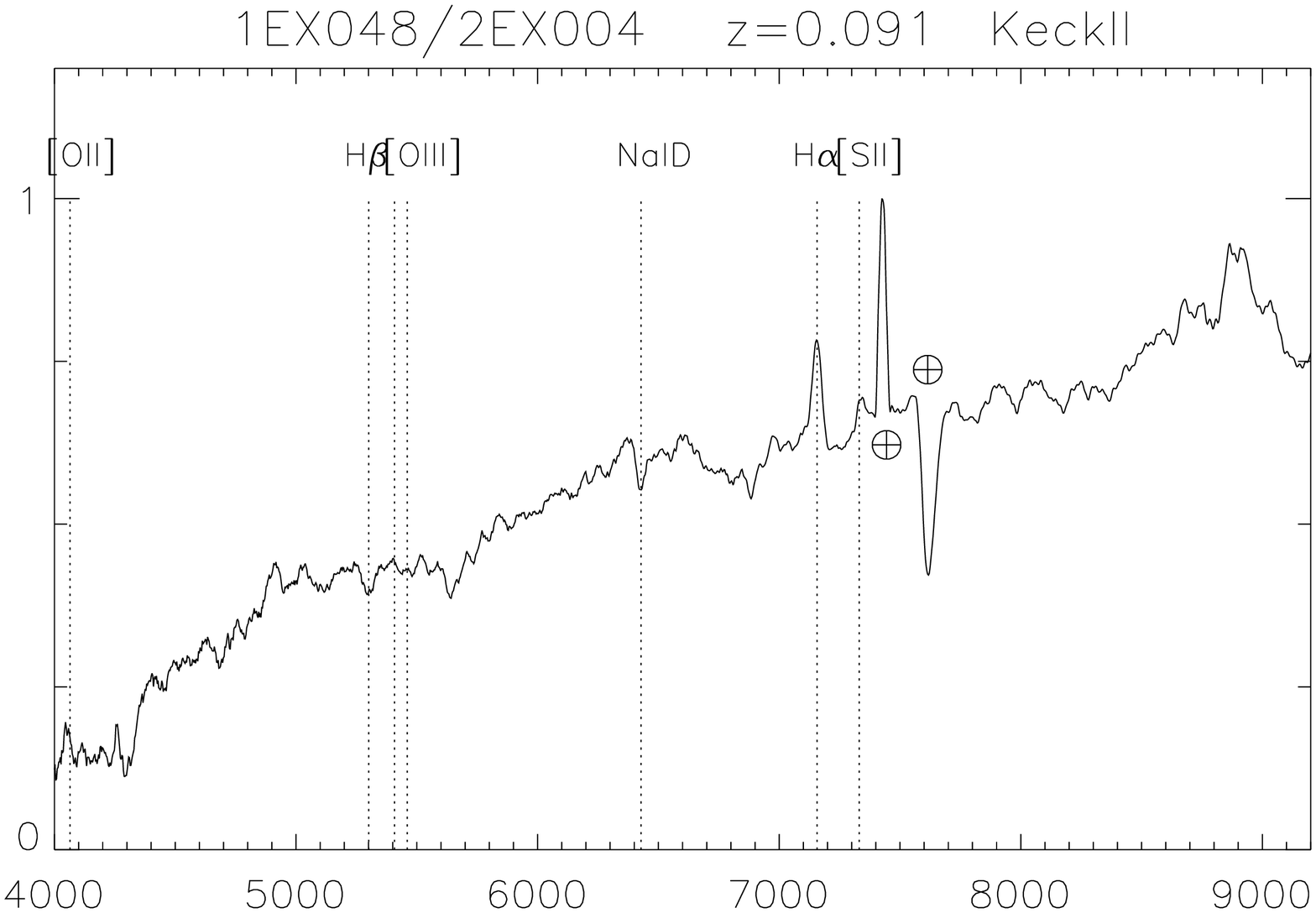}} \\
      \resizebox{65mm}{!}{\includegraphics{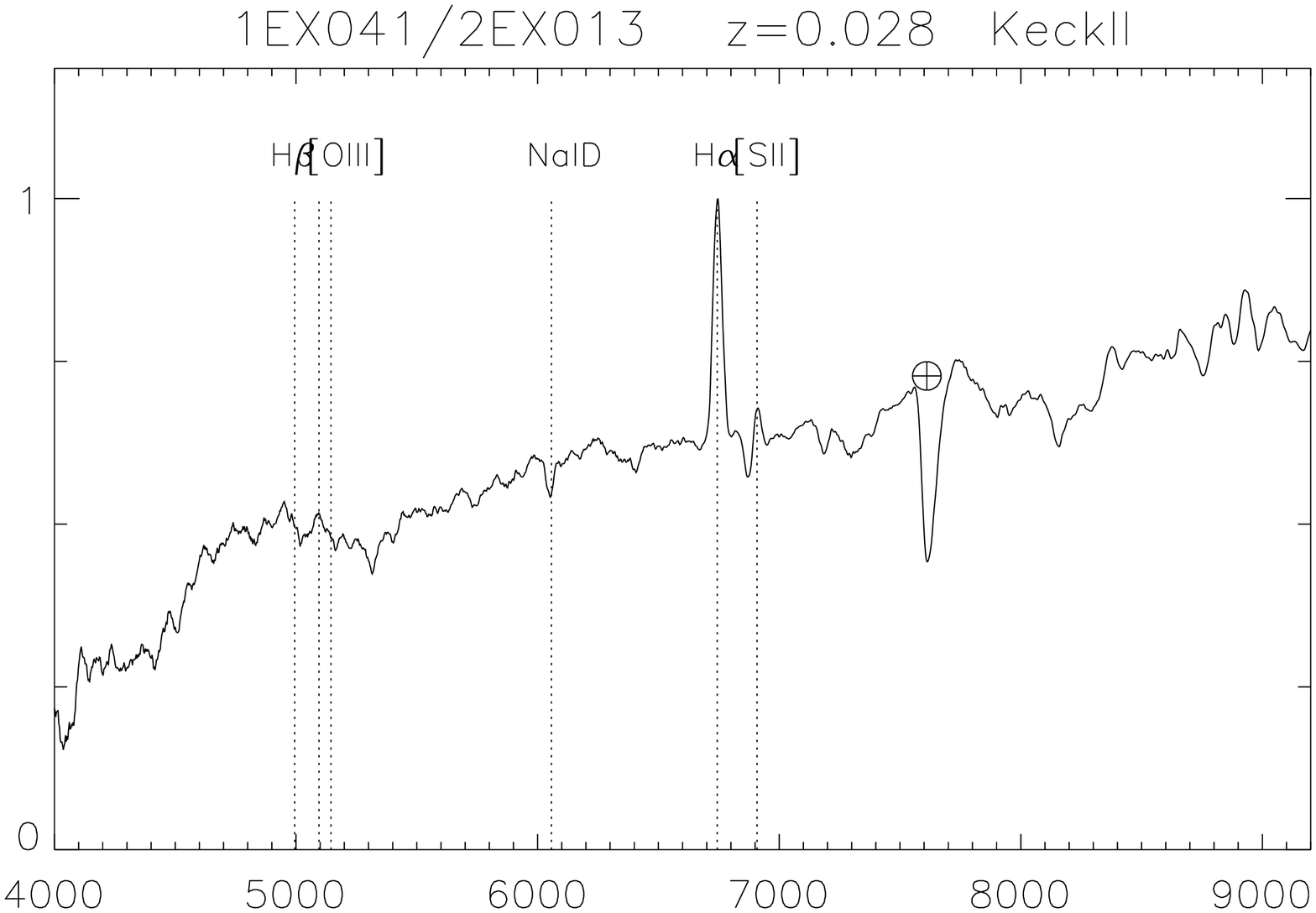}} &
      \resizebox{65mm}{!}{\includegraphics{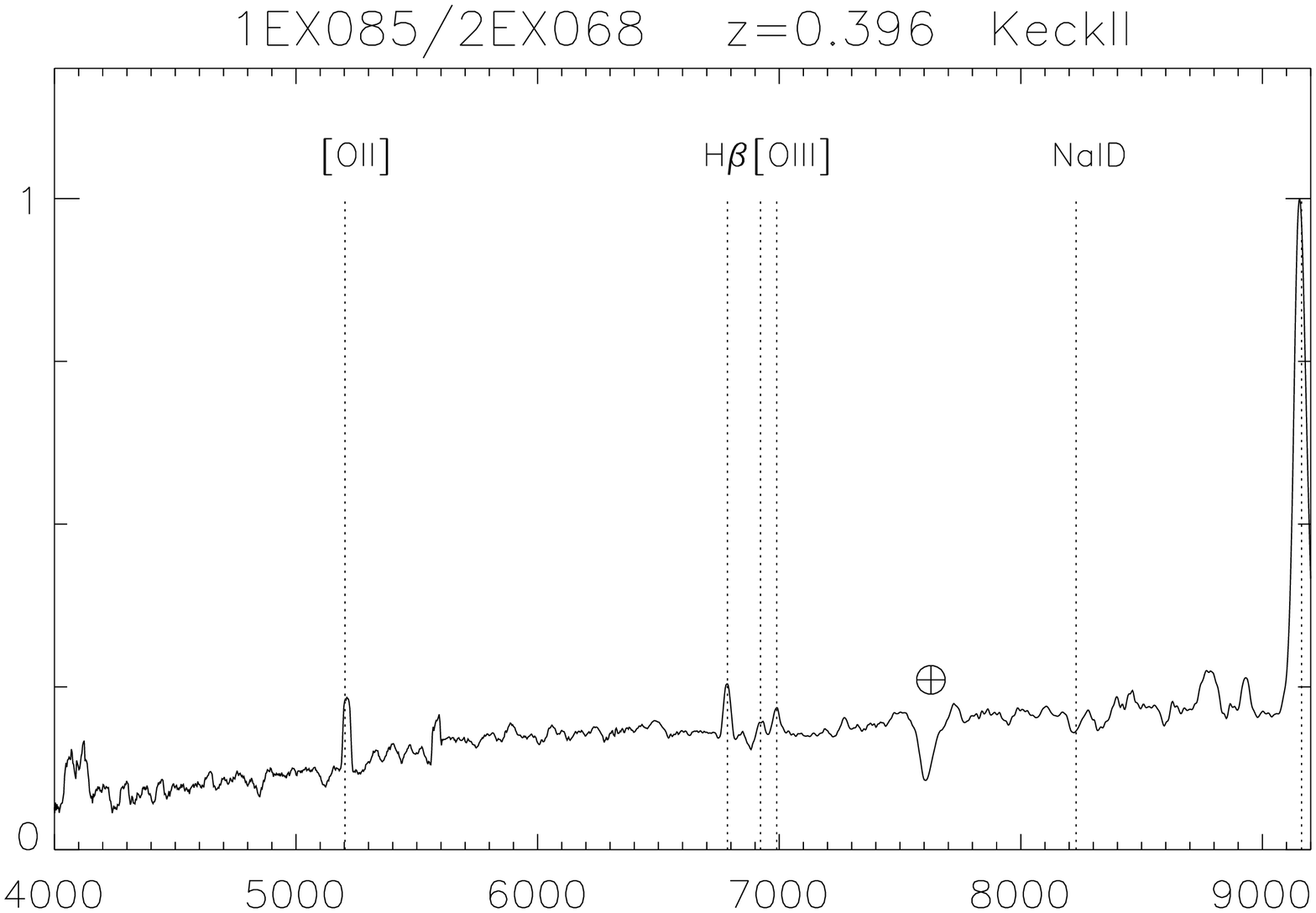}} \\
      \resizebox{65mm}{!}{\includegraphics{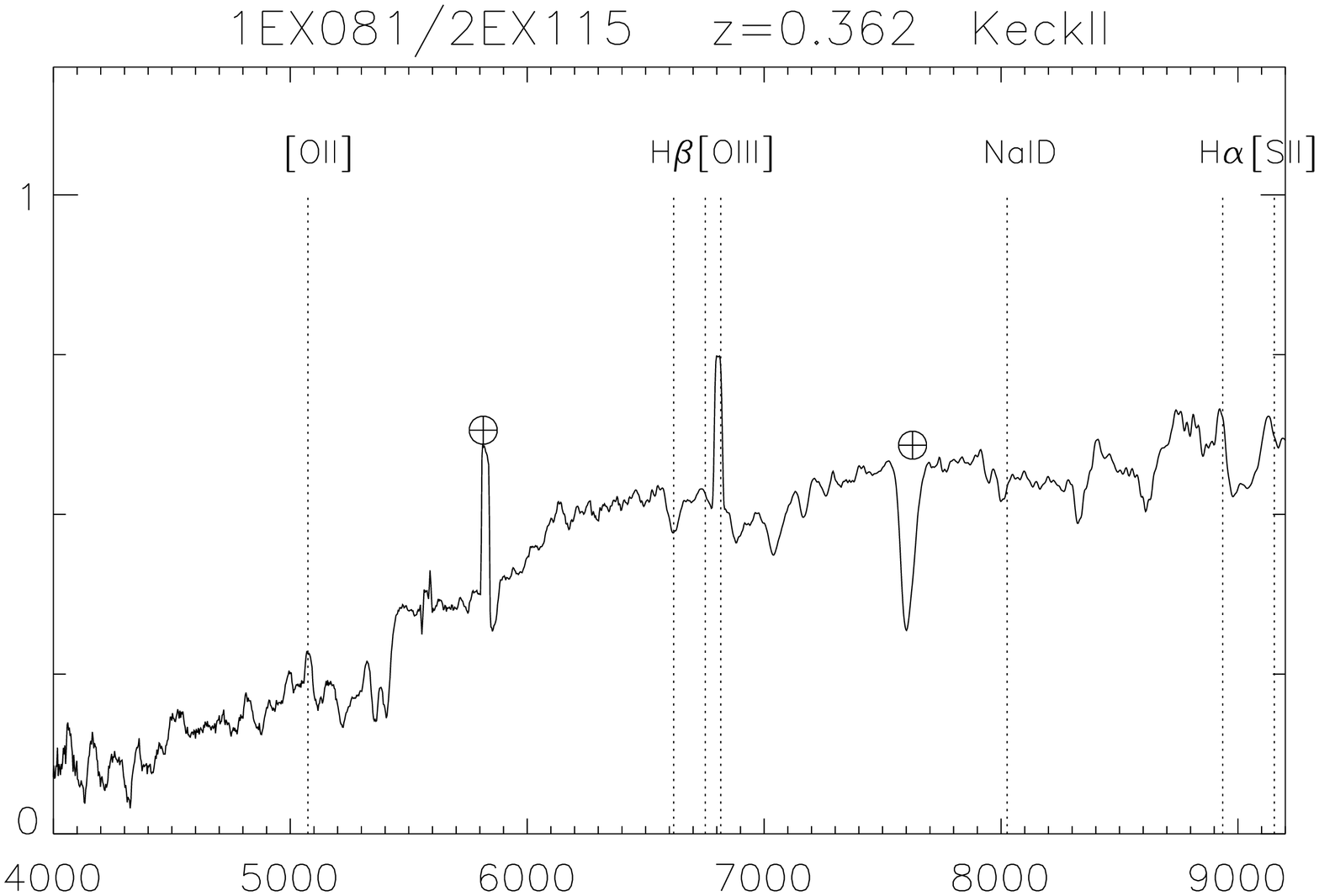}} &
      \resizebox{65mm}{!}{\includegraphics{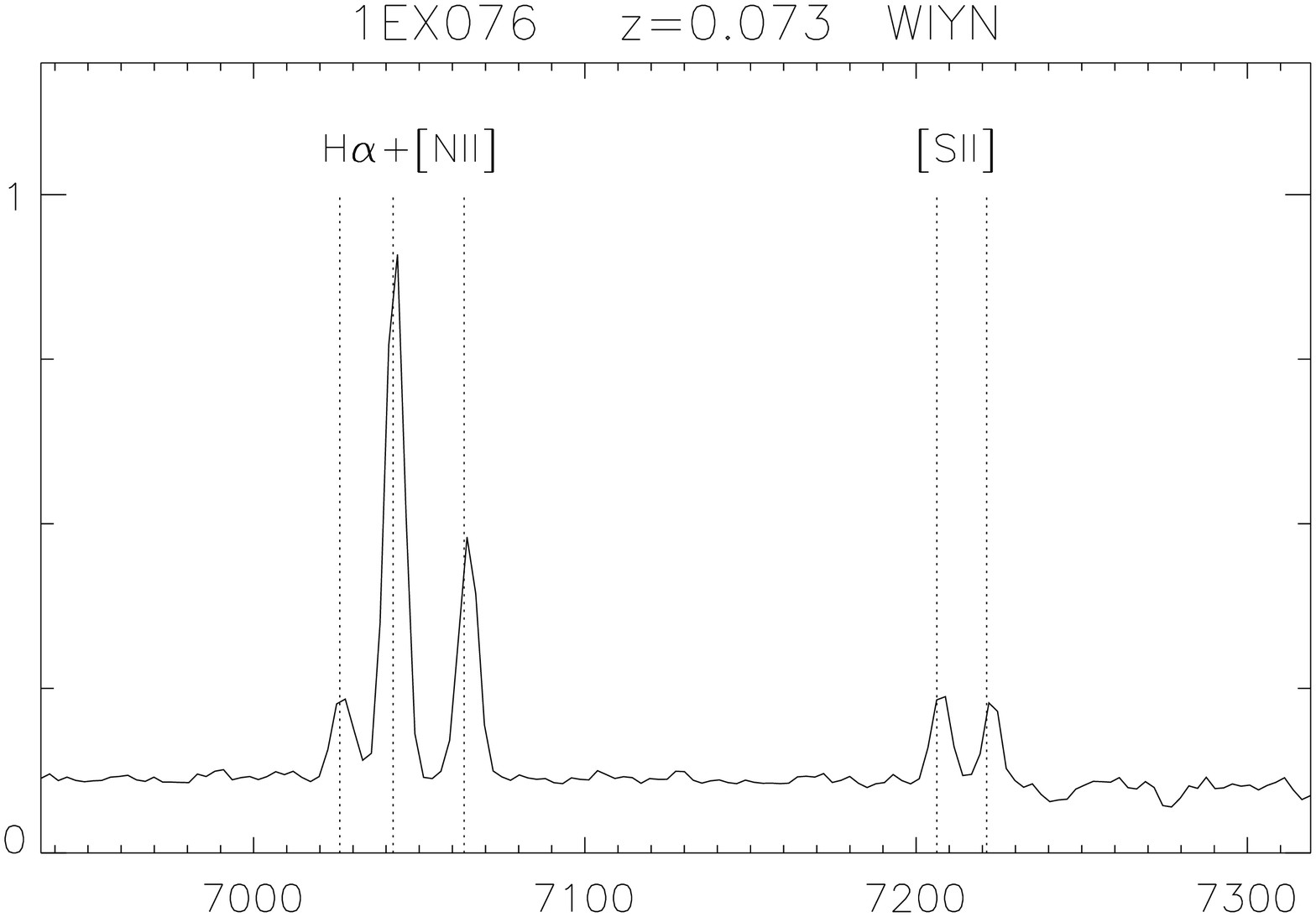}} \\      
      \resizebox{65mm}{!}{\includegraphics{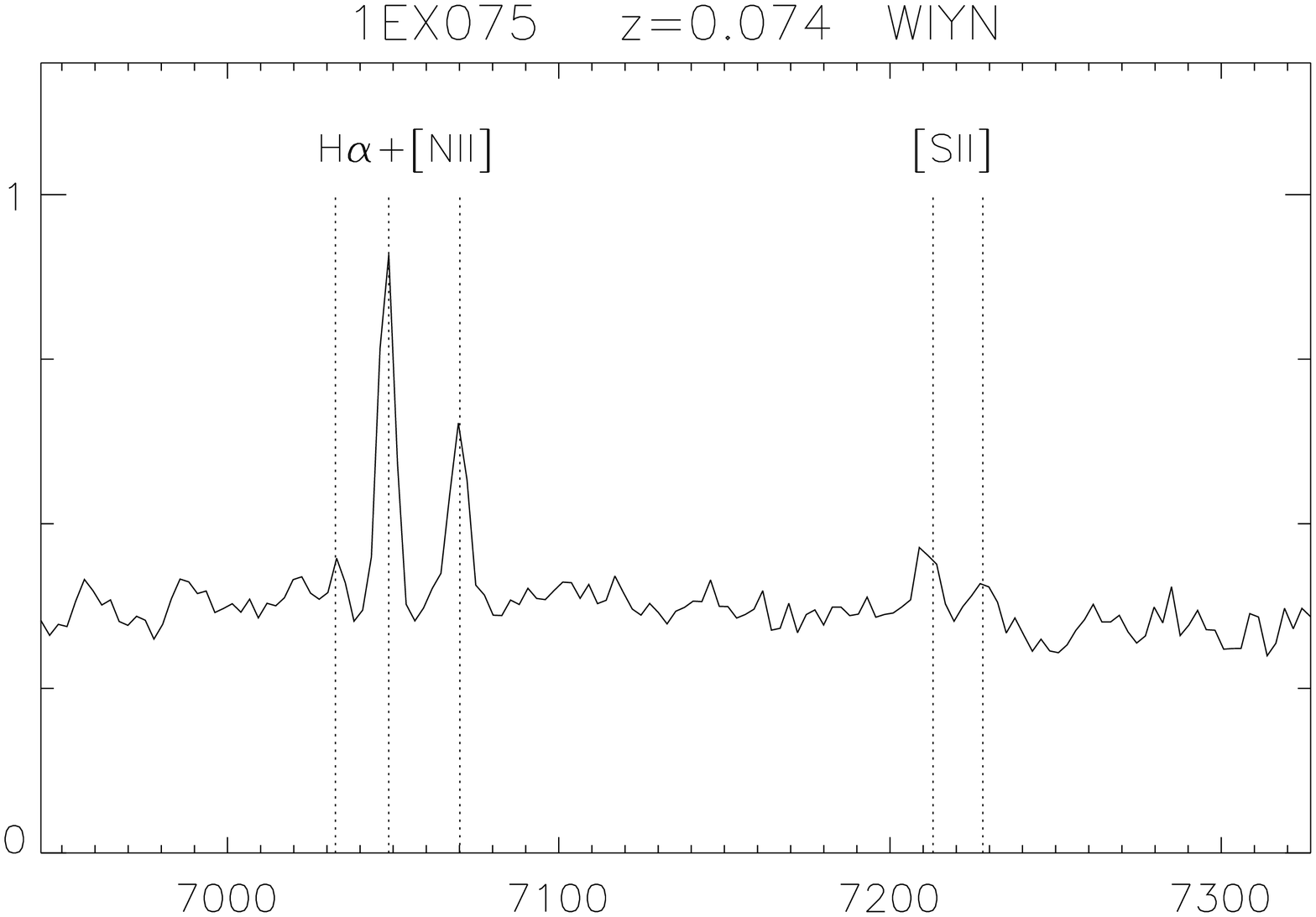}} &      
      \resizebox{65mm}{!}{\includegraphics{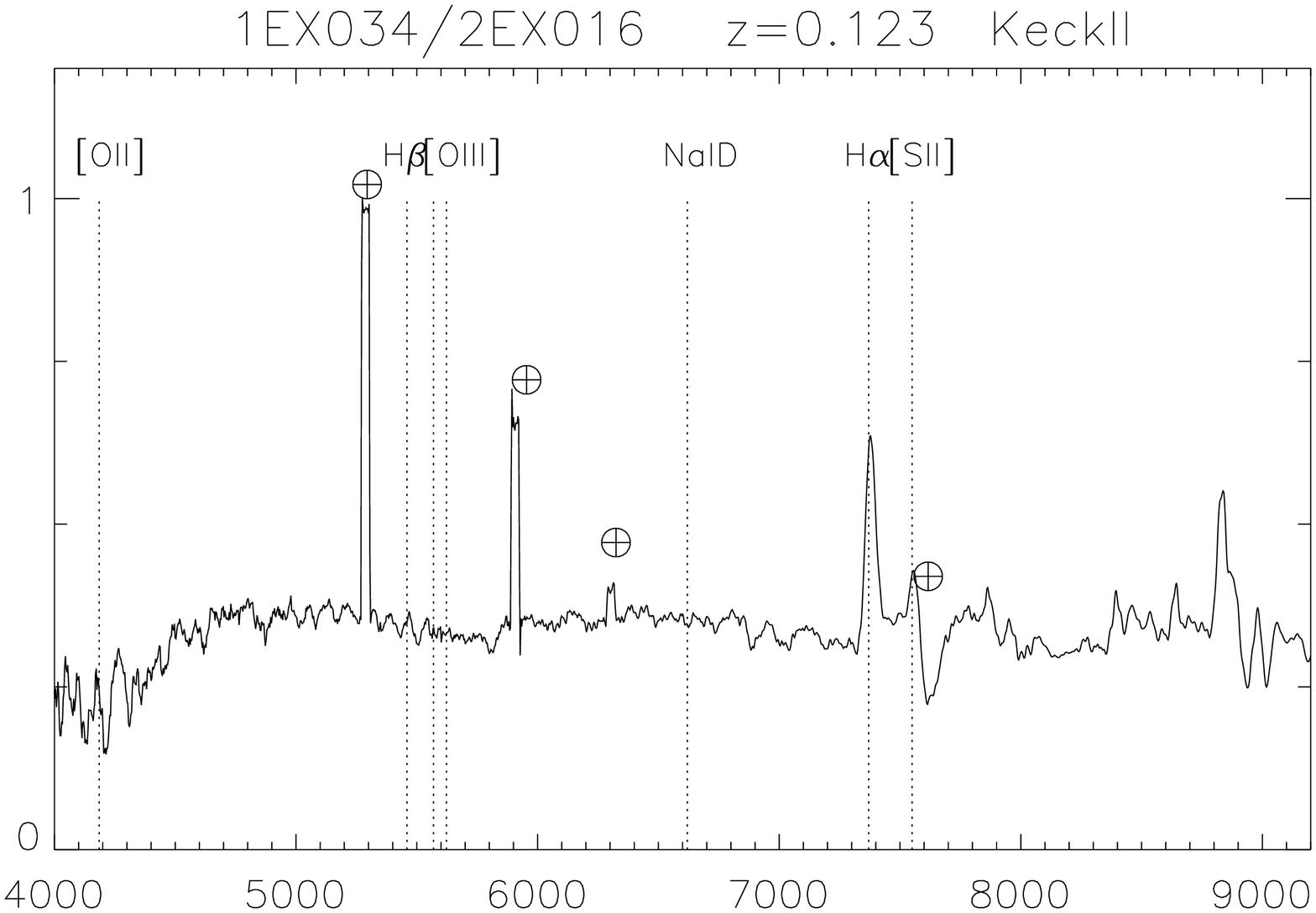}} \\
    \end{tabular}    
    \caption{Optical spectra of \ISO~FIR sources. The wavelength
      is given in \AA ngstrom and vertical axis is an arbitrary
      unit. Regions with bad pixels, residuals of night sky emission
      and atmospheric absorptions are marked with $\oplus$.}
    \label{fig:spe}
  \end{center}
\end{figure}

\clearpage
\begin{figure}
  \figurenum{9}
  \begin{center}
    \begin{tabular}{ccc}
      \resizebox{65mm}{!}{\includegraphics{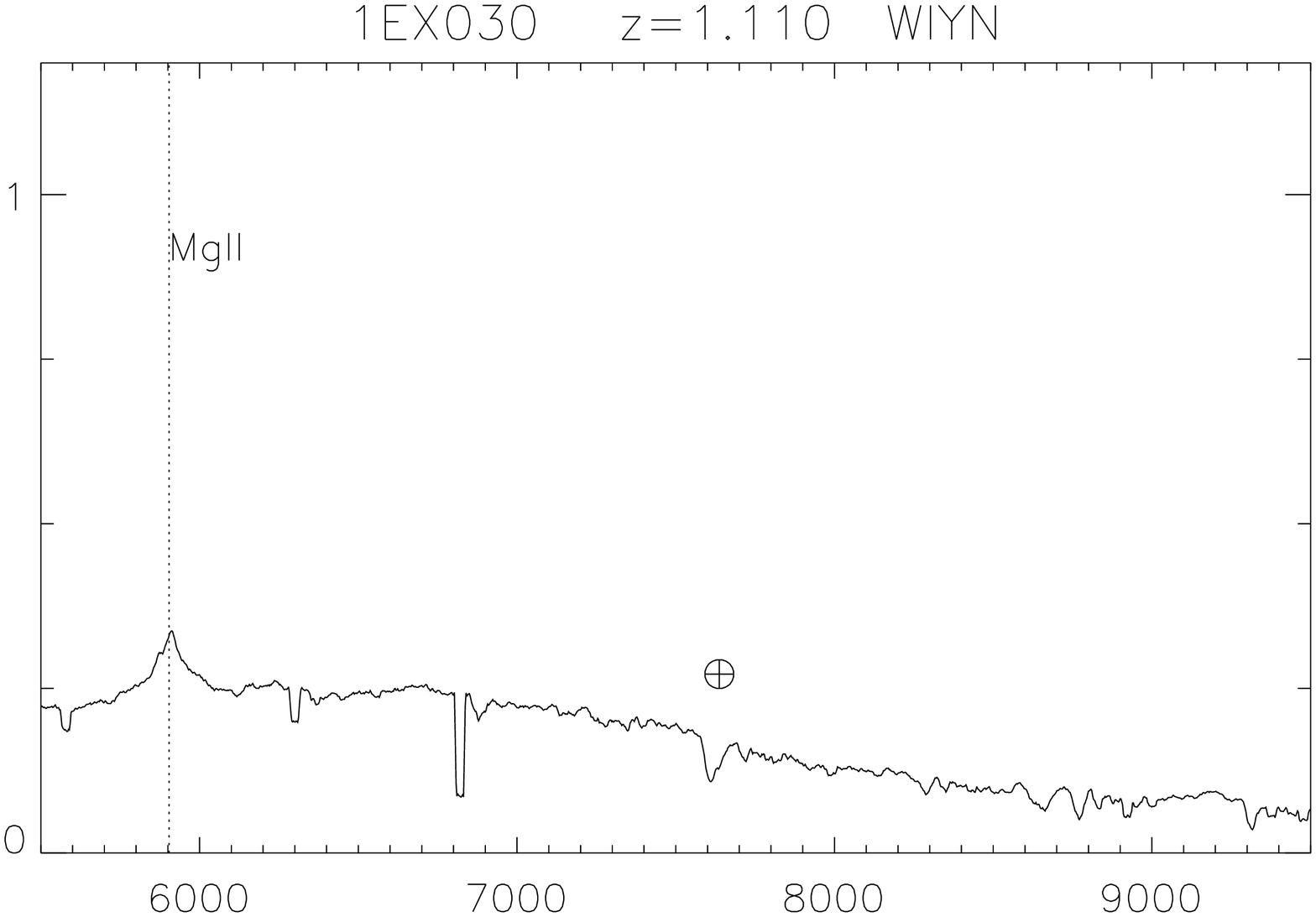}} &
      \resizebox{65mm}{!}{\includegraphics{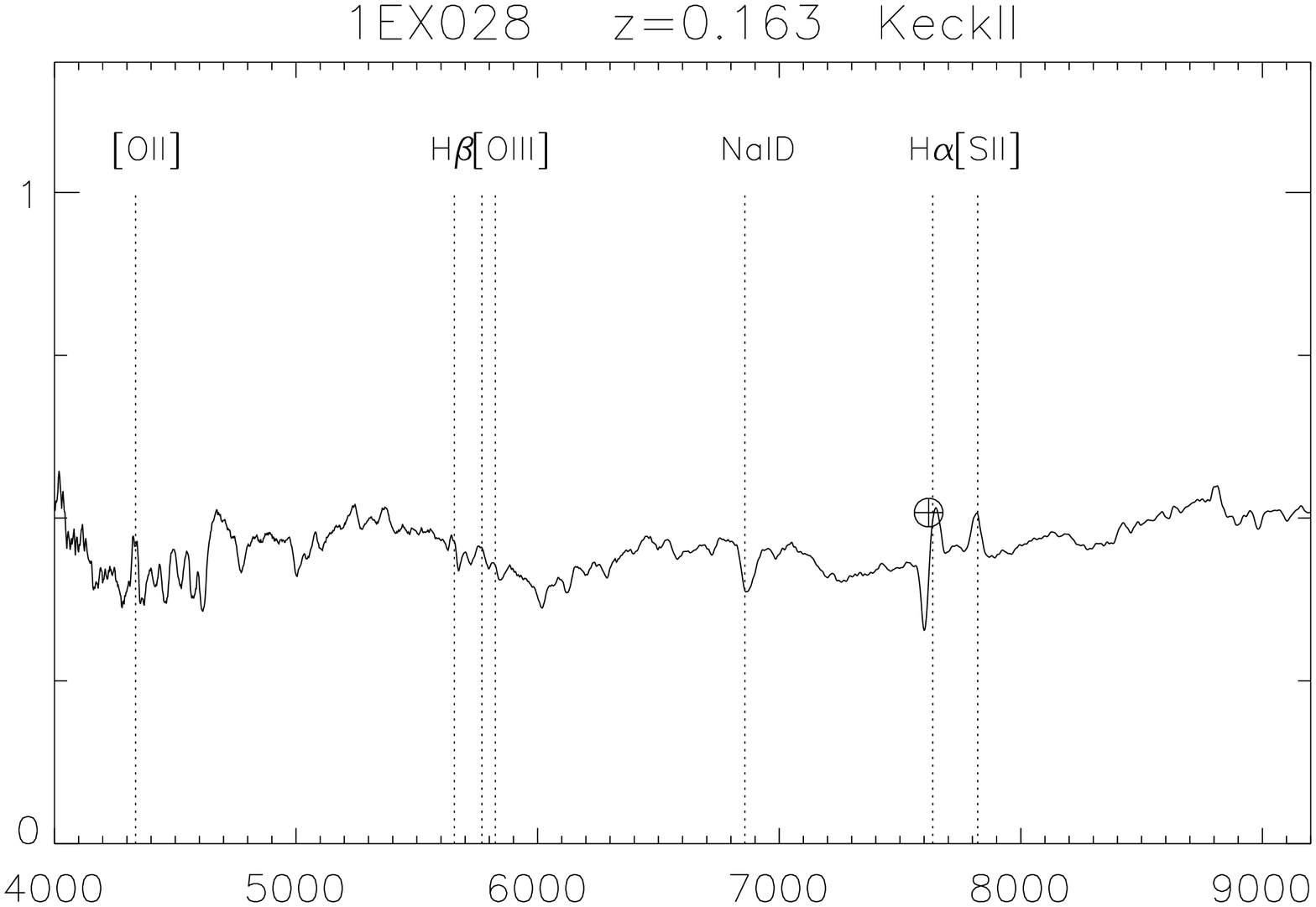}} \\
      \resizebox{65mm}{!}{\includegraphics{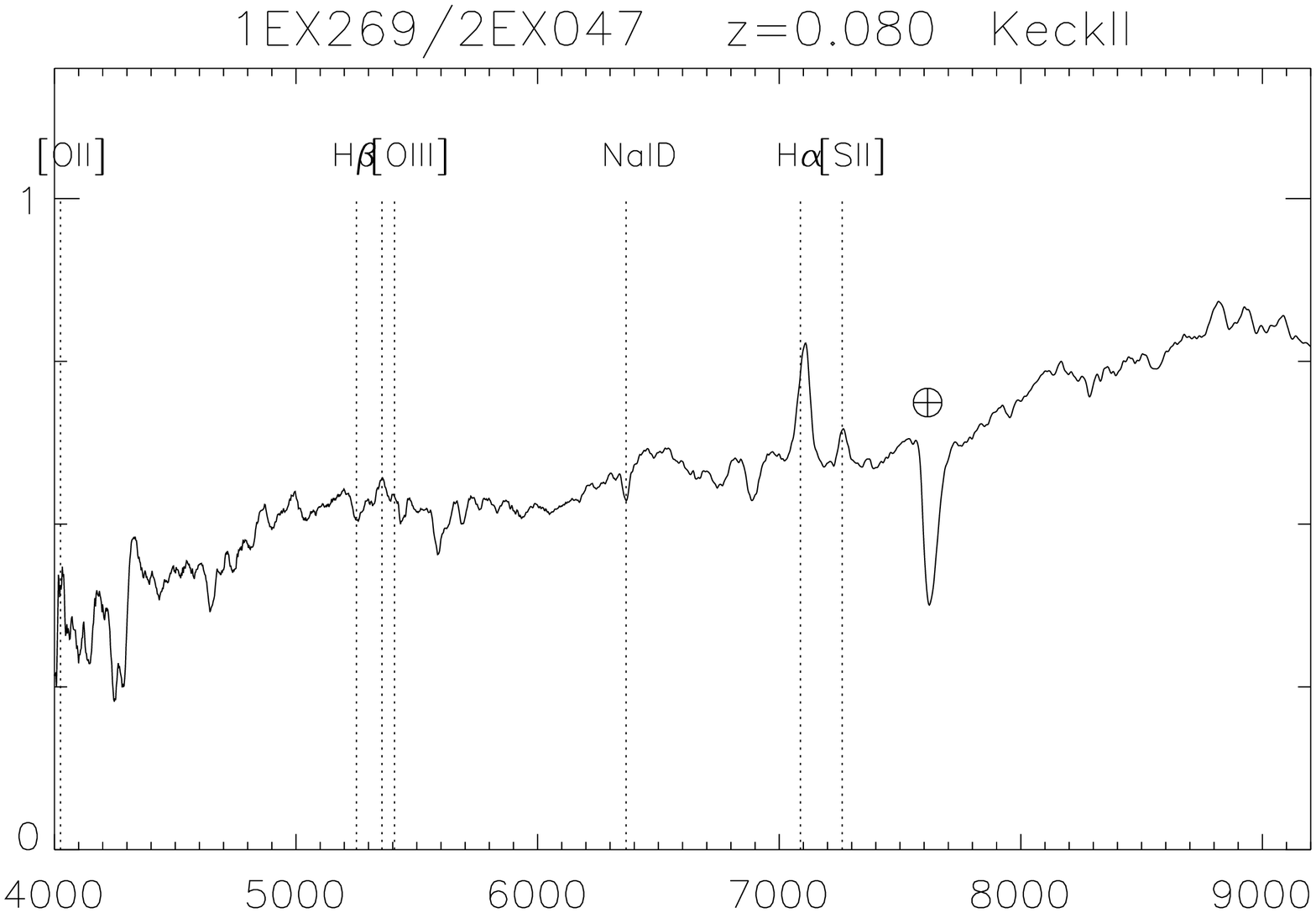}} &
      \resizebox{65mm}{!}{\includegraphics{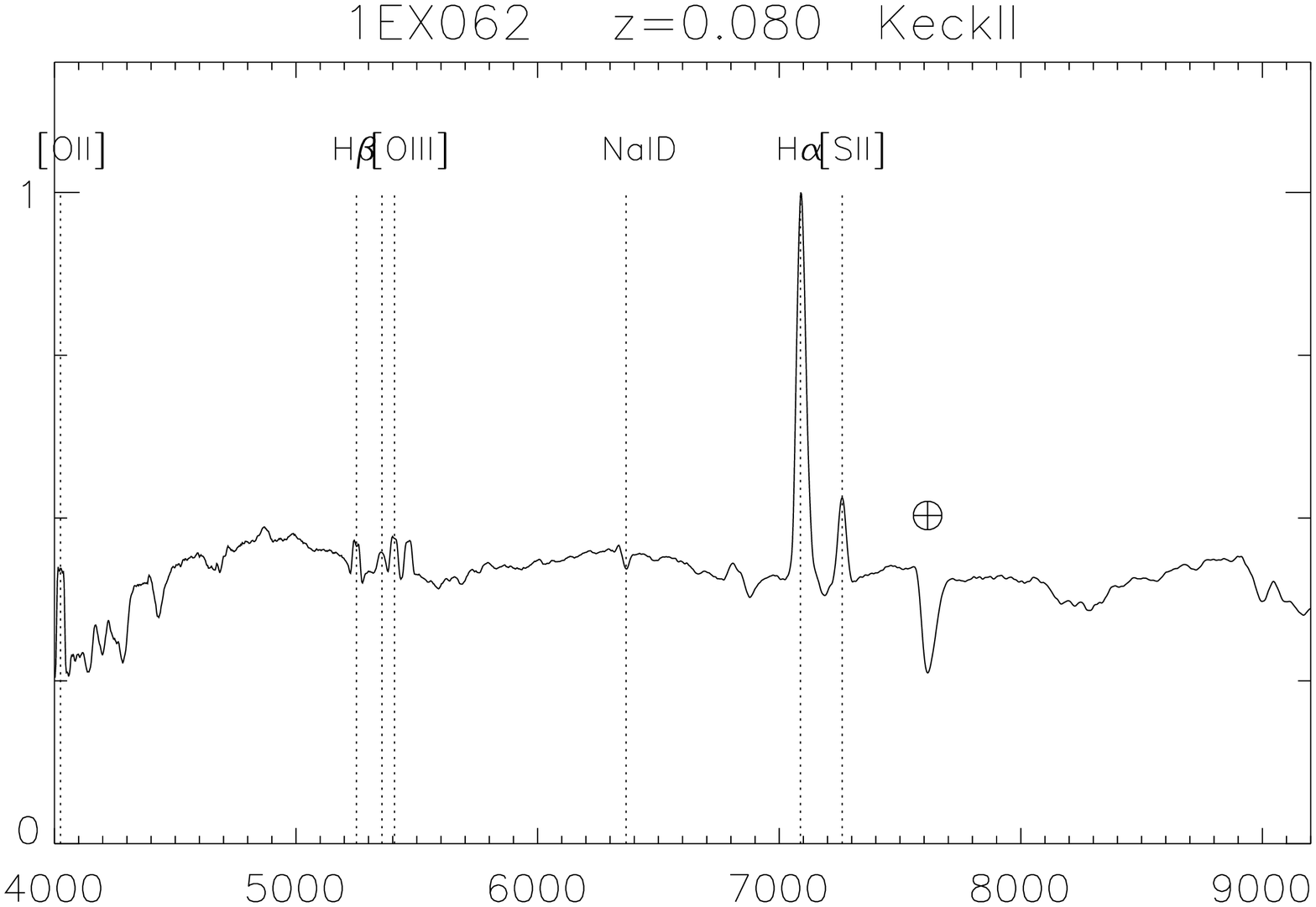}} \\
      \resizebox{65mm}{!}{\includegraphics{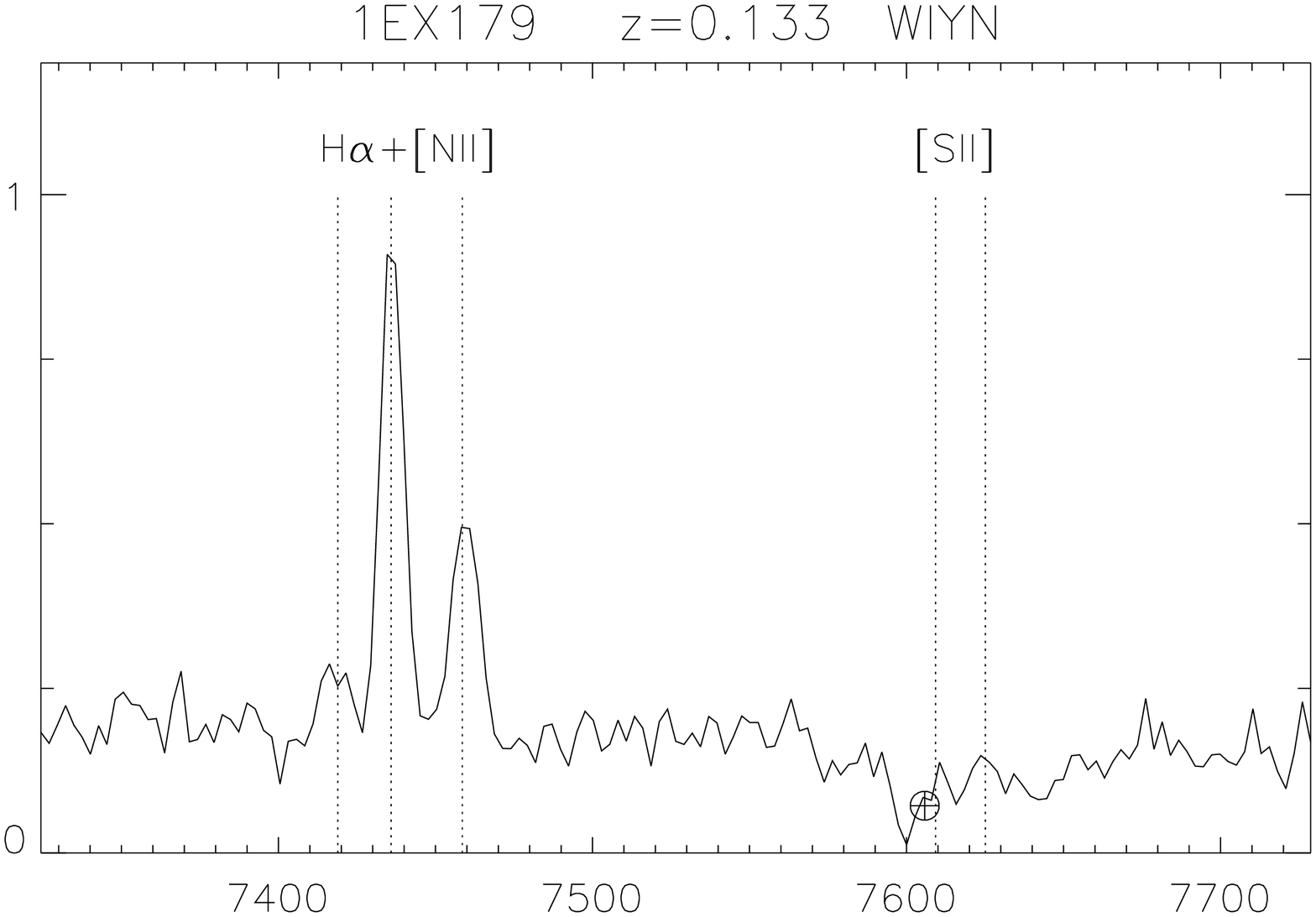}} &
      \resizebox{65mm}{!}{\includegraphics{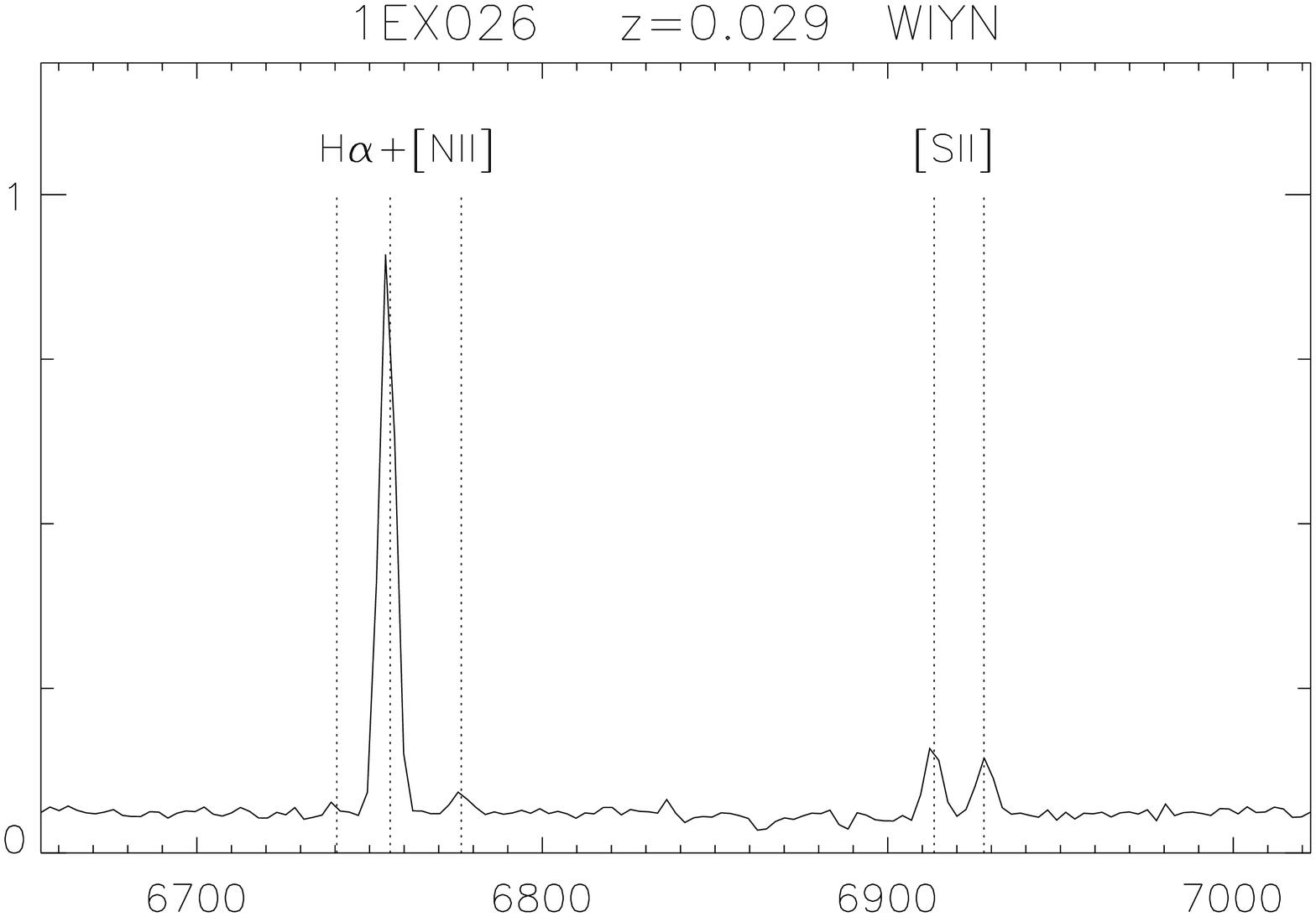}} \\
      \resizebox{65mm}{!}{\includegraphics{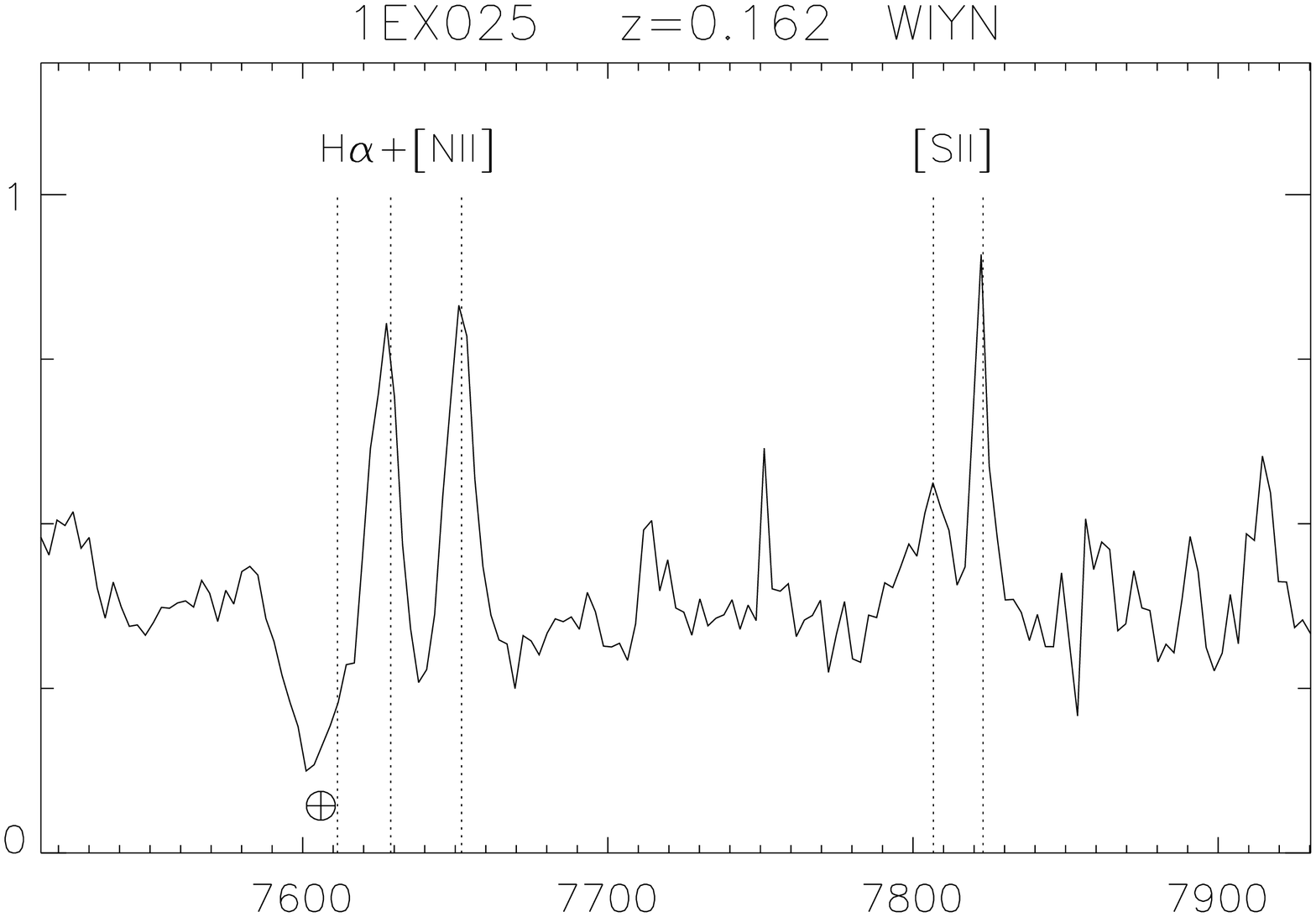}} &
      \resizebox{65mm}{!}{\includegraphics{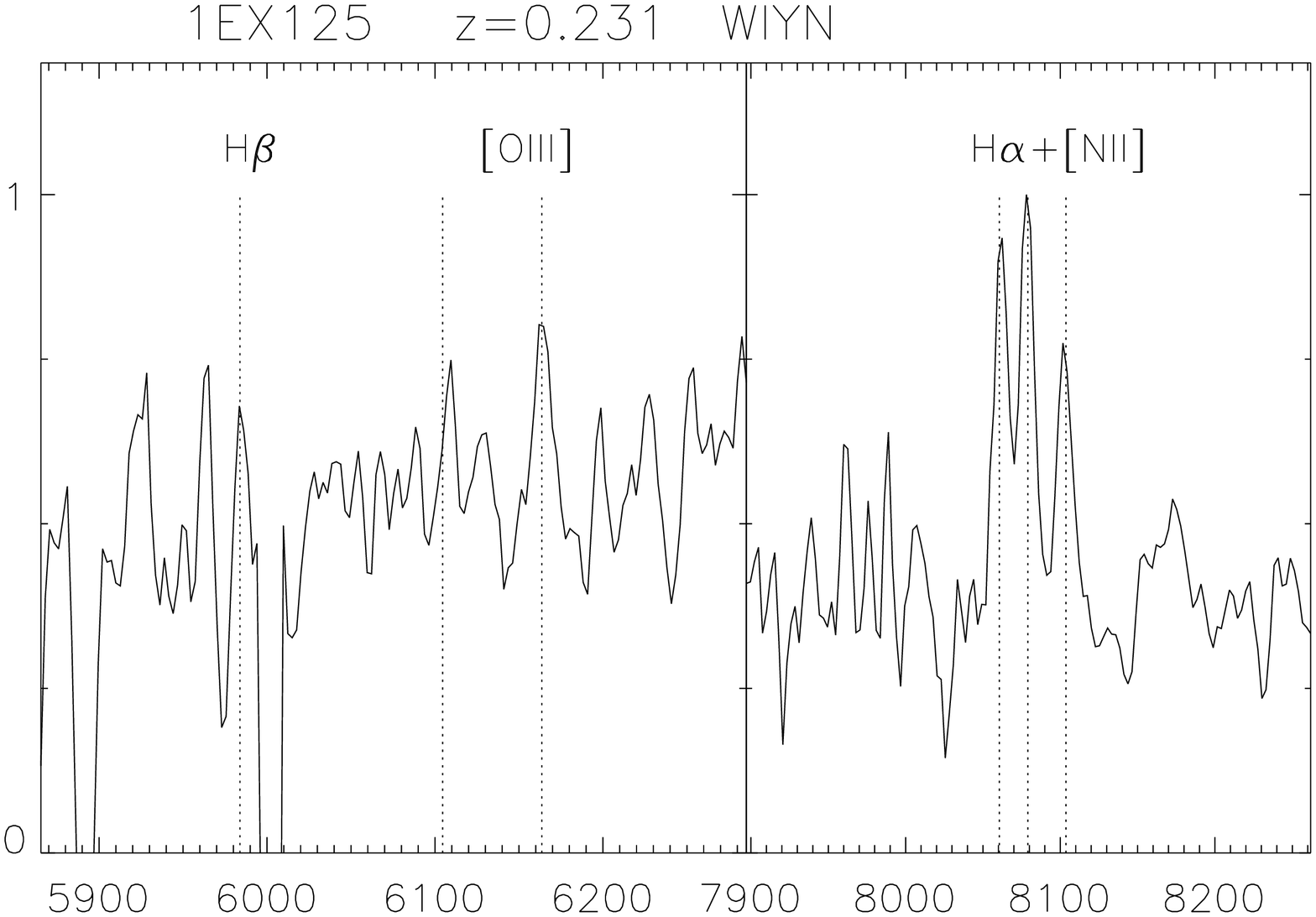}} \\
    \end{tabular}
    \caption{Continued}
  \end{center}
\end{figure}

\clearpage
\begin{figure}
  \figurenum{9}
  \begin{center}
    \begin{tabular}{ccc}
      \resizebox{65mm}{!}{\includegraphics{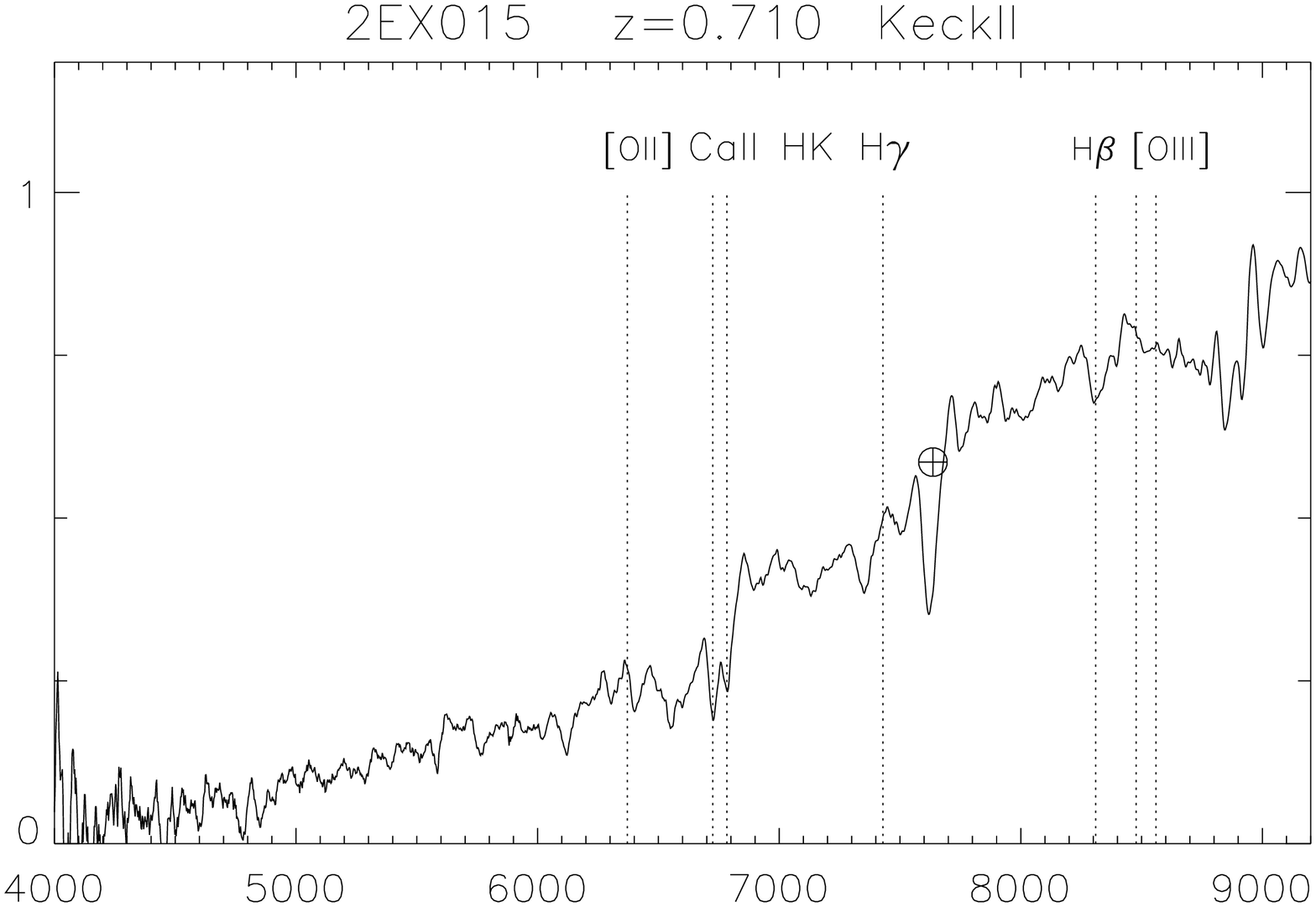}} &
      \resizebox{65mm}{!}{\includegraphics{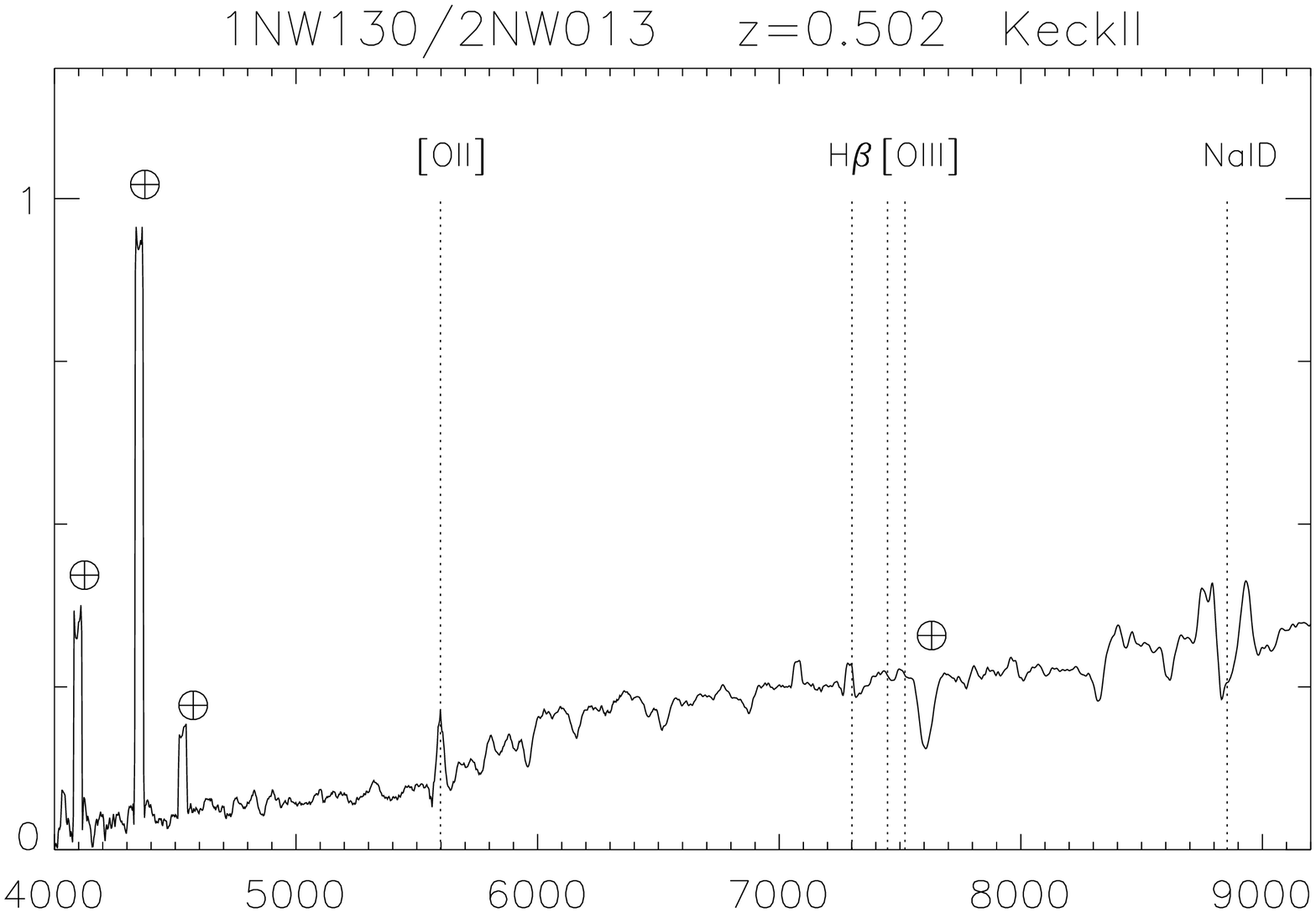}} \\
      \resizebox{65mm}{!}{\includegraphics{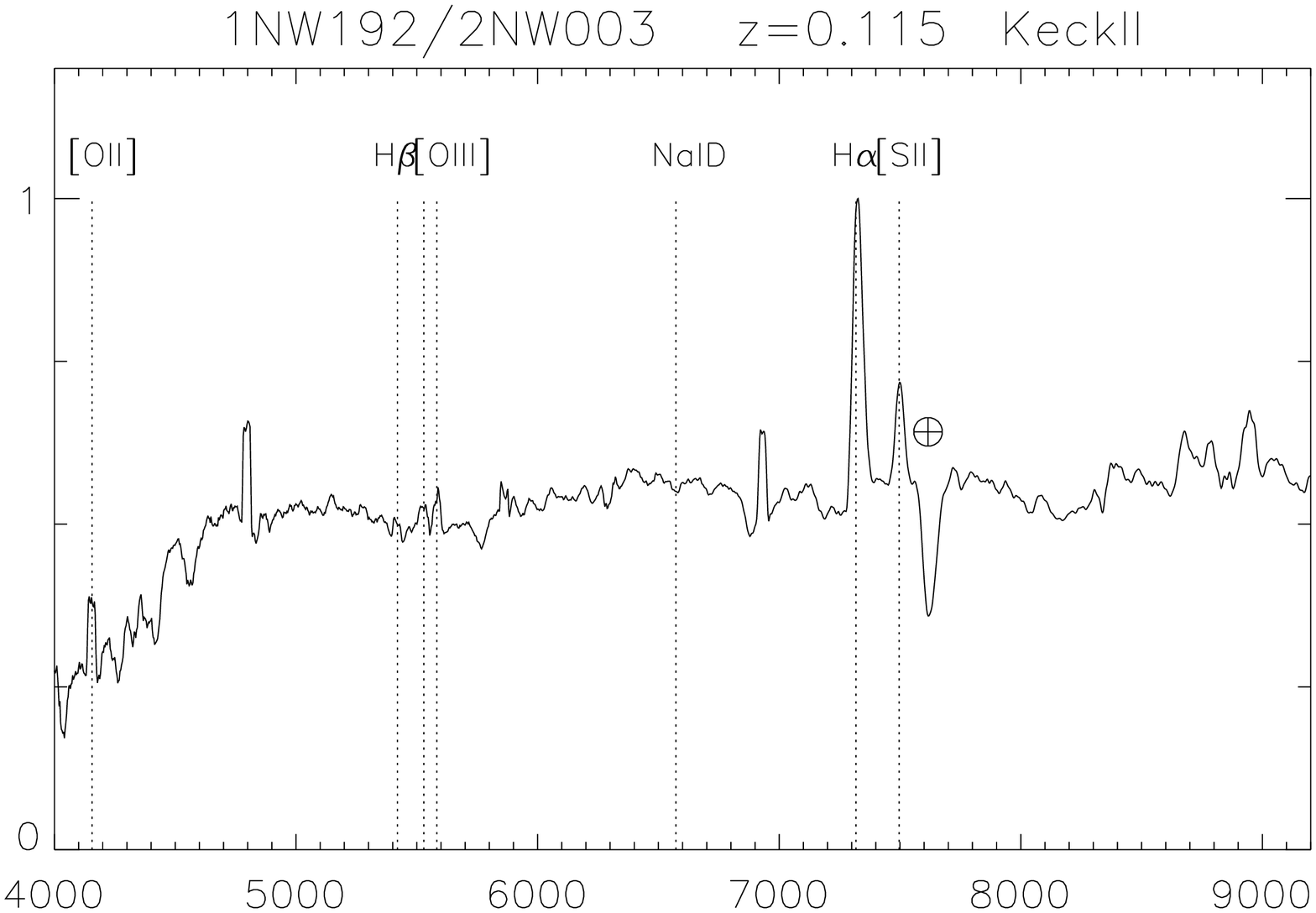}} &
      \resizebox{65mm}{!}{\includegraphics{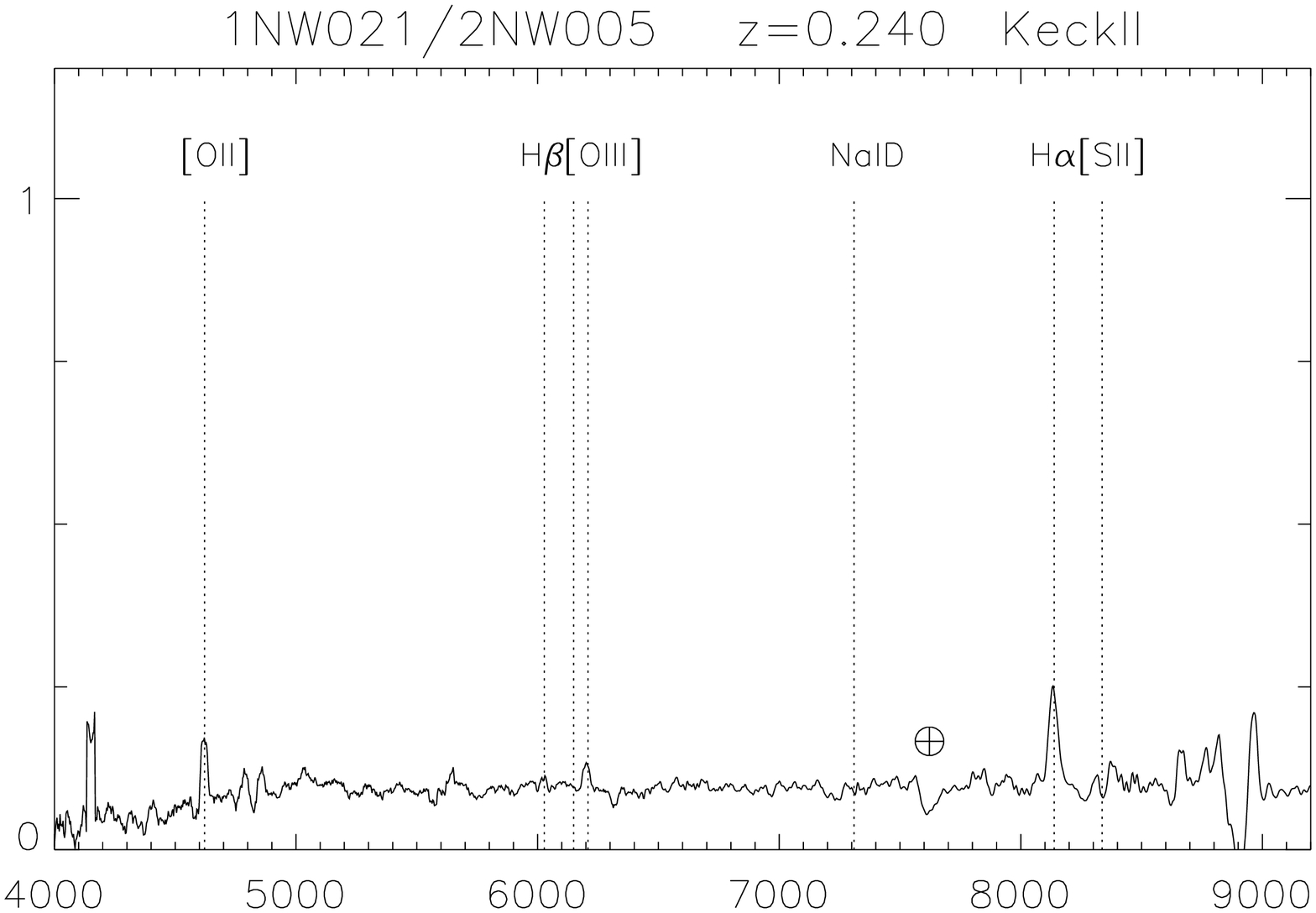}} \\
      \resizebox{65mm}{!}{\includegraphics{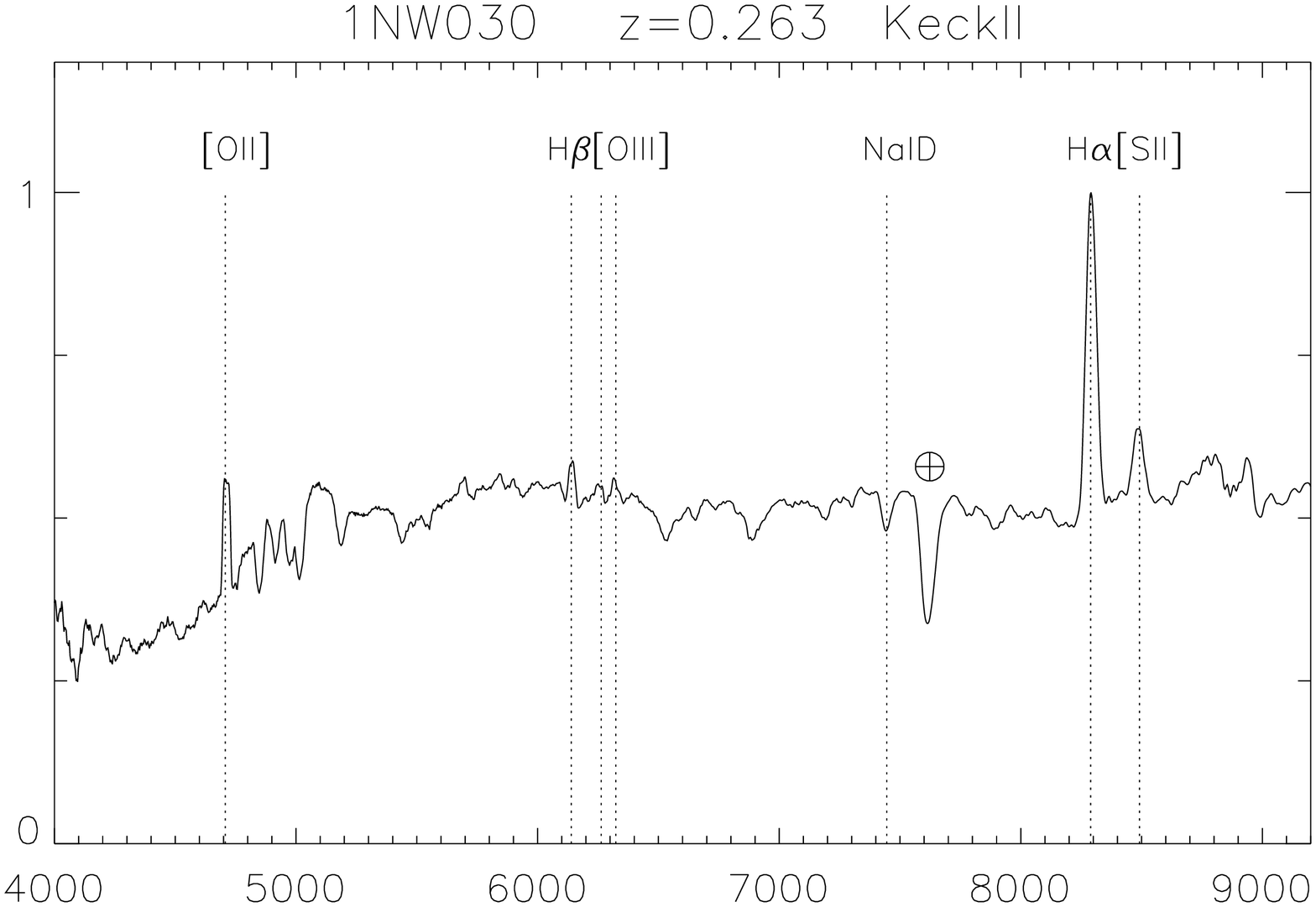}} &
      \resizebox{65mm}{!}{\includegraphics{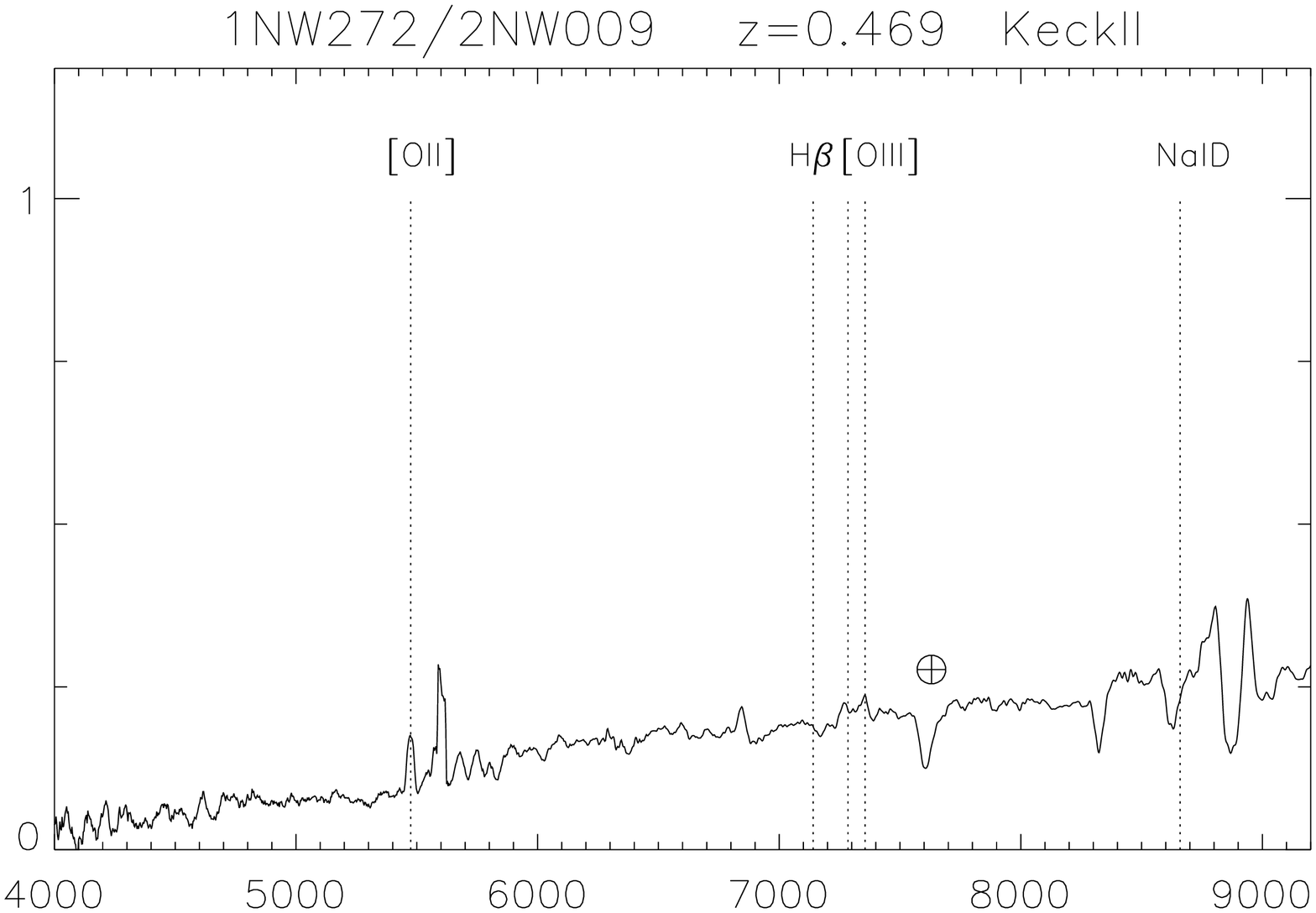}} \\
      \resizebox{65mm}{!}{\includegraphics{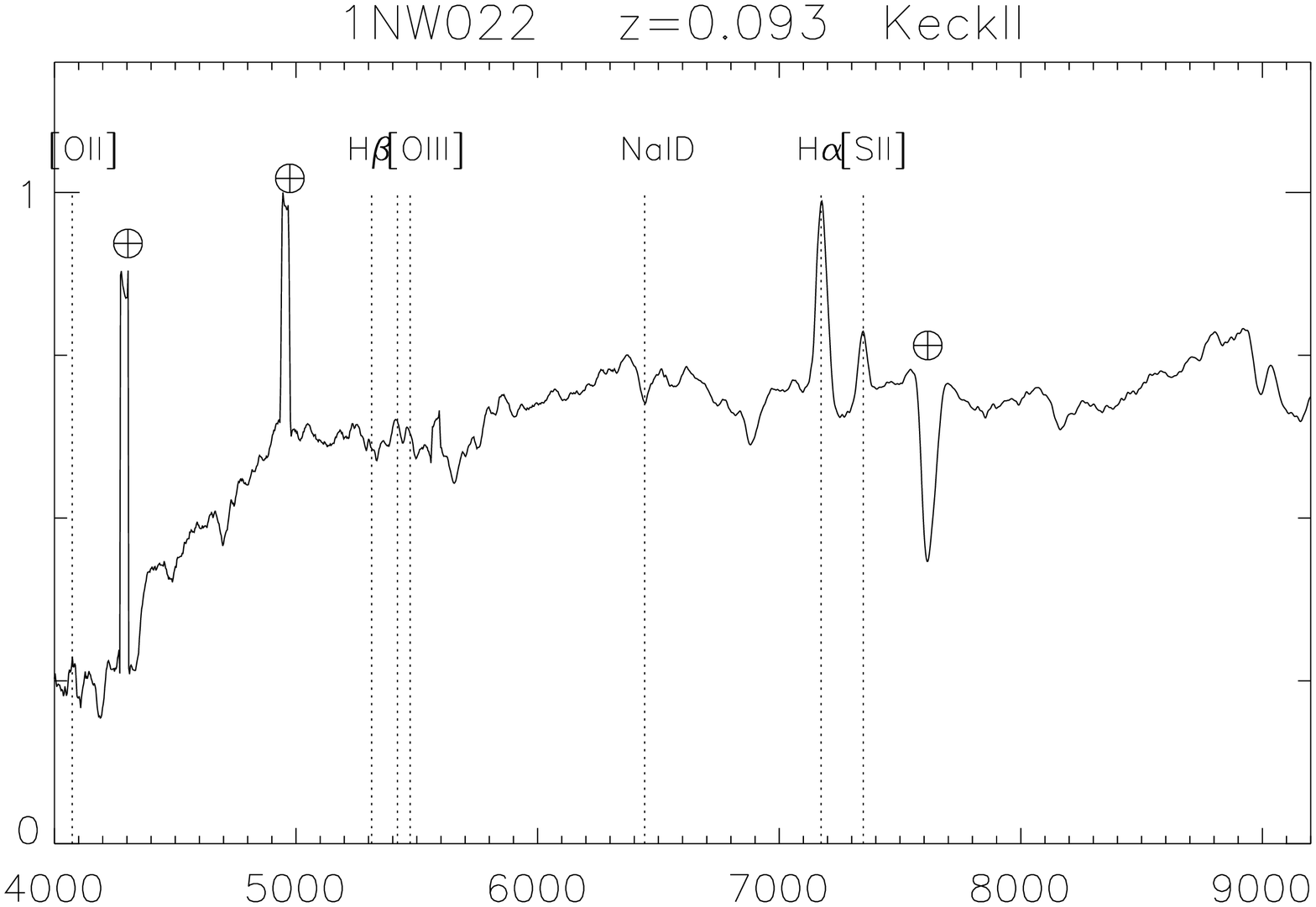}} &
      \resizebox{65mm}{!}{\includegraphics{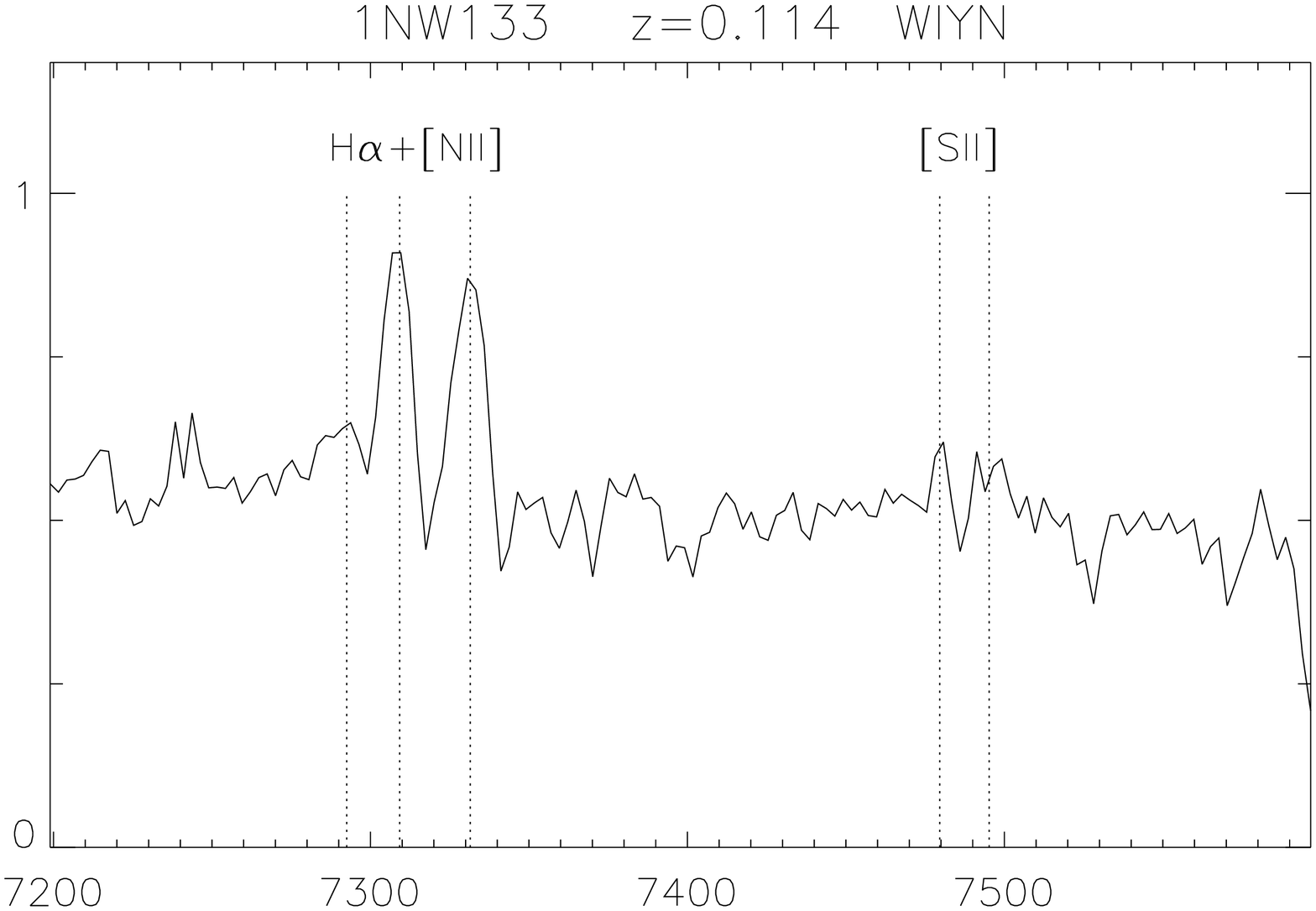}} \\
    \end{tabular}
    \caption{Continued}
  \end{center}
\end{figure}

\clearpage
\begin{figure}
  \figurenum{9}
  \begin{center}
    \begin{tabular}{ccc} 
      \resizebox{65mm}{!}{\includegraphics{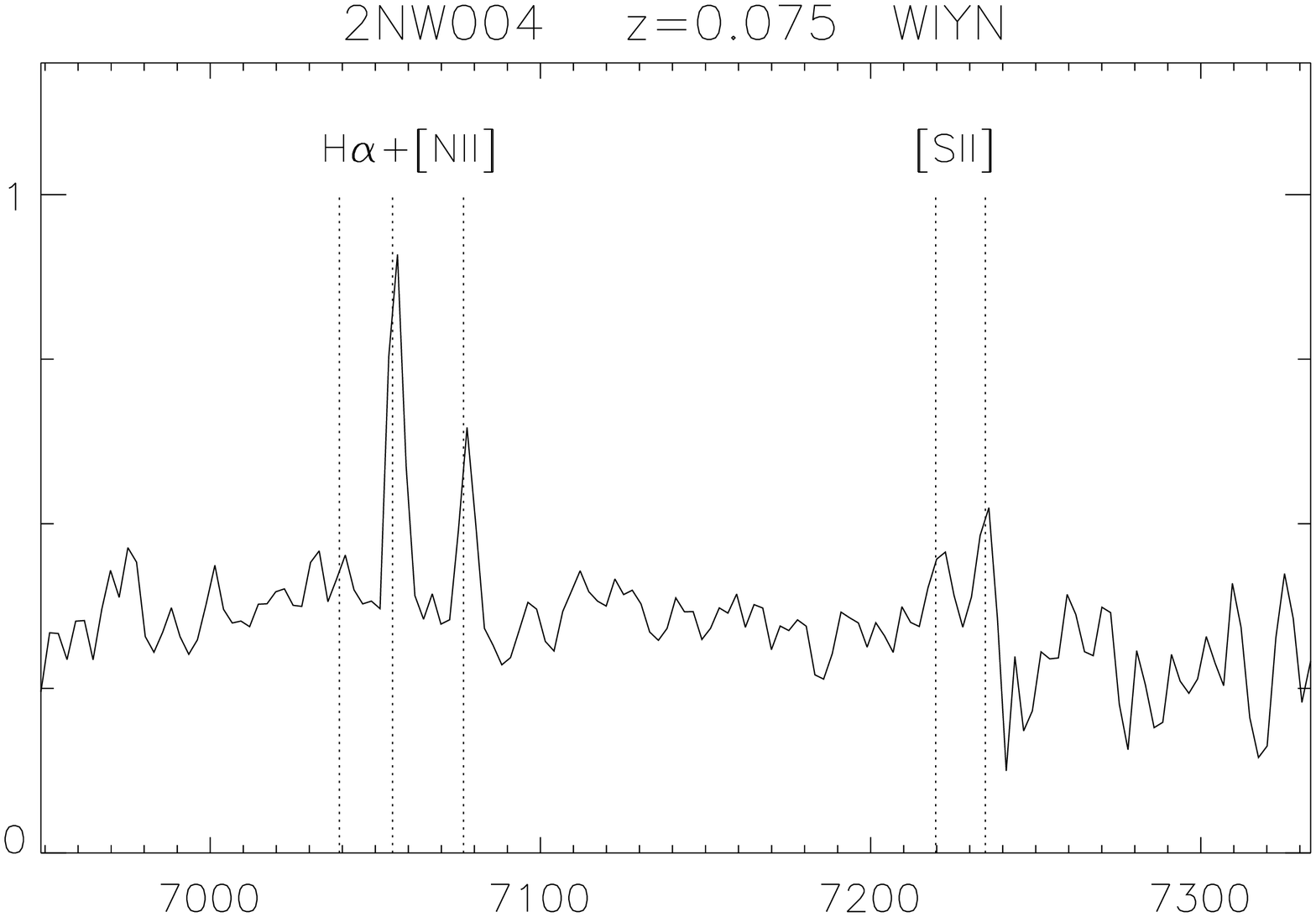}} \\
    \end{tabular}
    \caption{Continued}
  \end{center}
\end{figure}
\clearpage

\begin{figure}[htbp]
  \plotone{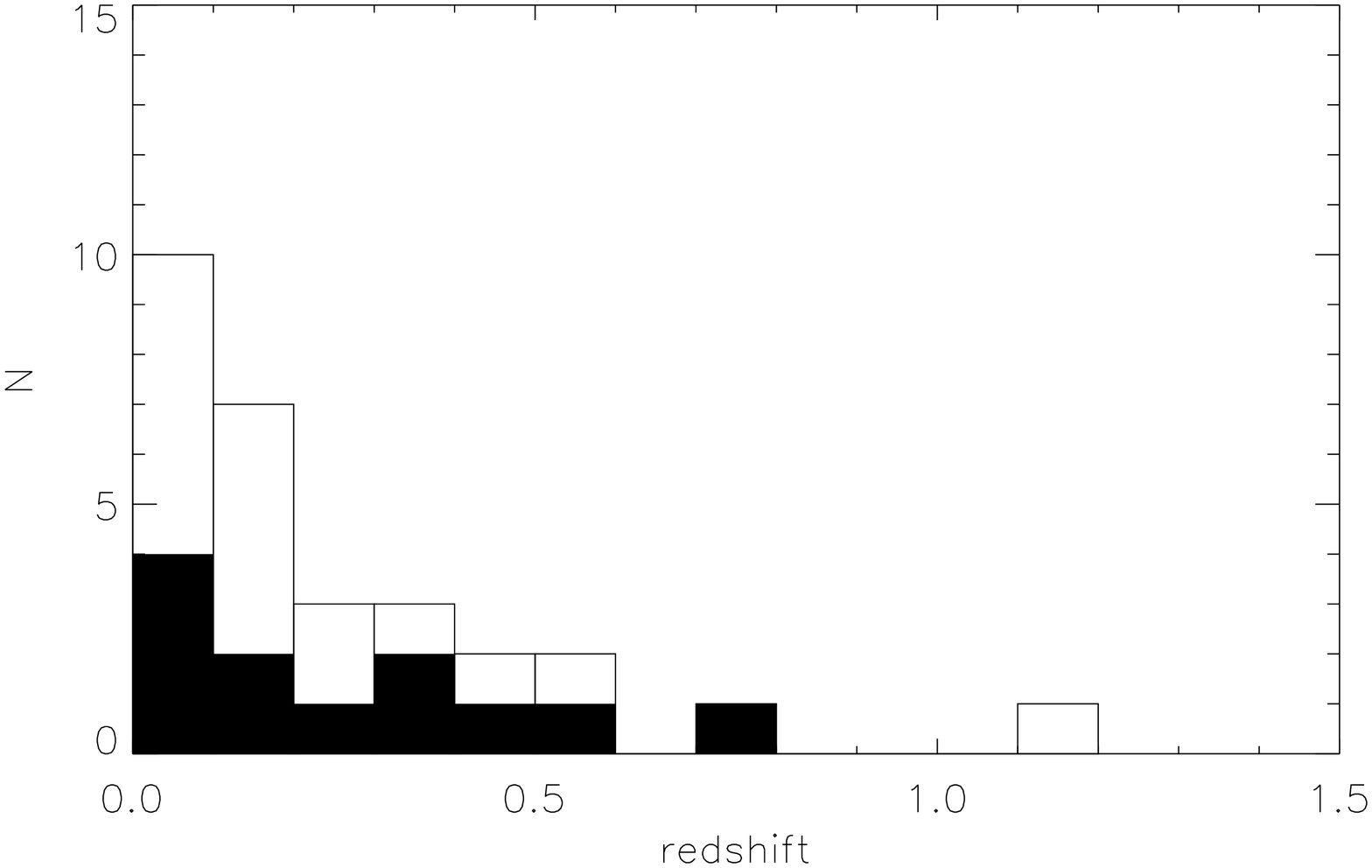}
  \caption{Redshift distributions of $ISO$ FIR sources in the Lockman Hole.
    The open histogram show all the sources while the shaded histogram plots 
    the 170\micron~detected sources. 
  } 
  \label{fig:hist}
\end{figure}

The redshift distributions of all FIR sources and 170\micron-detected 
FIR sources 
are shown in Figure~\ref{fig:hist} by open and filled histograms. 
Although more distant sources are expected at 170\micron~than at 
90\micron~because of the 
strong k-correction brightening for dust emission, there are 
no significant differences in the redshift distribution between 
the sources detected at 170 \micron\ and the sources detected only
at 90 \micron.  Seven out of twelve (58\%)
170\micron-detected source and nineteen out of twenty seven (70\%)
90\micron-only-detected sources lie 
at $z<0.3$. This may be attributed to the small number statistics. 
\citet{patris} reported that 95\% (20/21) of 170\micron\ sources brighter 
than 200 mJy found in their FIRBACK southern Marano fields 
are at $z<0.3$.
Our 170\micron\ sources are slightly more distant than theirs although
their flux limit and their cumulative number density down to this limit are 
similar to ours. 
Their
radio survey is shallower using a larger observing beam, and 
these and other aspects of their source identification procedure
might have introduced a bias toward optically bright foreground
sources.  

\begin{figure}[htbp]
  \plotone{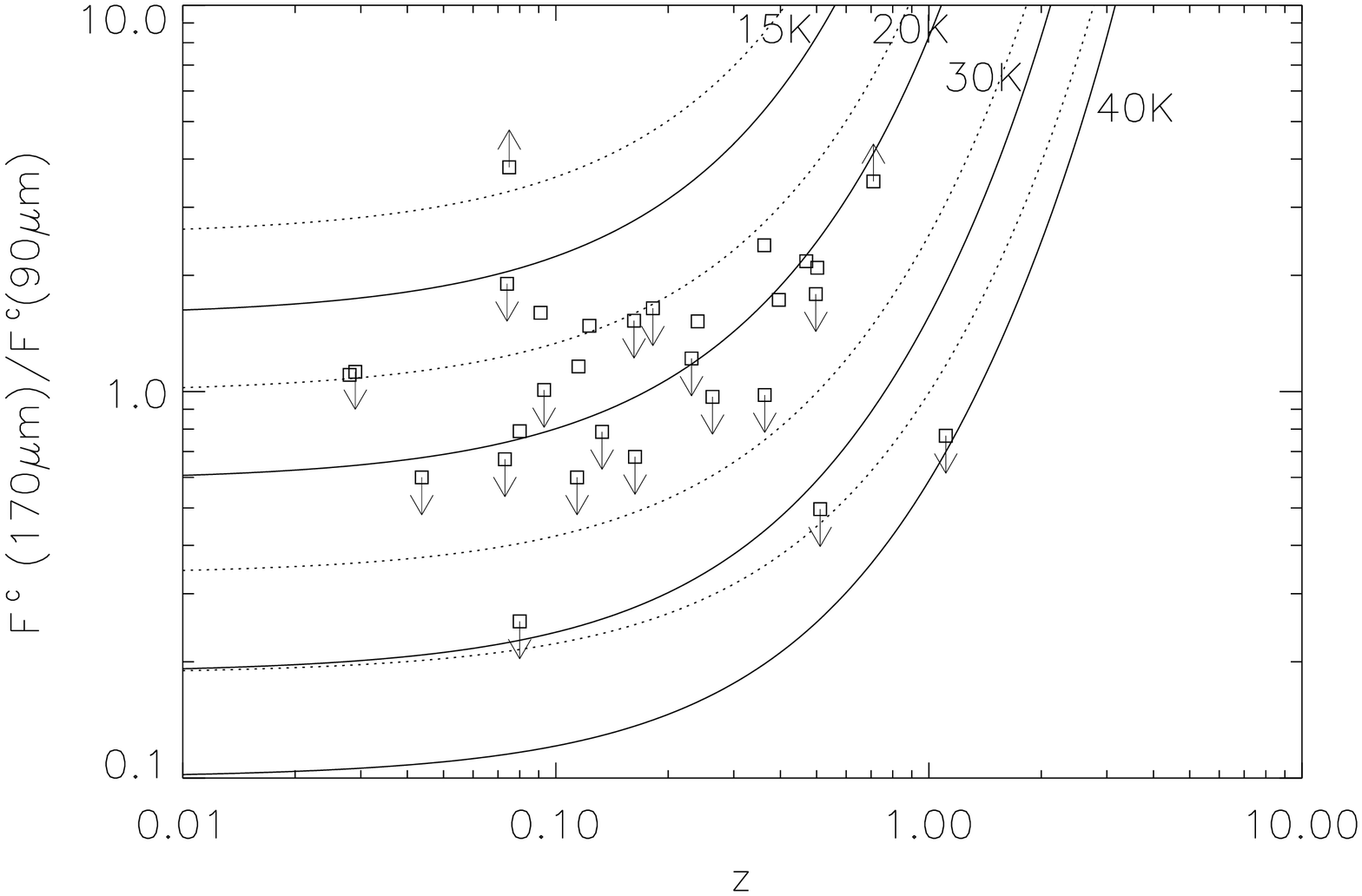}
  \caption{Relation of redshift versus flux ratio, 
   $F^C(170\mu \mathrm{m})/F^C(90 \mu \mathrm{m})$.  Solid and dotted lines 
   show the expectation from a single temperature blackbody multiplied by an 
   $\lambda^{-\beta}$ emissivity. The calculations are made for temperatures of
   15K, 20K, 30K and 40K. Solid lines plot the flux ratios for an 
   $\lambda^{-2}$ emissivity, and dashed lines an $\lambda^{-1}$ emissivity.}
  \label{fig:temp}
\end{figure}

The redshift versus IR color relation is plotted in Figure \ref{fig:temp} 
for our $ISO$ FIR sources together with expectation from greybodies 
with $\lambda ^{-1}$ and $\lambda ^{-2}$ emissivities at $z = 0-10$. 
Most of $ISO$ FIR sources with 170\micron~detection have a dust 
temperature ranging from 20-30 K for $\lambda ^{-1}$ and 15-25 K for
$\lambda ^{-2}$, which is consistent with the 170\micron/90\micron\ 
color temperature 
distribution of 74 ELAIS sources reported by \citet{heraudeau}. 


The FIR flux, $F_{FIR}(40\mu\mathrm{m}-500\mu\mathrm{m})$, can be
estimated 
from the gray body fitting with the observed $F^C(90\mu m)$ and
$F^C(170\mu m)$, the temperature from Figure \ref{fig:temp}  
and the assumption of $\lambda ^{-2}$ emissivity.
Adopting the different dust emissivity, $\lambda ^{-1}$, the FIR flux,
$F_{FIR}$, increase 15 percents on average. This assumption will not
change the main conclusion of this paper.   
It should be noted that
the detection limits (43 mJy at 90\micron~ and 102 mJy at 170\micron~
after the correction for the flux bias) are substituted into the 
undetected band for the objects detected only in one band.
 
\begin{figure}[htbp]
  \plotone{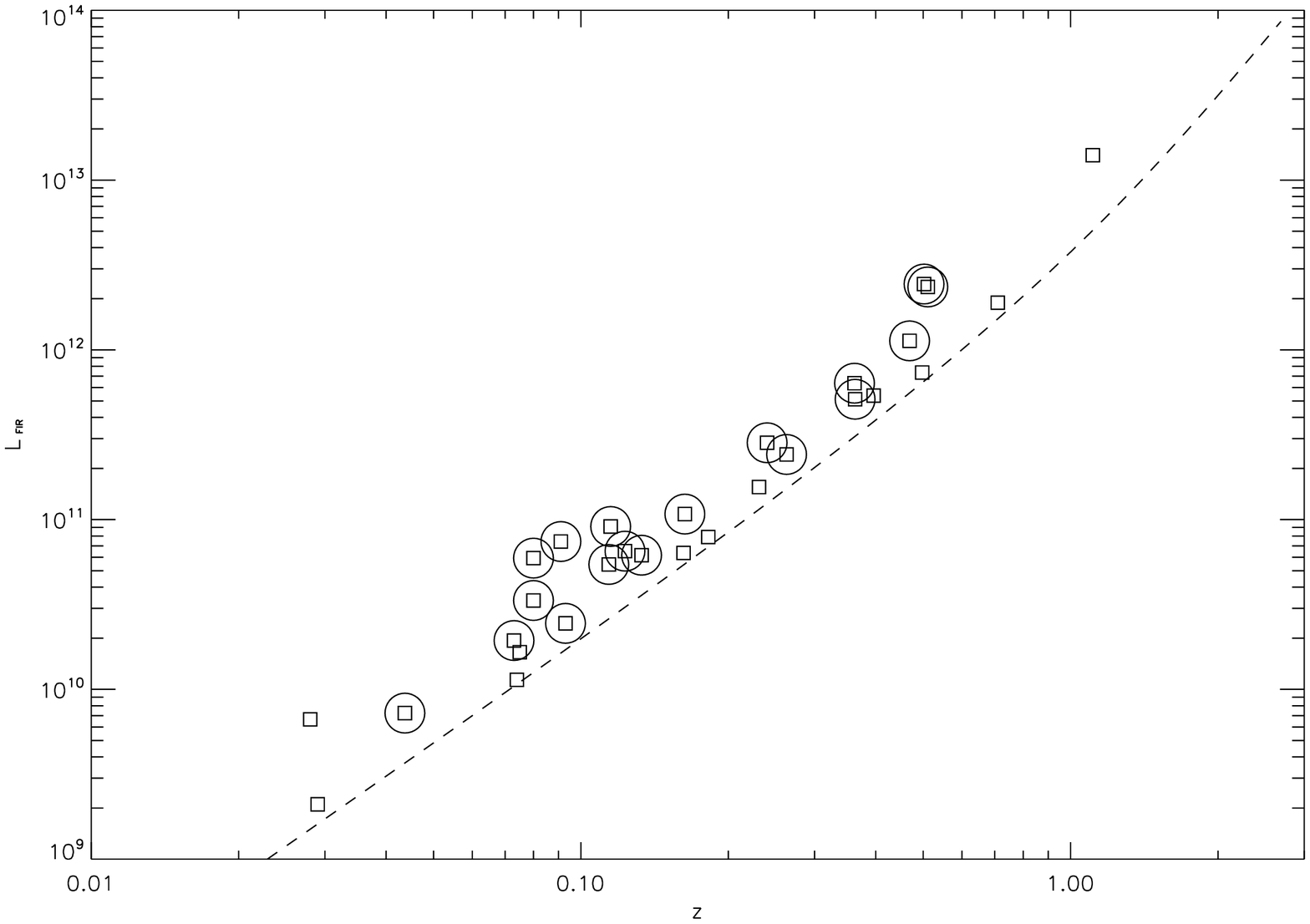}
  \caption{Relation of $L_{FIR}$ versus redshift. 
    Squares with no circle denote sources which are not used to derive 
    the luminosity function of \ISO~FIR sources, because they are
    either too faint 
    or outside the redshift bins. The dashed line presents our
    detection limits of the \ISO~FIR observations.}
  \label{fig:lir}
\end{figure}

The FIR luminosity, $L_{FIR}$, is then obtained as,
\begin{equation}
  \label{eq:lfir}
  L_{FIR}=4\pi D_L^2 \times F_{FIR},
\end{equation}
where $D_L$ is the luminosity distance. The resultant FIR luminosity
is given in Table \ref{tab:id} and plotted in Figure \ref{fig:lir} 
as a function of redshift. Our sample consists of 24 sources with  
$L_{FIR}< 10^{12} \mathrm{L}_{\odot}$, four with 
$L_{FIR}= 10^{12-13} \mathrm{L}_{\odot}$, and one with 
$L_{FIR}> 10^{13} \mathrm{L}_{\odot}$. 
In the $IRAS$ bright source catalog, only six out of 324 sources have 
$L_{FIR} > 10^{12} \mathrm{L}_{\odot}$ \citep{soifer}. Thus, a fraction of 
$L_{FIR} > 10^{12} \mathrm{L}_{\odot}$ sources are 10 times greater in
our $ISO$ 
sample than in the IRAS bright source catalog. All the $IRAS$ bright
sources lie 
within $z \le 0.1$, while, only 10 of 29 sources are at $z\sim 0.1$ in
our \ISO~ sample. 
It is noted here that our spectroscopy project is not complete yet, and  
all the sources with $L(90\mu\mathrm{m})/L(R) > 50$ are left unobserved. It is 
generally agreed that sources with $L(90\mu\mathrm{m})/L(R) > 50$ belong to 
a population of ULIRGs with 
$L_{FIR} > 10^{12} \mathrm{L}_{\odot}$. Hence, our sample should contain 
a greater fraction of $L_{FIR} > 10^{12} \mathrm{L}_{\odot}$ sources 
than that derived from our current spectroscopic knowledge.

\clearpage

\clearpage

\clearpage

\clearpage
\clearpage

\subsection{Luminosity functions}


\begin{deluxetable}{lrrr}
  \tablecolumns{4}
  \tablewidth{400pt}
  \tablecaption{Luminosity function of $ISO$ FIR sources.\label{tab:lf}}
  \tablehead{ & \colhead{$z=0.03-0.10$} & \colhead{$z=0.10-0.30$} & \colhead{$z=0.30-0.60$}}
  \startdata
  $L_{FIR}$ $[10^{10} L_{\odot}]$ & 0.7-7.4 & 5.4 - 28 & 51 -240 \\
  N\tablenotemark{a} & 6 &  7 & 5 \\
  $\frac{d\Phi}{d \log L}$\tablenotemark{b} [$\mathrm{Mpc}^{-3}\ \mathrm{dex}^{-1}$] & 
  $(1.9\pm0.9) \times 10^{-3}$ & $(4.6\pm1.3) \times 10^{-4}$ & 
  $(4.6\pm1.3) \times 10^{-5}$ \\
  \enddata
  \tablenotetext{a}{The number of objects used to derive the  space
    densities.}
  \tablenotetext{b}{Error is from the Poisson statistics.}
\end{deluxetable}


Here we derive luminosity functions of our \ISO~FIR sources having 
$F^C(90 \mu \mathrm{m}) \ge 85$ mJy. These new criterion is more
strict than those used to identify sources having $F^C(90 \mu
\mathrm{m}) \ge 43$ mJy or $F^C(170 \mu \mathrm{m}) \ge 102$ mJy. 
The reason for using the new criteria is to control the detection
limit of our \ISO~FIR sources simply and 
to avoid from the large correction for the completeness of the sample.
27 out of 44 galaxies identified with \ISO~FIR sources meet the new
criterion, and their redshifts of 21 sources are available.

The luminosity function, $d \Phi(L_{FIR})/dL_{FIR}$,
(i.e., the volume density of galaxies per unit luminosity range) 
is derived by
following the 
the $1/V_{max}$ method as described in \citet{schmidt} and \citet{eales}.
The volume density of galaxies with luminosity between $L$ and $L + dL$ 
is defined as, 
\begin{equation}
  \label{equ:lf}
  \frac{d \Phi(L_{FIR})}{dL_{FIR}} dL_{FIR}=\sum_j \frac{1}{p(F^C) V_j},
\end{equation}
where $p(F^C)$ is the detection probability for source with corrected flux $F^C$, 
and the summation is over all sources with luminosity between $L$ and $L + dL$ in 
the sample. $p(F^C)$ is obtained by combining the detection rate given in 
Figure~3 by \citet{kawara2} with Equ. (\ref{equ:flux}) for 
transforming the observed flux $F$ to the corrected $F^C$.
$V_j$ is defined as,  
\begin{equation}
  V_j=
   \int_{\Omega}\int_{z_{min}}^{z_{max}}\frac{d^2V}{d\Omega
    dz} dz d\Omega. 
\end{equation}
where $\Omega$ is the solid angle in this survey and $d^2 V/d \Omega dz$ is
the comoving volume element. $z_{max}$ is the maximum redshift defined by the 
limiting fluxes at the faintest end, namely, $F^C(90 \mu
\mathrm{m})=85$ mJy, 
while $z_{min}$ is the minimum redshift
by the flux limits at the brightest end which is set to 1 Jy for 90\micron.
The luminosity functions are derived for three redshift bins;  $z=0.03-0.10$,
$z=0.10-0.30$ and $z=0.30-0.60$. These redshift bins correspond to the
different FIR
luminosities, $L_{FIR}= 0.7-7.4 \times 10^{10} \mathrm{L_{\sun}}$, 
$L_{FIR}= 5.4-28 \times 10^{10} \mathrm{L_{\sun}}$, and $L_{FIR}=
51-240 \times 10^{10} \mathrm{L_{\sun}}$, respectively. This reason is
that  the luminosity of the
\ISO~FIR sources is following as a function of redshift (See
Fig. \ref{fig:lir}). The numbers of objects in each bin are 6, 7 and
5 and three sources which have $F^C(90\mu\mathrm{m}) \ge 85\mathrm{mJy}$
are out of these 
three bins: two of them are at $z<0.03$ and the other is at $z=1.1$. 
The luminosity function requires the following additional conditions;
$z_{max}$ = $z_u$ for $z_{max} > z_u$ and $z_{min}$ = $z_l$ for 
$z_{min} < z_l$ 
where $z_u$ and $z_l$ are the maximum and minimum redshifts of the specific 
redshift bin. 

\begin{figure}[htbp]
  \plotone{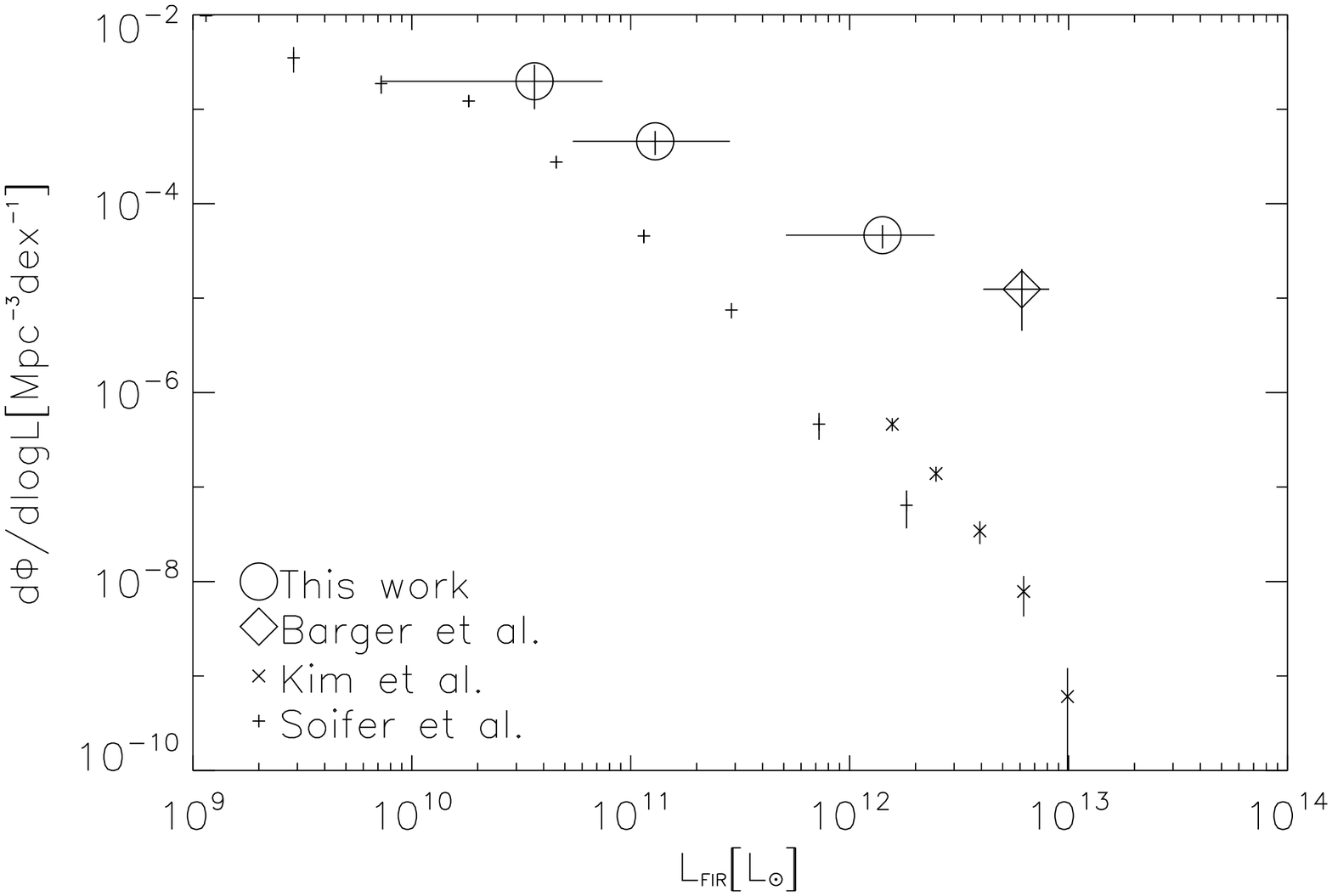}
  \caption{Comparison of the luminosity function (LF) of the $ISO$ FIR source 
   sample with other samples. Circles present the LF of the  $ISO$ FIR source 
   sample which is plotted in the redshift bins;
    $z=0.03-0.10$, $z=0.10-0.30$ and $z=0.30-0.60$ from left to
    right. Horizontal error bars represent the luminosity ranges of the sub-sample
    in the redshift bins and vertical error bars represent the Poisson errors. 
    Pluses show the LF of the \IRAS~bright galaxy sample at $<z> \sim 0.04$ 
    \citep{soifer}, small crosses the \IRAS~1Jy ULIRG sample at $<z> \sim 0.15$ 
     \citep{kim}, and diamonds the SCUBA source sample at $z=1-3$
     \citep{barger2}. In order to present the data of \citet{soifer},
     \citet{kim} and \cite{barger2}, the infrared luminosity,
     $L_{IR}$, is assumed to be $L_{FIR}$. 
  }
  \label{fig:lf}
\end{figure}

The luminosity function of our \ISO~FIR sources sample is given in Table
\ref{tab:lf}.
Figure \ref{fig:lf} compares the \ISO~FIR sources sample with other
galaxy samples. These result is calculated by summing up
$1/p(F^C) V_j$ of the sources 
having the redshift(Equ.~(\ref{equ:lf})). 
Our spectroscopic redshifts are obtained for 78\%(21/27) of this
sample 
and no correction by a factor of 1.3 (27/21) was not applied in
Table \ref{tab:lf} and Figure \ref{fig:lf}. The six \ISO~FIR sources
without redshift are not expected to be the same
populations as our 21 \ISO~FIR sources having redshift, because the
six sources without redshift tend to have higher
$L(90\mu\mathrm{m})/L(R)$ than those in redshift-measured source;
sources without redshift have the range $ 30 < L(90\mu\mathrm{m})/L(R)
< 1400$, while the
20 redshift-measured sources have $  1 \lesssim L(90\mu\mathrm{m})/L(R)
\lesssim 20$ and one source at z$=0.469$ have $L(90\mu\mathrm{m})/L(R) = 43.7$.

The comparison includes the \IRAS~bright galaxy sample by
\citet{soifer} with the mean redshift of $<z> \sim 0.04$, the \IRAS~1Jy 
ULIRG sample by \citet{kim} with $<z> \sim 0.15$, and the 
SCUBA galaxy sample by \citet{barger2} at $z=1-3$. Here we assume that
the infrared luminosity, $L_{IR}(8-1000\mu\mathrm{m})$, is nearly equal
with the FIR luminosity, $\L_{FIR}(40-500\mu\mathrm{m})$. 
The comparison shows a clear trend of the evolutionary effect; at a 
given luminosity, a greater density of galaxies for a higher redshift.
It is particularly clear that there are a rapid evolution in the 
ULIRG population toward high redshift; the space densities are 
$1 \times 10^{-7}\ \mathrm{Mpc^{-3}}$ at $<z> \sim 0.04$ \citep{soifer},
$5 \times 10^{-7}\ \mathrm{Mpc^{-3}}$ at $<z> \sim 0.15$ \citep{kim}, and 
$4.6 \times 10^{-5}\ \mathrm{Mpc^{-3}}$ at $z=0.3-0.6$. In other 
words, relative to the local universe, the space densities are 
$\sim$ 5 times greater at $<z> \sim 0.15$ and $\sim$ 460 times greater
at $z=0.3-0.6$. At the highest end of the FIR luminosity, the space 
densities are 1000 times greater at $z=1-3$ than the local Universe. 

There is uncertainty of flux calibration of our \ISO~FIR sources. Our
flux calibration was done with one IRAS source(UGC 06009) which have
$\sim 50$ percent flux errors\citep{kawara2}. This uncertainty
brought systematic luminosity shift to 0.3 dex in Figure
\ref{fig:lf}. However even if there were the 0.3 dex shift to lower
luminosity, the evolution of \ISO~FIR sources, especially at
$z=0.3-0.6$ still exists. In addition,
we compared the 90\micron~luminosity function of the ELAIS\citep{serjeant}.
\citet{serjeant} presents that the \ISO~90\micron~luminosity function of
the ELAIS is consistent with the IRAS with the assumption of pure
luminosity evolution of $(1+z)^3$. If we apply their pure luminosity
evolution of $(1+z)^3$ in our data, at least our luminosity function at
$z=0.03-0.1$ and $0.1-0.3$ are consistent with that of ELAIS and 
IRAS Bright galaxy sample. On the other 
hand, the luminosity function at $z=0.3-0.6$ 
show an luminosity excess by a factor of $\sim$2 
of that of IRAS Bright galaxy sample after luminosities
reduced by a factor of $(1+z)^3$.
This suggests much stronger
evolution to the \ISO~FIR sources at this redshift range where ELAIS
can barely observe.
It is noted again that 
our spectroscopic observations miss almost all sources with 
$L(90\mu\mathrm{m})/L(R) > 30$, many of which would belong to 
a population of ULIRGs with $L_{FIR} > 10^{12} \mathrm{L}_{\odot}$. 
The space density of \ISO~FIR sources with $L_{FIR} > 10^{12}$ ,which
is derived here, must be significantly underestimated. 

\clearpage
\clearpage

\subsection{Nature of \ISO~FIR sources}
\label{sec:nat}

The fraction of \ISO~FIR sources associated with an AGN  
is estimated from optical emission lines, radio continuum emission, and X-ray 
activity. The excitation diagnosis by \citet{veilleux} is applied to 
the spectra of the \ISO~FIR sources
for the optical spectral classification. One object (1EX030) is a quasar with 
$L_{FIR} = 10^{13.1} L_{\sun}$ at $z = 1.11$.  Two objects
(1EX025 \& 1NW133) are type II Seyferts or LINERs with 
$L_{FIR} = 10^{10.8-10.9} L_{\sun}$ at $z = 0.11-0.16$.  Two objects 
(1EX026 \& 1EX076) are HII galaxies with $L_{FIR} = 10^{9.4-10.4} L_{\sun}$ at 
$z = 0.03-0.07$.  The remaining galaxies are left unclassified because of the
insufficient wavelength coverage, spectral resolution, or quality of the
spectrum. Optical images
indicate that four galaxies and possibly another two are interacting
systems. Their optical properties are summarized in Table \ref{tab:id}. 

\begin{figure}
  \plotone{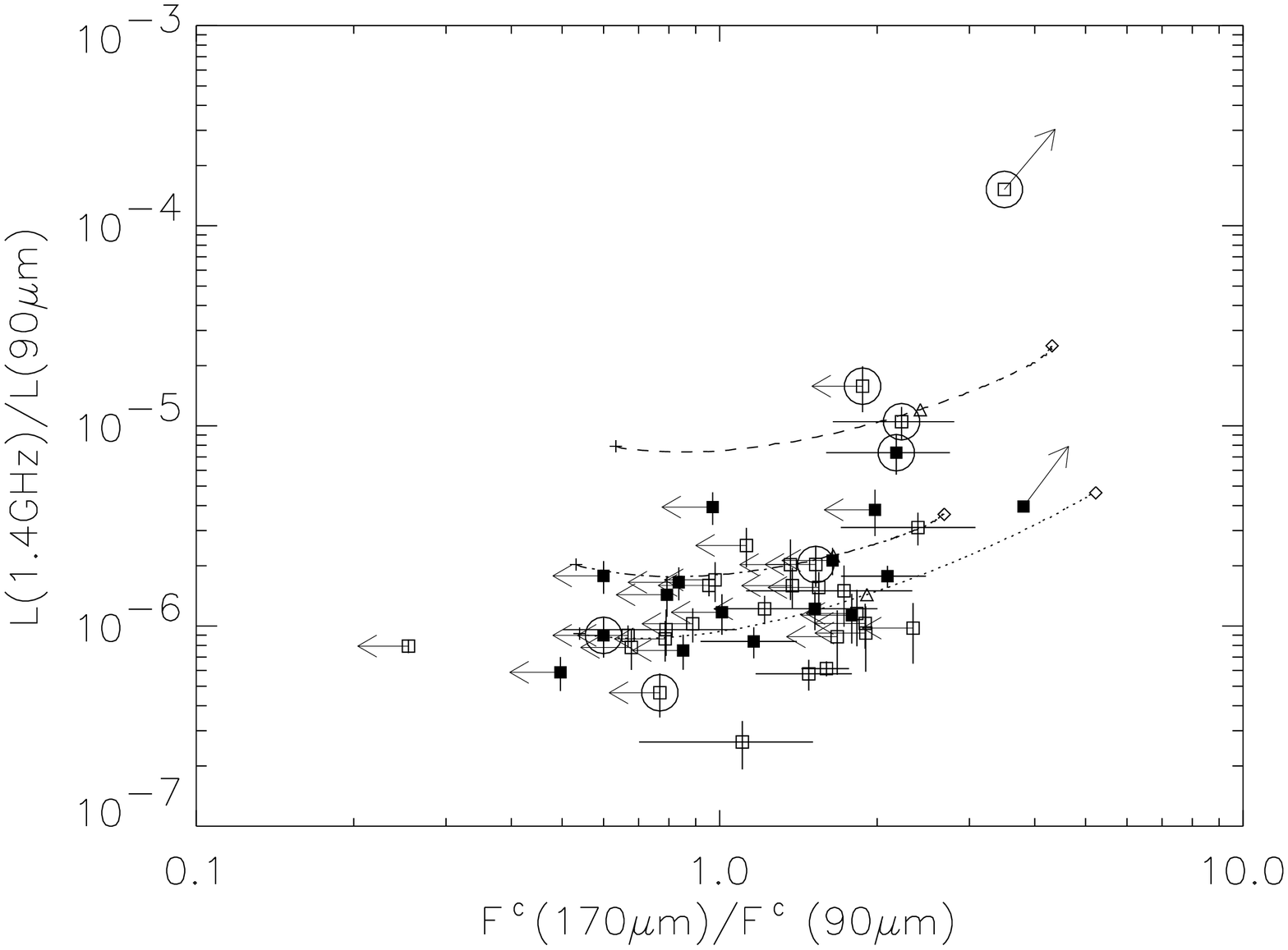}
  \caption{Relation of $F^C(170\mu m)/F^C(90\mu m)$ versus $L(1.4 GHz)/L(90\mu m)$.
   Open squares show sources in the LHEX field while filled squares present 
   those in the LHNW field. Seven large circles indicate galaxies showing AGN 
   signatures. Dashed, dotted and dash-dotted lines
   represent the loci of redshiftted SEDs of NGC~6240, Arp~220, and M82,
   respectively. The crosses, triangles, and diamonds on the loci indicate 
   z=0, 1.0 and 2.0. 
    \label{fig:radio}} 
\end{figure}

The relationship between 1.4 GHz and FIR emission is examined on the 
$L(1.4 GHz)/L(90\mu\mathrm{m})$ versus 
$F^{C}(170 \mu\mathrm{m})/F^{C}(90 \mu\mathrm{m})$ diagram as shown in 
Figure~\ref{fig:radio}\footnote{$L(1.4 GHz)$,
  $L(90\mu\mathrm{m})$ and $L(R)$ are monochromatic luminosities and
  are defined by $4\pi D_L^2 \nu
  f_{\nu}$ where $D_L$ denotes the luminosity distance to the object.}.
Open and filled squares represent sources lying in the LHEX and LHNW
fields, respectively. The dashed and dotted lines are the redshift loci 
of two different types of ULIRGs, NGC~6240 (with an AGN) and Arp~220 
(without an AGN) \citep[e.g.][]{smith},
while the dash dotted line shows that of starburst galaxy M82. 
This figure indicates that most of the \ISO~FIR sources are pure
star forming or star-formation dominated galaxies as they lie near 
the loci of Arp~220 and M82.  Three optically classified AGNs,
1EX030, 1EX025 \& 1NW133, also appear in the same area occupied
by star-formation dominated objects.
Three FIR sources, 1EX100, 1EX047/2EX036, and 1NW272/2NW009, are found 
near the locus of NGC~6240. 1NW272/2NW009 is classified as type II AGN with 
$L_{FIR} = 10^{12.2} L_{\sun}$ at $z = 0.47$.  No spectroscopic
data are available for the two other sources.  
The source 2EX015, which has the highest $L(1.4 GHz)/L(90\mu\mathrm{m})$, 
is identified with a powerful radio galaxy at $z = 0.710$.

As discussed so far, our sample contains at least seven galaxies
hosting an AGN and five ULIRGs. Three AGN-host galaxies have 
FIR luminosity characteristic of ULIRGs.  Thus, 60\% (3/5) of our 
ULIRGs are AGN galaxies, in agreement with that found in 
the IRAS sample \citep{sm}, and this suggests that the fraction of AGN 
galaxies with the ULIRG luminosity does not change much from the local
Universe to  $z \sim 0.5$. 
The deep X-ray survey have been conducted in the LHEX field by 
ROSAT and XMM-Newton satellites \citep[e.g.][]{lehmann, mainieri}.
If our seven AGNs have the X-ray to FIR luminosity ratio 
$L_{X}/L_{FIR}$ similar to those of AGN-associated ULIRGs such as NGC~6240 and 
Mrk~231, $2 \times 10^{-15}$ erg cm$^{-2}$ s$^{-1}$ and 
$6 \times 10^{-15}$ erg cm$^{-2}$ s$^{-1}$ are expected at the 0.5-2 keV and 
2-10 keV energy bands, respectively, for objects having 100 mJy at 100\micron.
Out of our seven AGN galaxies, only one source, 1EX030, identified as
a quasar, has been detected in the X-ray. 
Two undetected sources lie within the ROSAT deep survey area with a flux 
limit of $10 \times 10^{-15}$ erg cm$^{-2}$ s$^{-1}$, and it is not 
surprising that these galaxies were not detected by ROSAT. 
The powerful radio galaxy 2EX015 at $z = 0.710$ and another
AGN candidate 1EX110 (which has no optical spectrum yet) are found 
within the XMM-Newton survey area 
with a limiting flux of $1.4 \times 10^{-15}$ erg cm$^{-2}$ s $^{-1}$ in the 
2-10 keV energy bands. The X-ray non-detection of these two objects may be
attributed to a Compton thick absorber with a column density 
of $3\times 10^{24}$ cm$^{-2}$, which is equivalent to $A_V \sim 1200$ mag. 
The two remaining AGN sources are outside the X-ray survey area.  
It is interesting that the fraction of AGN host galaxies detected 
by the deep X-ray surveys is rather small.  This is also a nice
demonstration that the
combination of FIR and radio observations offers a powerful
tool to find heavily obscured AGNs which might be difficult to be
found in the X-ray.

\begin{figure}
  \plotone{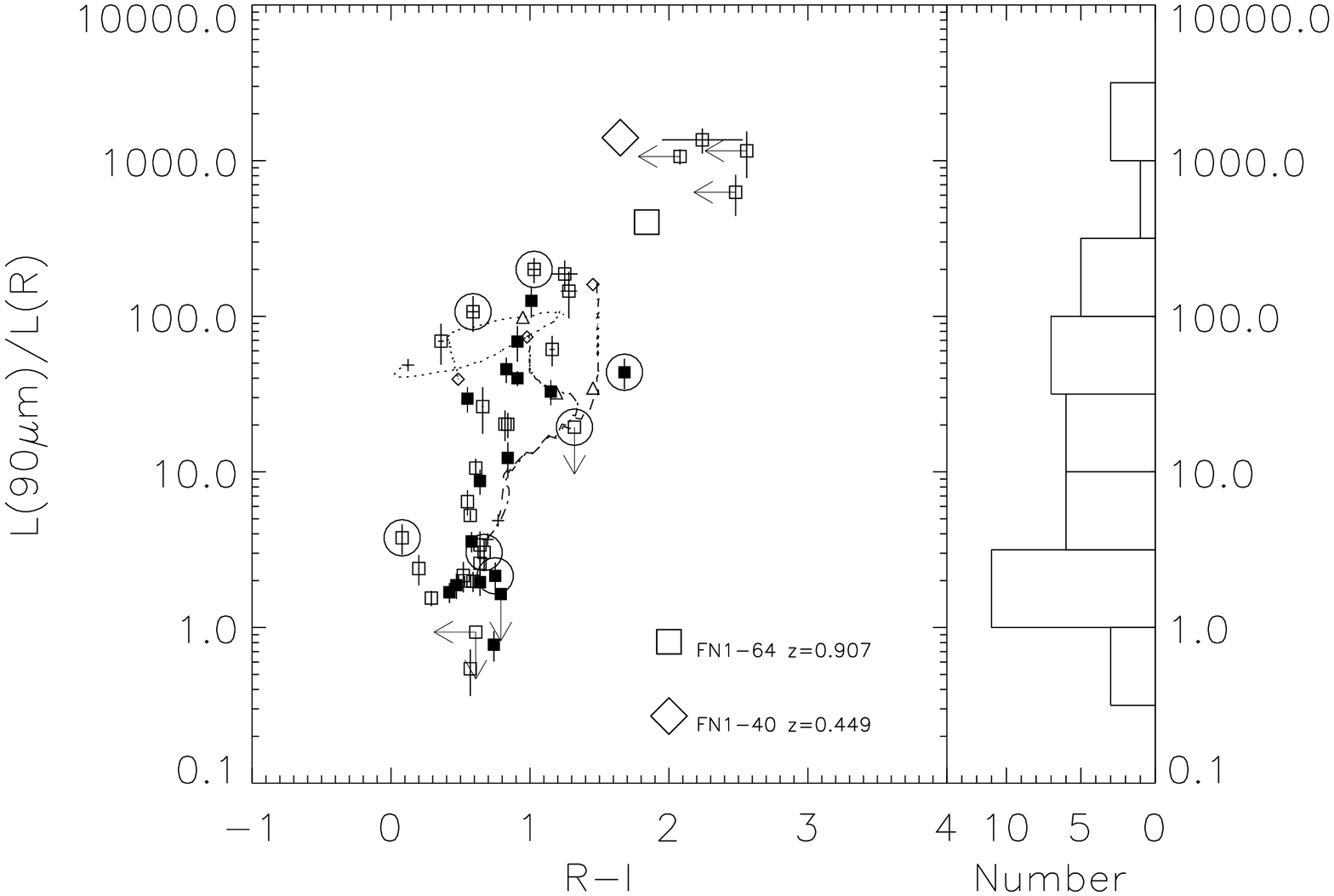}
  \caption{(left) 90\micron~luminosity relative to the optical $R$-band
    luminosity plotted against optical $R-I$ color. Symbols are the
    same as in Figure~\ref{fig:radio} but with
    a large square and diamond which show the infrared luminous
    galaxies, FN1-40 and FN1-61, from the FIRBACK survey
    \citep{chapman}. 
    (right) The number distribution as a function of $L(90\mu m)/L(R)$ is 
   shown to the right.  \label{fig:cratio}}
\end{figure}

The 90\micron~ luminosity relative to the optical R-band luminosity, 
$L(90\mu\mathrm{m})/L(R)$, is plotted against optical color , $R-I$, in
Figure \ref{fig:cratio}. Two large symbols (square and diamond) are 
FIRBACK 170\micron~sources, 
FN1-40 at $z=0.449$ and FN1-64 at $z=0.907$ \citep{chapman}.  
The loci expected from the SEDs of NGC~6240, Arp~220, and M82 
at $z=0-2$ are shown using dashed and dotted lines.
This figure demonstrates that the \ISO\ succeeded in detecting 
objects with a wide range of
$L(90\mu\mathrm{m})/L(R)$ ratio, from 1 to 1000, and show a bimodal
distribution of $L(90\mu\mathrm{m})/L(R)$ for our \ISO~FIR
sources. One peak with $1 < L(90\mu\mathrm{m})/L(R) < 3$ appears to
represent the population of normal star forming galaxies in nearby universe
because of their brightness in the optical. The other peak with
FIR excess represents infrared dominant sources like Arp~220 whose luminosity
is nearly entirely reprocessed by dust. 
In \citet{rodighiero3} which reduced same ISO data in the LHEX and
made identification with radio and 15\micron~sources(See Appendix for
details), 
nine of 11 sources were fitted
with the SEDs of M82 or M51 which are FIR moderate galaxies, while two
others have a Arp~220 SED. The number of Arp~220-like sources seem to
be small, however their SED-template fitting was done for the sources
whose redshifts are available. Thus there might be a bias to
optically bright sources. 

Figure \ref{fig:cratio} also shows that there is possibly a third
class of objects with an extreme excess of FIR luminosity,
$L(90\mu\mathrm{m})/L(R) > 500$.  Although 
ULIRGs are known to have the largest intrinsic excess of the FIR luminosity 
relative to the optical luminosity, their extreme optical and FIR colors
cannot be adequately explained by simply redshifting the observed
SEDs of ULIRGs or starburst galaxies.
Therefore, the five such objects, one from FIRBACK and four from 
our survey, may represent a new population of extreme FIR-excess
galaxies. 
If the extreme values of $L(90\mu\mathrm{m})/L(R)$ are intrinsic, the 
small optical luminosity may imply that their stellar system has not 
been fully developed or their entire stellar structures are heavily 
obscured by dust.

Four such objects in the Lockman Hole fields are located on the outside of
the field where the submillimeter and millimeter observations with the
SCUBA, Bolocam and MAMBO\citep{scott,greve,laurent} were performed.
In future, the measurements in such wavelength will give
the additional information to them. In addition, one of the Spitzer
Space Telescope Legacy Programs, SWIRE\citep{lonsdale}, cover these
fields in the wavelength from 3.6\micron~to 160\micron~and will
release their data and help to understand four such objects.  

\clearpage

\clearpage
\clearpage

\section{Summary}

By exploiting the FIR-radio correlation, we have performed the
Likelihood-Ratio analysis to identify the far-infrared
sources that have been found in an area of $\sim 0.9$ deg$^2$ during
the \ISO~deep far-infrared survey in the Lockman Hole. 
New observations have been conducted to construct the catalogs of radio
and optical objects, which include a deep VLA 1.4 GHz observations,
optical $R-$ \& $I-$band imaging on the Subaru 8m and UH2.2m
telescopes, and optical spectroscopy on the Keck II 10m and WIYN 3.5m
telescopes.  
A summary of the results presented in this paper is as follows:
\begin{itemize}
\item  Our samples of 116 and 20 sources are
  selected with the criteria of $F^C(90\mu\mathrm{m}) \ge 43$ mJy and
  $F^C(170\mu\mathrm{m}) \ge 102$ mJy, respectively. 
  Our 1.4 GHz radio sample includes a total of 463 sources. 
\item In order to remove positional coincidence by chance,
  we calculate the Likelihood-Ratio and their reliability. 
  As a results, 44 FIR sources are identified with radio sources.
\item Optical identification of the 44 FIR/radio association is
  conducted using accurate radio positions.  The dispersion in
  the difference between the radio and optical position is
  0\farcs6.  
\item Dust temperature derived from the FIR color ranges between 15 and 30 K. 
\item Spectroscopic redshifts have been obtained for 29 out of 44 identified
  sources. There are no significant differences in the redshift
  between 170\micron-detected sources and 90\micron-only-detected
  sources.
\item  24 (out of 29) FIR galaxies with redshifts have $L_{FIR} <
  10^{12}\mathrm{L}_{\odot}$, four with $L_{FIR} =
  10^{12-13}\mathrm{L}_{\odot}$, and one $L_{FIR} >
  10^{13}\mathrm{L}_{\odot}$. 
\item The luminosity functions are calculated using the $1/V_{max}$
  method. The
  space density of the our sample galaxies at $z = 0.3-0.6$ is $4.6\times
  10^{-5}$ Mpc$^{-3}$, which is 460 times as high as that at the local
  universe. A rapid evolution in the ULIRG population is suggested. 
\item Most of \ISO~FIR sources have $L(1.4 GHz)/L(90\mu\mathrm{m})$
  similar to
  star-forming galaxies Arp~220 and M82, indicating star formation is
  the dominant source of their FIR and radio emission in these
  galaxies.
\item Our FIR sample contains at least seven AGNs, which are
  classified either from optical emission lines,
  excess in radio emission, or X-ray activity.
\item 60\%(3/5) of our ULIRGs are AGN galaxies, implying that the
  percentage of AGN galaxies with ULIRG luminosity does not change
  significantly between $z=0$ and $z\sim 0.6$.
\item Five of the seven AGN galaxies are within the ROSAT X-ray survey
  field, and two are within the XMM-Newton survey fields. X-ray
  emission has been detected in only one source, 1EX030, which is
  optically classified quasar. If our AGN galaxies have
  $L_{X}/L_{FIR}$ similar to NGC~6240, a ULIRG hosting
  an AGN, then none of our AGN galaxies would have been detected by ROSAT. The
  non-detection by XMM-Newton 2-10 keV band implies a very
  thick absorption column density of $3 \times 10^{24} \mathrm{cm}^2$
  or $A_V \sim 1200$
  mag obscuring the central source of the two AGN galaxies.  The
  combination of FIR and radio observations would provide a powerful
  tool to find heavily obscured AGNs which might be difficult to be
  found in the X-ray.
\item Several sources show an extreme FIR luminosity
  relative to 
  the optical $R$-band, $L(90\mu\mathrm{m})/L(R) > 500$. Such extreme values 
  cannot be explained from the redshifted SEDs of ULIRGs and may imply a new 
  population of galaxies where an extreme activity of star formation in an 
  undeveloped stellar system. If so, we might be looking at the formation of 
  bulges or ellipticals.
\end{itemize}

\acknowledgments
We wish to thank the staff of the Subaru Observatory,
NOAO, NRAO, Keck Observatory and the UH88 telescope for their assistance and
hospitality during the several observing runs in which collected data
for this paper. This research made use of the NASA/IPAC Extragalactic
Database (NED) which is operated by the Jet Propulsion Laboratory,
California Institute of Technology, under contract with the National
Aeronautics and Space Administration. This paper is based on
observations with \ISO, an ESA project with instruments funded by ESA
member states and with the participation of ISAS and NASA.

\appendix
\section{Comparison with Rodighiero et al.\label{sec:app}}

\citet{rodighiero1} and \citet{rodighiero2} have reduced our 90\micron~data in the
LHEX and LHNW fields, by using their own method, a parametric
algorithm that fits the signal time history of each detector pixel. 
They then identified the singularities induced by cosmic
ray impacts and by transient effects in the detectors and extract real
sky sources. The numbers of their sources with signal-to-noise ratio
S/N $>$ 3 are 36 and 28 in the LHEX and LHNW, respectively, while our
catalogs consist of 116 and 107 sources having S/N $>$ 3\citep{kawara2}. 
Both catalogs are made with the different method.
Thus the reason of this difference is expected to be in the reduction
and source extracting methods, but is unclear. 

\begin{figure}
  \plotone{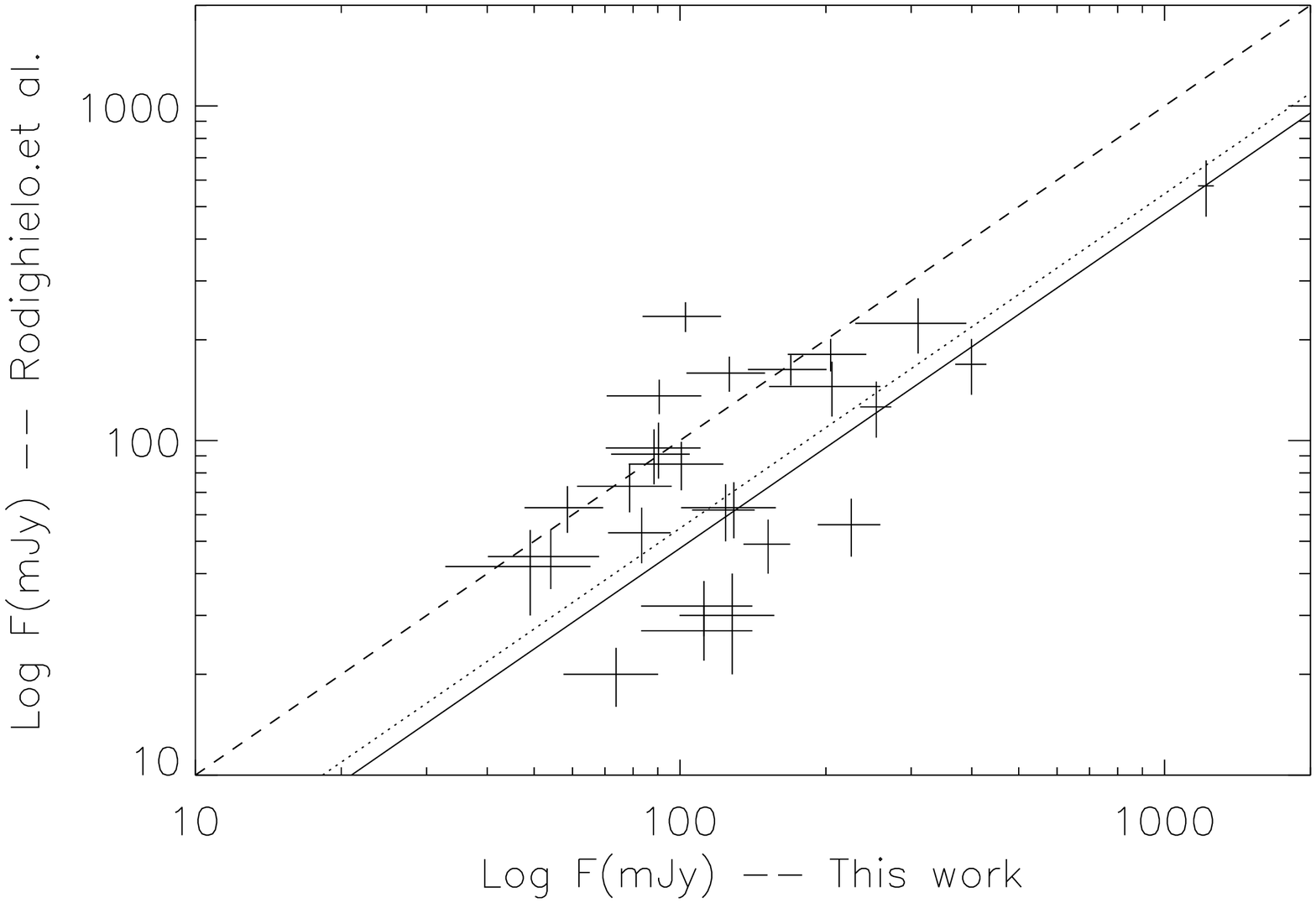}
  \caption{Comparison of our 90\micron~fluxes with
    \citet{rodighiero1} and \citet{rodighiero2}.  
    A solid line denote the ratio of 2.1 which comes from the flux
    difference of the IRAS source. The mean ratio of ours with those in
    \citet{rodighiero1} and \citet{rodighiero2} is 1.8(dotted
    line) and a dashed line presents the ratio for 1.0.
    \label{fig:crossid}}
\end{figure}

In Table \ref{tab:crossid}, their 15 LHEX and 11 LHNW sources have
a counterpart within 30\arcsec~in our catalogs. Figure
\ref{fig:crossid} compares the Rodighiero et al. flux with ours. One
IRAS source, UGC 06009, is reported as ex003(577 $\pm$ 110 mJy) in
\citet{rodighiero1}, while we used this source for the flux
calibration of the IRAS 100\micron~measurement(1218 mJy). The
difference of a factor of 2.1 between ours and Rodighiero et al. might
be expected and the mean ratio of ours with those in
\citet{rodighiero1} and \citet{rodighiero2} is 1.8. 
As shown in Figure \ref{fig:crossid}, there are some sources in our
catalog having $\sim$100 mJy which show
much fainter fluxes in theirs, and these sources could make the
deviation of the flux ratio bigger.

\citet{rodighiero3} have done the identification work of
their 36 90\micron~sources\citep{rodighiero1} in the LHEX field using
the association with the shallow radio\citep{ruiter} and
15\micron~sources
\citep{fadda,rodighiero4}. 
17 of them are identified with radio and 15\micron~
sources, all of which are optically identified. 14 of the 17 optically
identified
sources are found in our catalog\citep{kawara2} and we made same
optical identification in this paper. However
only one(ex002/1EX100) of their 19 optically unidentified sources is found
in \citet{kawara2} and is succeeded to identify in this work. 
It is unclear why most of their unidentified sources are not found in
\citet{kawara2}. 

\begin{deluxetable}{llrllr}
\tablecaption{Identification with sources in Rodighiero et al.\label{tab:crossid}}
\tabletypesize{\scriptsize}
\tablewidth{300pt}
\tablecolumns{6}
\tablehead{\multicolumn{3}{c}{Rodighiero et al.} &
  & \multicolumn{2}{c}{Kawara et al.}\\ \cline{1-3} \cline{5-6}
\colhead{ID\tablenotemark{a}} & \colhead{Name} & \colhead{Flux(mJy)} & &
  \colhead{Name} & \colhead{Flux(mJy)\tablenotemark{b}}
}
\startdata
ex001 & LHJ105324+572921  &     95 $\pm$  18 & & 1EX026 &   90 $\pm$  20\\
ex003\tablenotemark{c} & LHJ105349+570716  &    577 $\pm$ 110 & & 1EX023 & 1218 $\pm$  44\\
ex004 & LHJ105052+573507  &    224 $\pm$  42 & & 1EX041 &  310 $\pm$  80\\
ex007 & LHJ105300+570548  &    169 $\pm$  32 & & 1EX062 &  399 $\pm$  29\\
ex008 & LHJ105041+570708  &    126 $\pm$  24 & & 1EX048 &  254 $\pm$  19\\
ex009 & LHJ105254+570816  &    145 $\pm$  27 & & 1EX269 &  206 $\pm$  53\\
ex012 & LHJ105113+571415  &     91 $\pm$  17 & & 1EX081 &   88 $\pm$  16\\
ex019 & LHJ105318+572130  &     63 $\pm$  12 & & 1EX179 &  129 $\pm$  29\\
ex020 & LHJ105127+573524  &     49 $\pm$   9 & & 1EX076 &  152 $\pm$  17\\
ex022\tablenotemark{d} & LHJ105223+570159  &     32 $\pm$   6 & & 1EX148 &  112 $\pm$  29\\
ex025 & LHJ105206+570751  &     62 $\pm$  12 & & 1EX034 &  124 $\pm$  18\\
ex027 & LHJ105328+571404  &     53 $\pm$  10 & & 1EX125 &   83 $\pm$  12\\
ex028\tablenotemark{d} & LHJ105226+570222  &     27 $\pm$   5 & & 1EX148 &  112 $\pm$  29\\
ex029 & LHJ105132+572925  &     45 $\pm$   9 & & 1EX100 &   54 $\pm$  14\\
ex036 & LHJ105323+571451  &     20 $\pm$   4 & & 1EX126 &   74 $\pm$  16\\
nw001 & LHJ103521+580034 &  235 $\pm$  24 & & 1NW055 &  103 $\pm$  19\\
nw002 & LHJ103606+574715 &  163 $\pm$  17 & & 1NW023 &  169 $\pm$  31\\
nw003 & LHJ103604+574815 &  181 $\pm$  20 & & 1NW092 &  205 $\pm$  38\\
nw006 & LHJ103515+573330 &  136 $\pm$  16 & & 1NW025 &   91 $\pm$  20\\
nw008 & LHJ103318+574925 &  159 $\pm$  19 & & 1NW021 &  126 $\pm$  23\\
nw011 & LHJ103530+573139 &   73 $\pm$  12 & & 1NW043 &   79 $\pm$  17\\
nw012 & LHJ103409+572715 &   63 $\pm$  10 & & 1NW221 &   59 $\pm$  11\\
nw016 & LHJ103610+574330 &   85 $\pm$  14 & & 1NW022 &  101 $\pm$  22\\
nw020 & LHJ103249+573719 &   56 $\pm$  11 & & 1NW192 &  226 $\pm$  33\\
nw026 & LHJ103316+573136 &   30 $\pm$  10 & & 1NW031 &  128 $\pm$  28\\
nw027 & LHJ103327+574539 &   42 $\pm$  12 & & 1NW114 &   49 $\pm$  16\\
\enddata
\tablenotetext{a}{Source numbers with prefix ex or nw. The number
  follows the order of the appearance in their catalogs. The prefix,
  ex , presents sources in \citet{rodighiero1} and the prefix, nw,
  presents sources in \citet{rodighiero2}.}
\tablenotetext{b}{These fluxes are corrected with
  Equ.~(\ref{equ:flux}). }
\tablenotetext{c}{The IRAS source, UGC 06009, which is used for our
  flux calibration.}
\tablenotetext{d}{ex022 and ex028 may be blending with each other.}
\end{deluxetable}

\end{document}